# Analytic results in conformal field theory

Proceedings of the 28$^e$ rencontre Itzykson

IPhT Saclay, 10-12 September 2024

---

**Speakers:** Balt van Rees, Emilio Trevisani, Colin Guillarmou, Petr Kravchuk, Ingo Runkel, Xin Sun, Philine van Vliet, Bernardo Zan, Volker Schomerus, Gregory Korchemsky, Yifei He

*Speakers whose names are underlined contributed a text to this collection.*

**Scientific organizers & proceedings editors:** Dalimil Mazac and Sylvain Ribault

**Conference website:** https://indico.in2p3.fr/event/32728/

**Acknowledgements:** The conference was supported by IPhT Saclay, and by the ERC Starting Grant 853507.

---

Since the 1980s, many exact results have been discovered in $2d$ CFT, from critical exponents to correlation functions to complete solutions of certain models. In $d > 2$, there is a wealth of numerical results as well as promising analytic approaches, but comparably fewer exact answers.

The aim of this conference was to review the most promising analytic methods and results in CFT in any dimension. In particular we tried to understand to which extent the $2d$ methods can be extended to $d > 2$, and what is missing to exactly solve $d > 2$ CFTs.

---

# Contents

# Foreword by the editors

The study of conformal field theory (CFT) has been one of the most fertile areas of theoretical and mathematical physics over the past four decades. Since the 1980s, two-dimensional conformal field theories have provided a wealth of exact results, ranging from critical exponents and correlation functions to the complete solution of entire classes of models. These advances were made possible by the rich mathematical structure of 2d CFT, which offers a high degree of solvability and has inspired developments across statistical mechanics, string theory, probability theory, and pure mathematics.

In dimensions greater than two, the situation is at once more challenging and more tantalizing. A wide array of numerical results have been obtained through the conformal bootstrap method, and significant analytic progress has been achieved in special settings. However, exact solutions remain scarce. This contrast between the abundance of exact results in 2d and the comparative scarcity in higher dimensions motivates the question:

> **To what extent can the methods of 2d CFT be extended to higher dimensions, and what new tools are required to exactly solve higher-dimensional conformal theories?**

It was in this spirit that the conference Exact Results in Conformal Field Theory, held at the Institut de Physique Théorique on 10-12 September 2024, brought together physicists and mathematicians working on conformal field theory in both two and higher dimensions. The aim was to review promising analytic approaches, exchange ideas across communities, and highlight connections between physics and mathematics that may prove essential in the search for new exact results.

The talks spanned a wide range of topics: from conformal correlators in higher dimensions, to the analytic structure of multi-particle spectra, to the bootstrap analysis of nonlocal models such as the long-range Ising model. Connections to random matrices, integrability, supersymmetry, and quantum groups were also explored, as well as rigorous probabilistic constructions of 2d CFT and recent breakthroughs in percolation theory. Together, these talks painted a vivid picture of a field that is simultaneously consolidating well-established methods and venturing into new territory.

The present volume collects the written contributions associated with the conference. They reflect the diversity of approaches and interests represented at the meeting, while also highlighting common themes. **Philine van Vliet** introduces the bootstrap analysis of the long-range Ising model and its role as a laboratory for nonlocal CFTs. **Emilio Trevisani** describes how Parisi–Sourlas supersymmetry can be harnessed to uplift solvable models to higher dimensions, opening the door to an infinite family of exactly solvable CFTs. **Ingo Runkel** presents the role of symmetry and chiral topological field theories in the structural understanding of 2d CFT. **Bernardo Zan** clarifies the appearance of quantum groups as genuine global symmetries in the continuum and their relation to minimal models. **Yifei He** analyzes logarithmic operators in bulk percolation CFTs and resolves longstanding puzzles around the $c \to 0$ limit. **Xin Sun** and collaborators report the exact derivation of the backbone exponent and crossing probabilities in planar percolation, connecting probability, integrability, and conformal field theory. Finally, **Volker Schomerus** constructs $d > 2$ conformal blocks from integrable systems that emerge from 2d WZNW models whose target spaces are conformal groups.

These texts strive to review recent results in a concise and non-technical way, so that they can be understood by scholars from different communities, and contribute to answering our question. The text by Schomerus also includes new results that did not appear previously.

We hope that this collection will serve not only as a record of the conference but also as a resource and inspiration for future work in the field.

# BibTeX code

```
@inproceedings{vli25,
    author = "van Vliet, Philine",
    title = "{Bootstrapping the long-range Ising model}",
    booktitle = "{Proceedings of the 28e rencontre Itzykson:
                Analytic results in conformal field theory}",
    editor = "{Mazac, Dalimil and Ribault, Sylvain}"
    year = "2025",
    eprint = "2510.09868",
    url = "https://arxiv.org/abs/2510.09868",
}

@inproceedings{tre25,
    author = "Trevisani, Emilio",
    title = "{Parisi--Sourlas dimensional uplift}",
    booktitle = "{Proceedings of the 28e rencontre Itzykson:
                Analytic results in conformal field theory}",
    editor = "{Mazac, Dalimil and Ribault, Sylvain}"
    year = "2025",
    eprint = "2510.09868",
    url = "https://arxiv.org/abs/2510.09868",
}

@inproceedings{run25,
    author = "Runkel, Ingo",
    title = "{Symmetry TFT, chiral TFT, and 2d CFT}",
    booktitle = "{Proceedings of the 28e rencontre Itzykson:
                Analytic results in conformal field theory}",
    editor = "{Mazac, Dalimil and Ribault, Sylvain}"
    year = "2025",
    eprint = "2510.09868",
    url = "https://arxiv.org/abs/2510.09868",
}

@inproceedings{zan25,
    author = "Zan, Bernardo",
    title = "{Quantum group symmetric conformal field theories}",
    booktitle = "{Proceedings of the 28e rencontre Itzykson:
                Analytic results in conformal field theory}",
    editor = "{Mazac, Dalimil and Ribault, Sylvain}"
    year = "2025",
    eprint = "2510.09868",
    url = "https://arxiv.org/abs/2510.09868",
}
```

# Bootstrapping the long-range Ising model

**Philine van Vliet**

*Laboratoire de Physique, École Normale Supérieure,*
*Université PSL, CNRS, Sorbonne Université, Université Paris Cité,*
*24 rue Lhomond, F-75005 Paris, France*

*E-mail:* `philine.vanvliet@phys.ens.fr`

Abstract: While the short-range Ising model is well known and well studied, its long-range cousin, the long-range Ising model (LRI), has remained more mysterious. Nevertheless, the critical point of the LRI forms a natural starting point for the study of nonlocal CFTs, and it can be analyzed using a variety of techniques, including the conformal bootstrap.
We give a concise introduction to recent progress on the LRI in two and three dimensions. The LRI can be interpreted as a conformal defect in an auxiliary, free scalar bulk CFT. We find relations between OPE coefficients and constraints on the spectrum that are derived from requirements of analyticity of the correlators and conformal blocks. We show how these can be used in the conformal bootstrap, and in a perturbative setup, and the power they have. These relations are not limited to the LRI, and I will highlight some other interesting setups.

# Contents

# 1 Introduction

The critical point of the Ising model with short-range (nearest-neighbour) interactions is one of the most well-studied conformal field theories (CFTs), and one of the great successes of the nonperturbative conformal bootstrap. In two dimensions, an exact solution on the lattice was found for any temperature by Onsager [1] and later a more elegant solution was constructed by Kaufman [2]. Its critical point was identified as the first unitary minimal model in [3]. In 3d the numerical conformal bootstrap [4] has given the most accurate predictions for scaling dimensions and operator product expansion (OPE) coefficients of low-lying operators to date [5–8]. In contrast, the critical points of the Ising model with long-range interactions remain much more mysterious. The long-range Ising model (LRI) does not have a single fixed point, but has a continuous line of fixed points parametrized by $\mathfrak{s}$, with $p/2 \leq \mathfrak{s} \leq \mathfrak{s}_*$ [9, 10]. Here, $p$ is the dimension of the LRI and $\mathfrak{s}_*$ can be expressed in terms of the critical exponents of the SRI, as we will see later. These fixed points correspond to a set of nonlocal, but unitary CFTs, one for each value of $\mathfrak{s}$. The nonlocal action gives rise to operators with protected dimensions, and an infinite number of relations between OPE coefficients. These exact results turn out to be highly constraining, allowing one to gain an extra order in perturbative computations, and can be used to pinpoint the LRI using the conformal bootstrap. I will show how they can be used together with the numerical conformal bootstrap to constrain scaling dimensions of conformal primaries in the LRI in two and three dimensions. This was originally done in [11], and this work provides a summary of some techniques and results presented in that paper.

There are other ways to study the LRI nonperturbatively besides the conformal bootstrap, such as using Monte Carlo. For the 2d LRI, for $\mathfrak{s} = 1.6$, there are Monte Carlo results [12, 13] determining the critical exponent $\nu$, which is related to the scaling dimension of the first unprotected spin-0 conformal primary. The second study [13] found a slightly higher value for this critical exponent, and while they still overlap within the error, it would be interesting to see if the conformal bootstrap agrees with one and not the other, or is even able to exclude one of the two results.

Studying the LRI also gives us more insight in other nonlocal CFTs. Examples are quantum field theories (QFTs) in anti-de Sitter spacetime (AdS), which are dual to CFTs on the boundary of AdS through the AdS/CFT correspondence. Without the presence of dynamical gravity in the bulk, the CFT on the boundary does not contain a stress-tensor. Another example are defect CFTs, where an extended object or defect breaks the conformal symmetry in the bulk while preserving a lower-dimensional conformal group on the defect. The breaking of translations in the bulk CFT results in the fact that there is no conserved stress-tensor on the defect, and the defect CFT is nonlocal. In many aspects, the LRI is one of the simplest nonlocal CFTs that exist, and hence forms a natural starting point for the study of nonlocal CFTs.

The layout is as follows. In section 2 the long-range Ising model is reviewed, and two different perturbative descriptions for either end of the critical interval of $\mathfrak{s}$, the mean-field theory end and the short-range Ising end, are discussed. Section 3 provides a derivation of the relations between OPE coefficients that follow from the nonlocal nature of the LRI, and how they lead to a protected odd-spin sector. In section 4 the results obtained with the numerical conformal bootstrap are shown, consisting of exclusion plots for the scaling dimension of the first spin-0 conformal primary. We conclude in section 5, and give some future directions and applications.

# 2 The long-range Ising model

The LRI in $p$ dimensions is described by the following hamiltonian:

$$H = -J \sum_{i,j} \frac{\sigma_i \sigma_j}{|i-j|^{p+\mathfrak{s}}} \ , \quad J > 0 \ , \quad \sigma_i = \pm 1 \ , \tag{2.1}$$

where $\sigma_i$ is the spin field on the lattice site $i$. The long-range interaction is parametrized by $\mathfrak{s}$, and we find nontrivial critical behaviour for $p/2 \leq \mathfrak{s} \leq \mathfrak{s}_*$. Both ends of this critical range of $\mathfrak{s}$ admit a weakly-coupled, perturbative description. Close to $p/2$ we can describe the LRI as a mean-fied theory (MFT) with $\phi^4$ interaction, while close to $\mathfrak{s}_*$ the LRI can be interpreted as a generalized free field (GFF) coupled to the short-range Ising model (SRI), which only has nearest-neighbour interactions, in $p$ dimensions. We will elaborate on these descriptions below.

## 2.1 The MFT end

The description of the LRI close to the MFT point was first introduced by Fisher, Ma and Nickel [9]. It is analogous to Wilson and Fisher's perturbative description of the SRI [14] with a local action and $\phi^4$ interaction. The long range interaction now changes this into a nonlocal action. The parameter $\mathfrak{s} = \frac{p+\varepsilon}{2}, \varepsilon > 0$ is defined close to $p/2$ for small values of $\varepsilon$. The action for a generalized free field $\phi$ consists of a nonlocal kinetic term, and a local $\phi^4$ interaction:

$$S = \mathcal{N}_{\mathfrak{s}} \mathcal{N}_{-\mathfrak{s}} \int d^p\tau_1 d^p\tau_2 \frac{\phi(\tau_1)\phi(\tau_2)}{|\tau_{12}|^{p+\mathfrak{s}}} + \int d^p\tau \frac{\lambda}{\sqrt{4!}} \phi^4 \ , \tag{2.2}$$

where the normalisation $\mathcal{N}_{\mathfrak{s}}$ is chosen such that $\phi$ is unit-normalized in the undeformed theory, and $\lambda \sim O(\varepsilon)$ at the (interacting) fixed point. From the action (2.2) it is easy to derive that $\phi$ has scaling dimension

$$\Delta_\phi = \frac{p-\mathfrak{s}}{2} \,. \tag{2.3}$$

Furthermore, using the nonlocal equations of motion (eom), one can show that there exists a *shadow relation* between $\phi$ and $\phi^3$, linking their dimensions such that

$$\Delta_{\phi^3} = \frac{p+\mathfrak{s}}{2} \,. \tag{2.4}$$

These dimensions are protected; they will not get renormalized when considering perturbations in $\varepsilon$. For $\phi$ this is due to the fact that local interactions cannot renormalize a nonlocal kinetic term [15]. The protected dimension of $\phi^3$ can be obtained from $\Delta_\phi$ through the nonlocal eom.

## 2.2 The SRI end

The description near the short-range Ising model took longer to complete. It was first thought that the upper value for $\mathfrak{s}$ was $\mathfrak{s}_* = 2$ [9]. Later it was found that it is actually slightly less than 2 [10]. The correct perturbative prescription was found in [16, 17] to be the short-range Ising model coupled to a GFF, with $\mathfrak{s}_* = p - 2\Delta_{\sigma^*}$, where $\Delta_{\sigma^*}$ is the dimension of the spin field $\sigma$ in the SRI. In two and three dimensions the value of $\Delta_{\sigma^*}$ is [8]

$$d = 2 : \Delta_{\sigma^*} = 1/8 \,, \quad d = 3 : \Delta_{\sigma^*} = 0.518148806(24) \,. \tag{2.5}$$

resulting in $\mathfrak{s}_* \sim 2$. At the SRI end the perturbative description is given in terms of a small parameter $\delta$: $\mathfrak{s} = p - 2(\Delta_{\sigma^*} + \delta)$, and the action is

$$S = S_{\text{SRI}} + \mathcal{N}_{\mathfrak{s}}\mathcal{N}_{-\mathfrak{s}} \int d^p\tau_1 d^p\tau_2 \frac{\chi(\tau_1)\chi(\tau_2)}{|\tau_{12}|^{p-\mathfrak{s}}} + \int d^p\tau g\sigma\chi \,, \tag{2.6}$$

where $g \sim O(\delta)$. We will denote this description by "SRI $+\chi$". Again we find a shadow relation due to the nonlocal equations of motion, this time between $\sigma$ and $\chi$, whose dimensions are protected as well:

$$\Delta_\chi = \frac{p+\mathfrak{s}}{2} \,, \quad \Delta_\sigma = \frac{p-\mathfrak{s}}{2} \,, \tag{2.7}$$

# 3 OPE relations

We have seen that the nonlocal eoms combined with a local interaction gives rise to a pair of shadow operators with protected dimensions. Besides that, they also give rise to relations between the OPE coefficients involving $\phi, \phi^3(\sigma, \chi)$ and a third conformal primary. These relations were first derived directly from the nonlocal eom in [18] for the case of scalar operators, and later on generalized to the spinning case in [19]. Here however we will follow an alternative derivation presented in [20], which holds more generally for defect CFTs and is also applicable to the LRI.

## 3.1 LRI as a defect

The LRI can be described as a defect CFT in the following way. The nonlocal kinetic term can be seen as a *local* kinetic term of a free field in a higher dimensional bulk. This is a known way of describing a generalized free field (GFF), also known as the Caffarelli-Silvestre trick [21], and the dimension of the bulk is then parametrized by the conformal dimension of the GFF. In our perturbative description an interaction term is also present, which will now become an interaction localized on a $p$-dimensional defect in the $d$-dimensional free bulk CFT [18]. Since we are studying the LRI in two and three dimensions, we will take $p = 2, 3$. The dimension $d$ and codimension $q = d - p$ are parametrized by $\mathfrak{s}$:

$$\text{MFT: } q = 2 - \mathfrak{s}\,, \quad \text{SRI } + \chi : q = 2 + \mathfrak{s}\,, \quad d = p + q\,. \tag{3.1}$$

The $p$-dimensional defect breaks the conformal symmetry of the bulk, but will preserve a $p$-dimensional conformal symmetry group $SO(p+1,1)$ on the defect, as well as rotations $SO(q)$ around the defect:

$$SO(d+1,1) \to SO(p+1,1) \times SO(q)\,. \tag{3.2}$$

The associated quantum number to the symmetry group of rotations $SO(q)$ is the *transverse spin* $j$. In this case, we have to remind ourselves that the LRI is not a true defect and the bulk CFT is just an auxiliary construction. The operators of the LRI are not charged under any transverse rotations, and we should set $j = 0$. In other words, the operators that appear in the spectrum of the LRI are all singlets under the global $SO(q)$ symmetry.

Besides the usual OPE between operators on the defect, which for scalars takes the form

$$\mathcal{O}_1(x_1)\mathcal{O}_2(x_2) = \sum_i \lambda_{12i}(x_{12}^2)^{-\Delta_1-\Delta_2+\Delta_k} C(x_{12}, \partial_2)\mathcal{O}_i(x_2)\,, \tag{3.3}$$

we now also have access to the *bulk-to-defect* OPE

$$\phi(x) = \sum_i b_i^\phi |x_\perp|^{\hat{\Delta}_i - \Delta} \hat{C}(|x_\perp|, \partial_\tau)\hat{\psi}_i(\tau)\,, \tag{3.4}$$

where a bulk operator in the presence of a defect is expanded in an infinite sum of local defect operators. The differential operators $C(x_{12}, \partial), \hat{C}(|x_\perp|, \partial_\tau)$ are fixed by conformal symmetry, and $\tau$ denotes the direction parallel to the defect, while $x_\perp$ denotes the perpendicular direction. The coefficients $b_i^\phi$ are the bulk-to-defect OPE coefficients. In the case of a free bulk CFT, the bulk-to-defect OPE is heavily restricted by the eom in the bulk, and the infinite sum reduces to a sum over two types of defect modes $\hat{\psi}^\pm$. The eom also fixes their dimensions:

$$\begin{aligned} \hat{\psi}_0^+ :& \quad \hat{\Delta}_0^+ = \Delta_\phi\,, \\ \hat{\psi}_0^- :& \quad \hat{\Delta}_0^- = \Delta_\phi + 2 - q\,. \end{aligned} \tag{3.5}$$

Looking back at the dimensions of the shadow pairs of operators found on either the MFT (2.3)–(2.4) or the SRI (2.7) side, we can idenfity $\hat{\psi}_0^+ = \phi, \chi$ and $\hat{\psi}_0^- = \phi^3, \sigma$.

## 3.2 Derivation of OPE relations

This general setup of the LRI as a defect CFT allows us to derive relations between OPE coefficients. Following [20], we start from a three-point function involving the bulk free field $\phi$ and two defect operators: $\langle \phi(x_1)\hat{\psi}_0^{\pm}(\tau_2)\widehat{O}^{(\hat{\ell})}(\infty)\rangle$[1]. One of the defect operators is the defect mode $\hat{\psi}_0^{\pm}$, the other defect operator $\widehat{O}^{(\hat{\ell})}$ is a traceless symmetric tensor of $SO(p)$ with parallel spin $\hat{\ell}$. Applying the bulk-to-defect OPE to $\phi$ results in a three-point correlation function of defect operators

$$\langle \phi(x_1)\hat{\psi}_0^{\pm}(\hat{x}_2)\widehat{O}^{(\hat{\ell})}(\infty)\rangle \overset{\text{bulk-to-defect OPE}}{\longrightarrow} \sum_{i=\pm} b_0^{\phi,i}\langle \hat{\psi}_0^{i}(\hat{x}_1)\hat{\psi}_0^{\pm}(\hat{x}_2)\widehat{O}^{(\hat{\ell})}(\infty)\rangle \ . \tag{3.6}$$

The former correlator has an expansion in conformal blocks, while the correlator on the right-hand side is kinematically fixed and is proportional to an OPE coefficient:

$$\langle \phi(\tau_1, x_{\perp,1})\hat{\psi}_0^{\pm}(\tau_2)\widehat{O}^{(\hat{\ell})}(\infty)\rangle = \sum_{\hat{\Delta}_0^k} \frac{b_k^{\phi}\lambda_{k\pm\widehat{O}}}{|x_{\perp,1}|^{\Delta_\phi+\hat{\Delta}_\pm-\hat{\Delta}_{\widehat{O}}}}\mathcal{F}^{\text{3pt}}_{\hat{\Delta}_0^k}(\chi) \times \ (\text{tensor structure})\ , \tag{3.7}$$

$$\langle \hat{\psi}_0^{i}(\tau_1)\hat{\psi}_0^{\pm}(\tau_2)\widehat{O}^{(\hat{\ell})}(\infty)\rangle = \frac{\lambda_{i\pm\widehat{O}}}{\tau_{12}^{\hat{\Delta}_i+\hat{\Delta}_\pm-\hat{\Delta}_{\widehat{O}}}} \times \ (\text{tensor structure})\ , \tag{3.8}$$

where the sum runs over defect conformal dimensions and spins $\hat{\Delta}_0^k$, $k=\pm$, and the blocks $\mathcal{F}^{\text{3pt}}_{\hat{\Delta},\hat{\ell}}(\chi)$ are given by

$$\mathcal{F}^{\text{3pt}}_{\hat{\Delta}_0^k}(\chi) = \chi^{-\frac{1}{2}(\hat{\Delta}_0^k+\hat{\Delta}_\pm-\hat{\Delta}_{\widehat{O}})}\times$$
$$ {}_2F_1\left(1-\frac{p}{2}+\frac{1}{2}(\hat{\Delta}_0^k+\hat{\Delta}_\pm-\hat{\Delta}_{\widehat{O}}-\hat{\ell}), \frac{1}{2}(\hat{\Delta}_0^k+\hat{\Delta}_\pm-\hat{\Delta}_{\widehat{O}}+\hat{\ell}), 1-\frac{p}{2}+\hat{\Delta}_0^k; -\frac{1}{\chi}\right)\ . \tag{3.9}$$

with the cross ratio given by $\chi = \frac{|\tau_{12}|^2}{|x_{\perp,1}|^2}$. Taking the limit $\chi\to\infty$ gives a discontinuity which is equivalent to the bulk-to-defect expansion of $\phi$. The blocks also have a discontinuity when $\tau_1=\tau_2$ or $\chi\to 0$, which is not physical when $x_{\perp,1}$ is not zero. Demanding that this discontinuity vanishes gives the following relations between OPE coefficients [20]:

$$\lambda_{++\widehat{O}} = \kappa_1(\hat{\Delta}_{\widehat{O}},\hat{\ell})\lambda_{-+\widehat{O}}\ , \quad \lambda_{--\widehat{O}} = \kappa_2(\hat{\Delta}_{\widehat{O}},\hat{\ell})\lambda_{-+\widehat{O}}\ , \tag{3.10}$$

where

$$\kappa_1(\hat{\Delta}_{\widehat{O}},\hat{\ell}) = -R(a_{\phi^2})\frac{\Gamma\left(\frac{4-q}{2}\right)\Gamma\left(\frac{\hat{\ell}+\hat{\Delta}_{\widehat{O}}}{2}\right)\Gamma\left(\frac{\hat{\ell}+p+q-2-\hat{\Delta}_{\widehat{O}}}{2}\right)}{\Gamma\left(\frac{q}{2}\right)\Gamma\left(\frac{\hat{\ell}+p-\hat{\Delta}_{\widehat{O}}}{2}\right)\Gamma\left(\frac{\hat{\ell}+2-q+\hat{\Delta}_{\widehat{O}}}{2}\right)}\ ,$$
$$\kappa_2(\hat{\Delta}_{\widehat{O}},\hat{\ell}) = -\frac{1}{R(a_{\phi^2})}\frac{\Gamma\left(\frac{q}{2}\right)\Gamma\left(\frac{\hat{\ell}+\hat{\Delta}_{\widehat{O}}}{2}\right)\Gamma\left(\frac{\hat{\ell}+p-q+2-\hat{\Delta}_{\widehat{O}}}{2}\right)}{\Gamma\left(\frac{4-q}{2}\right)\Gamma\left(\frac{\hat{\ell}+p-\hat{\Delta}_{\widehat{O}}}{2}\right)\Gamma\left(\frac{\hat{\ell}-2+q+\hat{\Delta}_{\widehat{O}}}{2}\right)}\ . \tag{3.11}$$

We have defined

$$R(a_{\phi^2}) \equiv b_0^{\phi,-}/b_0^{\phi,+}\ , \tag{3.12}$$

and this ratio can, through the defect bootstrap, also be expressed in terms of the one-point function coefficient $a_{\phi^2}$, defined through

$$\langle \phi^2(\tau, x_\perp)\rangle = \frac{a_{\phi^2}}{|x_\perp|^{\Delta_{\phi^2}}}\ . \tag{3.13}$$

[1]Correlation functions involving bulk and defect operators are evaluated in the presence of the defect.

We can then use $a_{\phi^2}$ to parametrize the infinite set of OPE relations.

While the derivation and definitions above in terms of $\phi, \phi^2$ used the language of the MFT description of the LRI, these relations equally hold for the SRI end. The two descriptions are related by a shadow transformation: one can go from one to the other by exchanging $\hat{\Delta}_0^+ \leftrightarrow \hat{\Delta}_0^-$. This transformation is achieved by considering $q \leftrightarrow 4-q$, or $\mathfrak{s} \leftrightarrow -\mathfrak{s}$. Applying this change to the OPE relations does however not immediately give the correct result. The bulk CFT is different for the two descriptions, and as a consequence we need to take different interpretations for the $\pm$ modes. The two descriptions correspond to different values of $a_{\phi^2}$, where for the SRI+$\chi$ description the "correct" parameter to interpolate between different values of $\mathfrak{s}$ would be $a_{\chi^2}$ instead of $a_{\phi^2}$. At the two endpoints of the critical range of $\mathfrak{s}$, which are $\mathfrak{s} = p/2$, $\varepsilon \to 0$ (MFT) and $\mathfrak{s} = \mathfrak{s}_*$, $\delta \to 0$ (SRI+$\chi$), the coefficients $b_0^{\phi,\pm}$ and $a_{\phi^2}$ have the following values:

$$\begin{aligned} &\text{MFT}: && b_0^{\phi,+} = 1\,, \quad b_0^{\phi,-} = 0\,, \quad a_{\phi^2} = 0\,, \\ &\text{SRI} + \chi: && b_0^{\phi,+} = 0\,, \quad b_0^{\phi,-} = 1\,, \quad a_{\chi^2} = 0 \;\leftrightarrow\; a_{\phi^2} = \begin{cases} -7/8\,, & p = 2\,, \\ -0.575\ldots\,, & p = 3\,. \end{cases} \end{aligned} \tag{3.14}$$

When using the numerical bootstrap, this gives us a range of $a_{\phi^2}$ to scan over.

### 3.3 Protected odd-spin sector

When we consider odd-spin operators there is an additional constraint that can be applied. Because of Bose symmetry, odd-spin operators cannot appear in the OPE of two identical operators and we find that

$$\lambda_{++\hat{\mathcal{O}}} = \lambda_{--\hat{\mathcal{O}}} = 0\,, \quad \lambda_{+-\hat{\mathcal{O}}} \neq 0\,. \tag{3.15}$$

Combining this with (3.10), one finds that $\kappa_1(\hat{\Delta}_{\hat{\mathcal{O}}}, \hat{\ell}) = \kappa_2(\hat{\Delta}_{\hat{\mathcal{O}}}, \hat{\ell}) = 0$ for $\hat{\ell}$ odd. This holds when the odd-spin operators have a certain protected conformal dimension [20]:

$$\Delta_{\hat{\mathcal{O}}}|_{\hat{\ell}\text{ odd}} = p + \hat{\ell} + 2n\,. \tag{3.16}$$

These dimensions will not get renormalized. This is a very powerful constraint on the spectrum of OPE coefficients, and will be of use in the numerical bootstrap discussed below. In addition we also remove the operator with $\hat{\ell} = 1, \Delta = p+1$ from the spectrum, because no such conformal primary exists at the MFT end nor the SRI end.

## 4 Numerical bootstrap

Our goal is to implement the OPE relations described above in the numerical conformal bootstrap for four-point correlation functions of defect operators, corresponding to the operators in the LRI. This will give exclusion plots for scaling dimensions of these operators, with the bounds being valid nonperturbatively. A comprehensive review of the numerical bootstrap can be found in e.g. [22]. The bounds are obtained using the semidefinite program solver SDPB [23,24]. The main idea is to use crossing symmetry to obtain bounds on the scaling dimensions of conformal primaries, and on OPE coefficients. Our starting point is the set of four-point functions of defect modes $\hat{\psi}_0^\pm$:

$$\langle \hat{\psi}_0^\pm(\tau_1) \hat{\psi}_0^\pm(\tau_2) \hat{\psi}_0^\pm(\tau_3) \hat{\psi}_0^\pm(\tau_4) \rangle\,, \tag{4.1}$$

where we take all possible inequivalent combinations containing an even number of plus and minus modes. The four-point functions are fixed by conformal invariance up to a function $\mathcal{G}(u,v)$ of the cross ratios $u, v$, for example

$$\langle \hat{\psi}_0^+(\tau_1)\hat{\psi}_0^+(\tau_2)\hat{\psi}_0^+(\tau_3)\hat{\psi}_0^+(\tau_4)\rangle = \frac{\mathcal{G}(u,v)}{|\tau_{12}|^{2\hat{\Delta}_+}|\tau_{34}|^{2\hat{\Delta}_+}}\,, \quad u = \frac{\tau_{12}^2\tau_{34}^2}{\tau_{13}^2\tau_{24}^2}\,, \quad v = \frac{\tau_{14}^2\tau_{23}^2}{\tau_{13}^2\tau_{24}^2}\,. \tag{4.2}$$

The function $\mathcal{G}(u,v)$ can be decomposed in terms of OPE coefficients and conformal blocks, which are known for any dimension $p$. Crossing symmetry can be depicted as

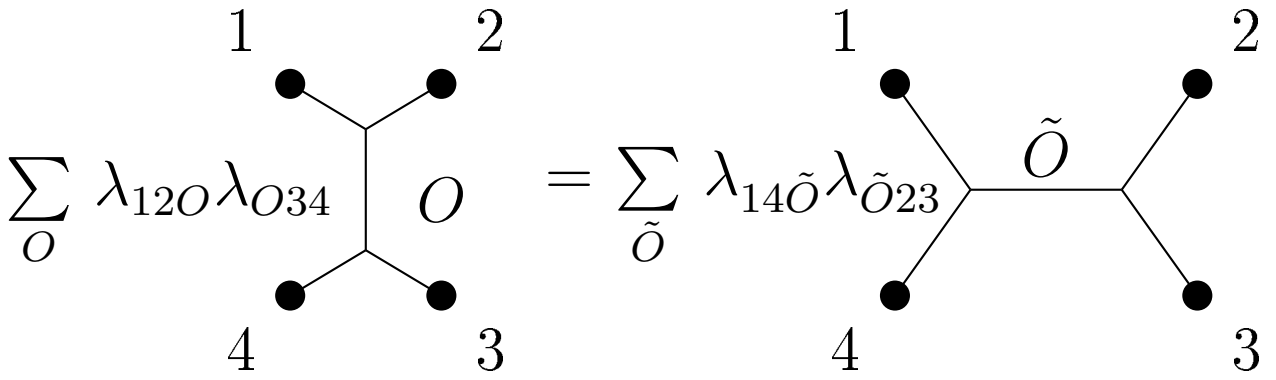

We will now put bounds on the conformal dimension $\hat{\Delta}_0$ of the lowest-lying scalar operator in the $\hat{\psi}_0^\pm \times \hat{\psi}_0^\pm$ OPE, as a function of the dimension of the first spin-2 operator $\hat{\Delta}_2$, the coefficient $a_{\phi^2}$, and the dimension of the external operators $\hat{\Delta}_+ = \Delta_\phi$. We include all the things discussed above: the OPE relations (3.10), the fact that the odd-spin operators appearing in $\hat{\psi}_0^\pm \times \hat{\psi}_0^\pm$ have protected dimension $\hat{\Delta} = p + \hat{\ell} + 2n$, and we demand that there is no spin-1 operator with $\hat{\Delta} = p + 1$. The resulting plots are shown below.

## 4.1 Results in two dimensions

We start with analyzing the LRI in two dimensions. Figure 1a shows the bound obtained on the scaling dimension of the first spin-0 operator appearing in the $\hat{\psi}_0^\pm \times \hat{\psi}_0^\pm$ OPE, as a function of the one-point function coefficient $a_{\phi^2}$, defined in (3.13). The shaded regions are allowed, while the white regions are disallowed. Three bounds are shown, for different values of $\Delta_{\hat{\phi}}$, which corresponds to different values of $\mathfrak{s}$ through (2.3). We have also imposed a gap in the spin-2 sector, meaning that we impose that the first spin-2 operator exchanged in the $\hat{\psi}_0^\pm \times \hat{\psi}_0^\pm$ OPE has a minimal dimension $\hat{\Delta}_2 = 2.4 - 2.2$. A higher gap produces a narrower shaded region. The gaps are chosen such that the regions do not overlap, and are relatively low. Since the LRI is nonlocal, it does not have a stress-tensor, which has $\Delta_T = p$. This puts the minimal gap in the spin-2 sector to be $\hat{\Delta}_2 > 2$.

We have also included predictions from analytic conformal bootstrap and perturbative computations in the small $\varepsilon$-expansion up to $O(\varepsilon^3)$, shown by the black dot and black line. The analytic bootstrap procedure and the results are detailed in [11].

For any $\mathfrak{s}$, the bounds show two sharp features known as kinks. These kinks are often a precursor for the presence of a CFT, and in our case we expect the LRI to be at the position of such a kink. However, only one of them should correspond to the LRI. This raises the questions which of the two represents the LRI, and what theory is described by the other kink?

To answer the first question, we vary some of the initial parameters, specifically the gap $\hat{\Delta}_2$ on the spin-2 sector. This is shown in figure 1b, for one of the shaded regions corresponding to $\Delta_\phi = 0.35$. We see that as we vary the spin-2 gap, the left kink moves to the right, while the rightmost kink stays stable. For a specific value $\hat{\Delta}_{2,*}$, the left kink even merges with the right kink and disappears completely. This indicates that it is the rightmost kink that describes the LRI, and we mark it with a cross in figure 1a. The remaining question of what the left kink might correspond to will be addressed in the conclusions.

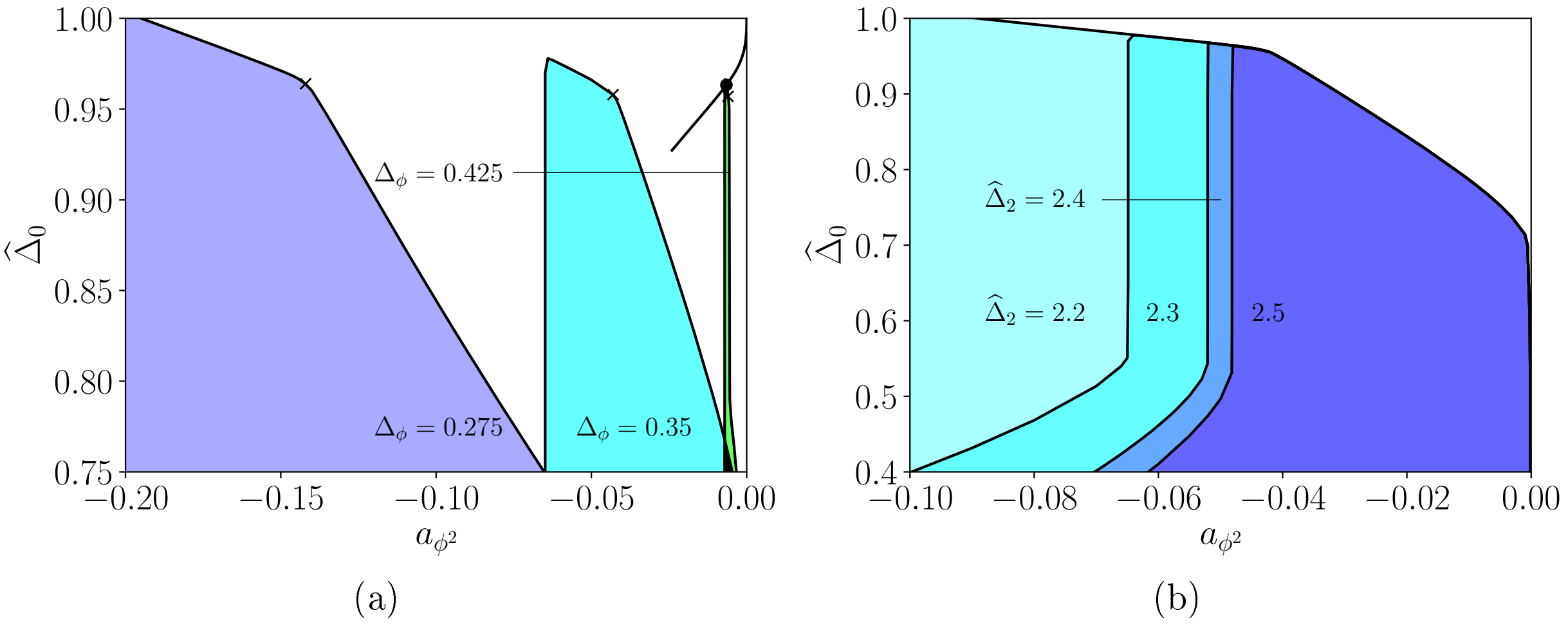

Figure 1: Bounds on the spin-0 gap $\hat{\Delta}_0$ as a function of $a_{\phi^2}$ for $p = 2$. In (a), the bounds are shown for different values of $\Delta_\phi$, with $\hat{\Delta}_2 \geq 2.4-2.2$ from narrow to wide respectively. In (b), $\Delta_\phi$ is set to 0.35, and the different regions correspond to different gaps $\hat{\Delta}_2$ in the spin-2 sector, causing the leftmost kink to move.

Figure 1 focuses on the regions where $\mathfrak{s}$ is close to the MFT end, with $a_{\phi^2}$ approaching zero. Let us now look at the bounds at the SRI end. Figure 2a shows the allowed region for $\Delta_\phi = 0.2$, which corresponds to $\mathfrak{s} = 1.6$. The other crosses correspond to the locations of kinks for different values of $\Delta_\phi$, still close to the SRI end. The black dots are predictions from conformal perturbation theory for small $\delta = (\mathfrak{s}_* - \mathfrak{s})/2$, further detailed in [11]. We have picked the value $\Delta_\phi = 0.2$, since for this value people have performed Monte Carlo simulations of the 2d LRI. There are two competing predictions, shown in the figure as [P-12] [12], and [APR-14] [13]. The latter group found a different value for $\hat{\Delta}_0$ than [12], which they attribute to strong finite size effects. We see that the kink lies just above the results of [12], indicating that the numerical bootstrap supports the results of [13] instead, assuming that the kink corresponds to the 2d LRI. However, it is not sufficient to exclude one of the two results. In order to do that, it would be necessary to obtain rigorous *lower bounds* on the dimension $\hat{\Delta}_0$ as well.

To get lower bounds as well as upper bounds, we have to assume a gap $\hat{\Delta}_0'$ on the second spin-0 operator which is larger than the dimension $\hat{\Delta}_0$ of the first spin-0 operator. We take the first operator to be the only relevant scalar, and vary its dimension, while keeping the second operator at $\hat{\Delta}_0' = 2$. The results are shown in 2b. One can see islands forming and becoming smaller when varying the spin-2 gap $\hat{\Delta}_2$. However, as we go on, we see that we start to exclude the kink that should indicate the LRI. Varying $\hat{\Delta}_2$ alone is not enough to find an island that includes the LRI solution, but excludes one of the Monte Carlo results. To get such a result, it is necessary to explore other avenues.

## 4.2 Results in three dimensions

Let us now move on to results for the three-dimensional LRI. We compute the same bounds as for the two-dimensional case: bounds on the spin-0 gap $\hat{\Delta}_0$ as a function of $a_{\phi^2}$. The results are shown in figure 3. We again find kinks, however here we just see a single kink for each value of $\Delta_\phi$ or $\mathfrak{s}$ and there is no confusion which one corresponds to the LRI. The location of the kinks agree with results from perturbation theory in small $\varepsilon$ close to the MFT end ($a_{\phi^2} = 0$), depicted by the black line and dot. The dot corresponds to $\Delta_\phi = 0.70363$.

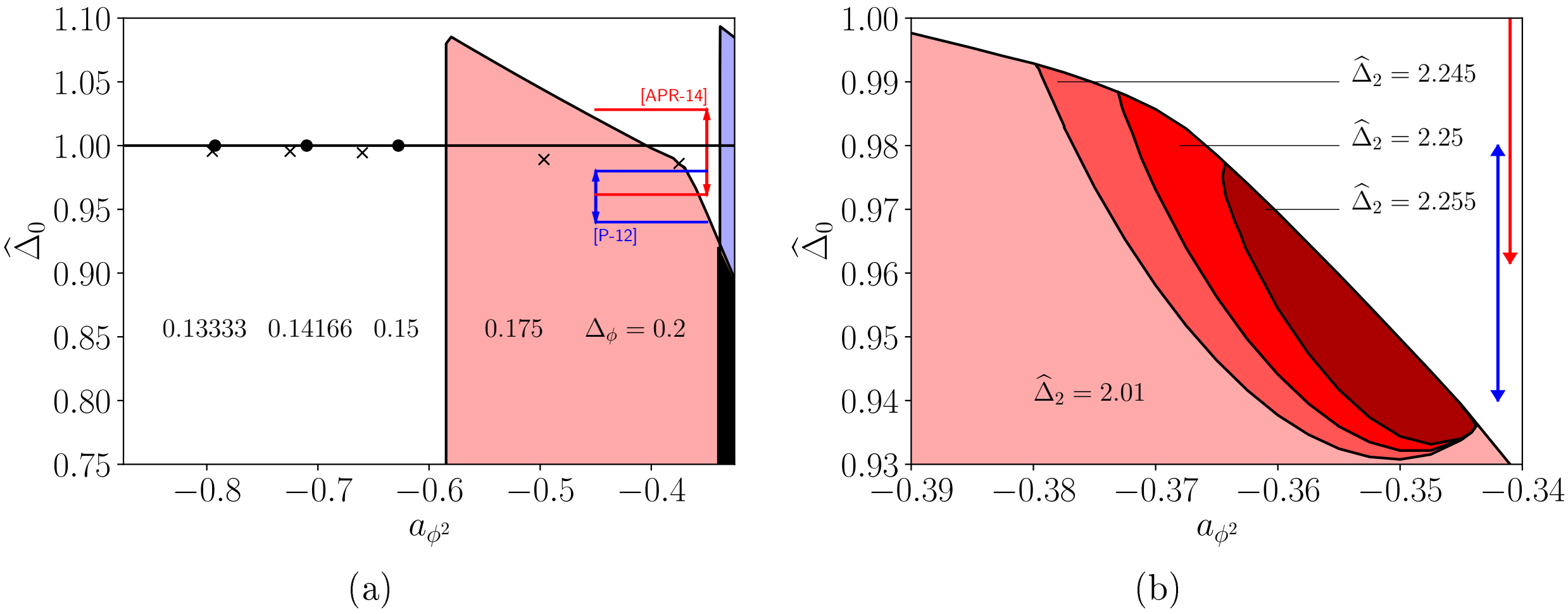

Figure 2: Bounds on the spin-0 gap for $\Delta_\phi = 0.2$ as a function of $a_{\phi^2}$ for $p = 2$, compared to Monte Carlo results from [12] (red) and [13] (blue). Kinks for different values of $\Delta_\phi$ close to the SRI end are shown in addition in (a) by black crosses, and the black dots are results from $O(\delta)$ conformal perturbation theory. In (b), instead of maximizing the gap $\hat{\Delta}_0$, we scan over the dimension of a single exchanged relevant scalar. The allowed region varies as a function of $\hat{\Delta}_2$.

The fact that we obtained kinks in two and three dimensions is nontrivial. In [19] the three-dimensional LRI was studied as well using among other methods the numerical bootstrap, but no kinks were found. The main addition here with respect to [19] is the scan over $a_{\phi^2}$, which allows us to include all OPE relations. In [19], $a_{\phi^2}$ was not included as an additional parameter that could be scanned over. Instead, a specific combination of OPE relations was considered in which $a_{\phi^2}$ drops out. It was found later in [25, 26] that scanning over $a_{\phi^2}$ is the better approach.

# 5 Conclusions and further applications

With the numerical bootstrap and the exact OPE relations, plus the fact that some operators have protected dimensions, we were able to find kinks in the bounds on the scaling dimension $\hat{\Delta}_0$ of the first spin-0 operator, that are in good correspondance with perturbative ($\varepsilon$-expansion, conformal perturbation theory) and nonperturbative (Monte Carlo) results. However, we are still left with the question what the other kinks appearing in the 2d LRI plots, figures 1 and 2, represent. A possible explanation, following a similar observation in [25, 26] for boundary CFTs, is that these kinks come from minimal models coupled to a GFF instead of the SRI. The observation that minimal models coupled to a GFF could lead to many more nonlocal CFTs, was already made in [17]. This explanation is further motivated by the fact that close to the SRI point, for $\alpha_{\phi^2} = -7/8$, the bound on $\hat{\Delta}_0$ becomes approximately $\frac{8}{3}\Delta_\phi + \frac{2}{3}$. This corresponds to the relation between two conformal primaries $\Delta_{(1,2)})$ and $\Delta_{(1,3)}$ for the $m$'th Virasoro minimal model. In addition, we see that in three dimensions only a single kink remains. This agrees with the upper critical dimensions of the minimal models, which lie between two and three. It would be interesting to construct such long-range models and compare predicitions from e.g. conformal perturbation theory with the additional kinks seen in figures 1 and 2.

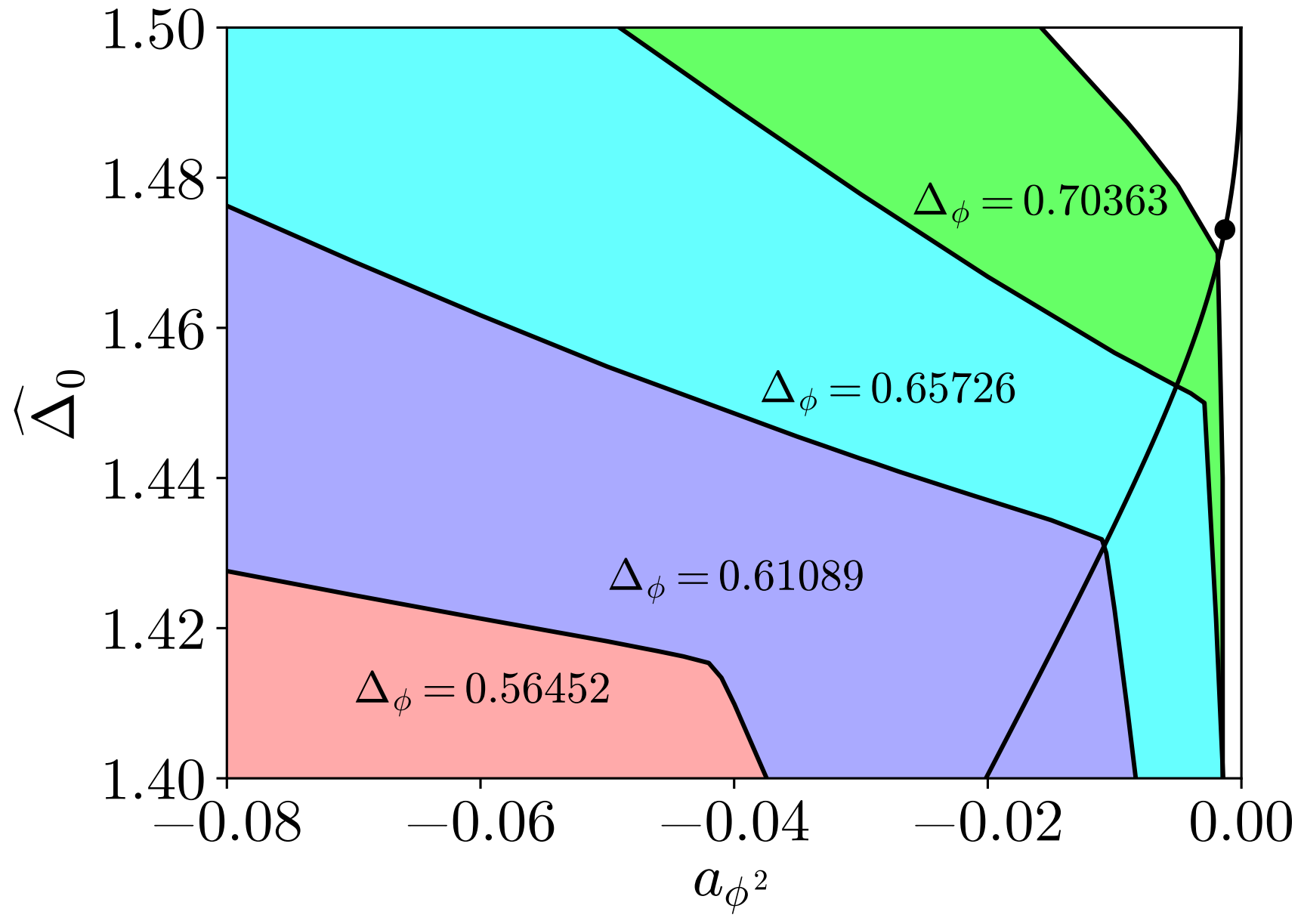

Figure 3: Bound on the spin-0 gap $\hat{\Delta}_0$ as a function of $a_{\phi^2}$ for $p = 3$. The colored regions represent different values of $\Delta_\phi$. The black line and dot indicate results from perturbation theory up to $O(\varepsilon^3)$. The dot corresponds to $\Delta_\phi = 0.70363$.

### 5.1 The 1d LRI

We have focused on the LRI in two and three dimensions, but what about the LRI in one dimension? There exists a critical line of nontrivial fixed points for $1/2 \le \mathfrak{s} \le 1$ for the 1d LRI, again corresponding to a family of unitary, nonlocal 1d CFTs [27–29]. The MFT description can be straightforwardly continued to the 1d case, and one can derive the OPE relations and use the analytic conformal bootstrap or perturbative techniques in the same way as described above to obtain information about the conformal data close to the MFT end. On the SRI end however we encounter a puzzle: there is no critical 1d SRI model and it has long been unclear how the weakly coupled description of the LRI would look like around this point.

Anderson and Yuval [30] gave a description of the 1d LRI at $\mathfrak{s} = 1$ in terms of a Coulomb gas of dilute, alternating kinks and antikinks. This picture was extended by Kosterlitz to $\mathfrak{s} < 1$ [31]. While this provides a correct picture of the physics for $\mathfrak{s}$ close to 1, it is a rather complicated model to do e.g. perturbative computations in. In a recent work [32] we proposed a field-theoretic description which is weakly coupled. In this framework, it becomes possible to push the perturbative computations for small $\delta = 1 - \mathfrak{s}$ using conformal perturbation theory. We also use the analytic conformal bootstrap to find predictions for scaling dimensions and OPE coefficients, which match the perturbative computations.

Having a weakly-coupled, field-theoretic description of the 1d LRI for both ends of the critical range of $\mathfrak{s}$ at our disposal, one could now use the numerical bootstrap including the OPE relations and information about protected operators to obtain bounds. These bounds can then be matched to the perturbative results in small $\varepsilon = \mathfrak{s} - 1/2$ and $\delta$.

### 5.2 Monodromy defects in free theory

In the case of the LRI the setup of a conformal defect in a free bulk CFT is auxiliary, and both $d$ and $q$ are noninteger. What about physical defects? The case of a boundary ($q = 1$) was already considered in [25, 26], where they derived OPE relations and performed the numerical bootstrap. Using the shadow relations described in section 3, where $q \leftrightarrow 4-q$, we see that there is a kinematical relation between this setup and a defect of codimension 3. While this would be an interesting thing to check, there is another value of $q$ that is left unexplored: $q = 2$. It turns out that for line and surface defects in a free bulk CFT, the only defects that would allow nontrivial dynamics for $q = 2$ are monodromy defects in $d \geq 4$ [20]. Consider a $\mathbb{Z}_2$ monodromy defect in $d = 4$, for which the free bulk field $\phi$ picks up a sign after a $2\pi$ rotation around the defect. This setup was studied before in [33, 34]. As was the case for the LRI, the eom in the free bulk CFT ensure that $\phi(x)$ can be expanded in two types of defect operators (the $\pm$ modes):

$$\phi(x) = \sum_{j,k} b_j^\phi |x_\perp|^{\hat{\Delta}_{\pm,j}-\Delta_\phi} C(x_\perp, \partial)\hat{\psi}^\pm_{k_j}(\tau)\ , \quad \hat{\Delta}_{\pm,j} = \Delta_\phi \pm |j|\ , \tag{5.1}$$

where the transverse spin $j$ is now half-integer. Furthermore, if we want a unitary theory, we are restricted to $j = \pm\frac{1}{2}$. One can study all correlators that contain combinations of $\hat{\psi}^\pm_{\pm 1/2}$ neutral under the $SO(q = 2)$ symmetry, leading to 18 crossing equations. Relations between OPE coefficients are derived in the same way as in section 3 and take the form

$$\lambda_{\hat{\psi}^+_{1/2}\hat{\psi}^+_{1/2}\hat{\mathcal{O}}}\lambda_{\hat{\mathcal{O}}\hat{\psi}^-_{-1/2}\hat{\psi}^-_{-1/2}} = \tilde{\kappa}_1\tilde{\kappa}_2\lambda_{\hat{\psi}^+_{1/2}\hat{\psi}^-_{1/2}\hat{\mathcal{O}}}\lambda_{\hat{\mathcal{O}}\hat{\psi}^+_{-1/2}\hat{\psi}^-_{-1/2}} = \lambda_{\hat{\psi}^+_{-1/2}\hat{\psi}^+_{-1/2}\hat{\mathcal{O}}}\lambda_{\hat{\mathcal{O}}\hat{\psi}^-_{1/2}\hat{\psi}^-_{1/2}}\ , \tag{5.2}$$

and similar for other combinations. The functions $\tilde{\kappa}_1, \tilde{\kappa}_2$ are of course different from those in (3.11), but are again ratios of gamma functions depending on $\hat{\Delta}_{\hat{\mathcal{O}}}, \hat{\ell}$ and $j$. We see immediately that there are more possible combinations, leading to more complicated OPE relations. On top of that, it seems that some relations are more fundamental than others. Take (5.2) as an example. The first equality is an OPE relation as we have seen before in (3.10), but we can relate the OPE coefficients on the furthest left and right directly to each other without the relative coefficients $\tilde{\kappa}_1, \tilde{\kappa}_2$. It would be interesting to see if such relations can be derived in a different way, for example directly from symmetry constraints, without using the eom in the bulk. This could then potentially be applied to a broader class of defects.

The similarities and differences between the LRI and the monodromy defect are also visible in the sector of protected operators. Just as for the LRI we find a protected sector of odd-spin operators. However, while for the LRI these were the only operators to appear in the OPE of two defect modes $\hat{\psi}^\pm_0$, for the monodromy defect unprotected odd-spin operators will appear too. Nevertheless, it is still possible to add the protected operators as an additional constraint to the numerical bootstrap, and try to find kinks and islands for the monodromy defect in a free $4d$ bulk [35].

# 6 Acknowledgements

I would like to thank the organisers of the "28e rencontre Itzykson: Analytic results in Conformal Field Theory" (Sept. 10–12, 2024) at the IPhT Saclay for the opportunity to present this work and for organising an inspiring meeting. I would also like to thank Connor Behan, Edoardo Lauria, Dalimil Mazáč and Sylvain Ribault for comments on the draft. This work was partially supported by the European Union (ERC, FUNBOOTS, project number 101043588).

# Parisi-Sourlas Dimensional Uplift

**Emilio Trevisani**

*Laboratoire de Physique Théorique et Hautes Énergies, CNRS & Sorbonne Université, 4 Place Jussieu, 75252 Paris, France*

*Department of Theoretical Physics, CERN, 1211 Meyrin, Switzerland*

*E-mail:* emilio.trevisani.et@gmail.com

Abstract: We review how Parisi-Sourlas supersymmetry can be used to uplift a model to higher dimensions. We show that any given scalar four-point function of a $\mathrm{CFT}_{d-2}$ defines 43 four-point functions related by supersymmetry in the uplifted $\mathrm{CFT}_d$, which are automatically crossing covariant and decomposable in conformal blocks. We then apply the uplift to generalized free theory, showing how the supersymmetry of the uplifted model enables us to bootstrap an infinite set of CFT data. We finally uplift the $2d$ minimal models to define solvable interacting CFTs in all even dimensions and we comment on interesting features of the resulting theories.

# Contents

# 1 Introduction

Parisi and Sourlas (PS) supersymmetry (SUSY) plays a crucial role in the study of disordered systems in statistical physics. For example, the RF Ising model in $d = 5$, and the RF $\phi^3$ model in $2 \leq d < 8$ have emergent SUSY at their IR fixed points [1–6]. Recently, a renormalization group (RG) argument, explained that the emergence of SUSY occurs in these cases because all SUSY-breaking interactions are irrelevant; conversely, in the RF Ising model in $d = 3, 4$, SUSY-breaking interactions are relevant and destabilize the SUSY fixed point [7–10] (see also [11–17] and references therein, for other approaches to this problem).[1] Other interesting applications of PS SUSY are discussed e.g. in [19–24].

In this note, we will focus on PS superconformal theories, referred to as PS CFTs, regardless of whether they arise from RF models. An interesting feature of all PS CFT is that they undergo dimensional reduction, meaning that a large part of the model can be described by a CFT in two fewer dimensions [1, 12, 25–29]. For example, the RF Ising model in $d = 5$ exhibits emergent PS SUSY, and through dimensional reduction, it is captured by the pure $3d$ Ising model. Since dimensional reduction applies to any PS CFT, it raises the question of whether a dimensional uplift is possible. Specifically, for a generic $\mathrm{CFT}_{d-2}$, is it always possible to construct a PS $\mathrm{CFT}_d$ that reduces to it? In the rest of the note we address this question, illustrated in the shaded area of figure 1. This will serve as an introduction to the work of [29, 30].

There are several compelling reasons to study this problem. First, some of the uplifted models are physically relevant, making it important to explore them further. Another motivation is that the uplift allows one to explore the space of consistent theories. E.g. it is widely believed that non-trivial unitary CFTs do not exist in dimensions greater than six. However, PS CFTs are non-unitary, and by iterating the uplift, they offer examples of non-trivial CFTs in arbitrarily high dimensions.

Additionally, the uplift can be applied to solvable lower-dimensional models, such as the minimal models. This leads to the definition of towers of solvable CFTs in all even dimensions, which give rise to fascinating mathematical structures worth exploring further.

[1] The upper critical dimension for the RF Ising and for RF $\phi^3$ is respectively $d = 6$ and $d = 8$. Above the upper critical dimensions these models are described by mean field theory but they still enjoy PS SUSY. A throughout discussion of these free theories is presented in [18].

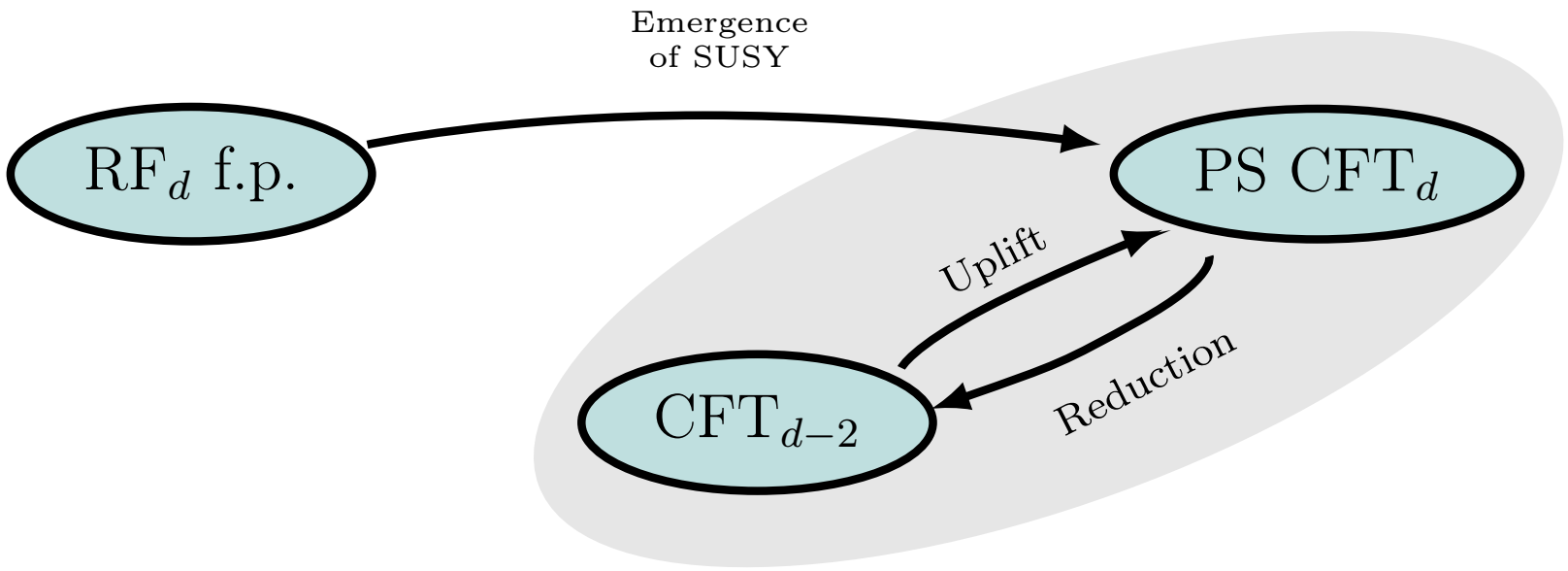

Figure 1: Diagram of relations between the fixed point of a random field theories in $d$ dimensions ($\mathrm{RF}_d$ f.p.), a PS $\mathrm{CFT}_d$ and a pure $\mathrm{CFT}_{d-2}$.

Finally, SUSY of the uplifted model can serve as a useful tool for addressing problems in the original theory. This applies both at the kinematical level, where SUSY provides relations for the conformal blocks (see section 2.3), and even at the dynamical level in some cases. For example, in section 3.1 we will show how PS SUSY can be used to bootstrap the conformal block decomposition of an infinite set of correlators in generalized free field theory (GFF).

The rest of the note is organized as follows. In section 2, we introduce PS SUSY and explain the idea of dimensional reduction and uplift. In section 3 we apply the uplift to GFF and minimal models. Finally, in section 4 we provide some concluding remarks and discuss open directions.

# 2 PS SUSY, reduction and uplift

## 2.1 PS SUSY

Let us first explain what is PS SUSY and how to easily achieve it. Roughly speaking this SUSY can be obtained by defining a theory on a superspace $\mathbb{R}^{d|2}$ with 2 fermionic scalar directions. The prototypical example is given by the following action

$$S = \int d^{d|2}y \left[\frac{1}{2}\partial_y^a \Phi(y) \partial_{ya}\Phi(y) + V(\Phi(y))\right] . \tag{2.1}$$

Here the index $a$ takes $d+2$ values $a = \mu, \theta, \bar{\theta}$ where $\mu = 1, \ldots, d$ and the superspace position is defined as $y^a = (x^\mu, \theta, \bar{\theta})$, $\partial_{ya} = (\partial_\mu, \partial_\theta, \partial_{\bar{\theta}})$. The indices of vectors can be lowered/raised using an orthosymplectic metric which is the usual identity metric in the first $d$ direction, while is equal to the symplectic metric $\left(\begin{smallmatrix} 0 & -1 \\ 1 & 0 \end{smallmatrix}\right)$ for the last two. One can expand the superfield $\Phi$ in components $\Phi(y) = \varphi(x) + \bar{\theta}\psi(x) + \theta\bar{\psi}(x) + \theta\bar{\theta}\omega(x)$. The variables $\theta, \bar{\theta}$ are Grassmann scalar and have dimension one, which implies that all components are scalars with $\varphi, \omega$ commuting and $\psi, \bar{\psi}$ anticommuting.

The focus of this work is CFT, where a Lagrangian description is often not available. Despite this, it is still possible to study CFTs with PS SUSY. These theories are invariant under the orthosymplectic group $OSp(d+1, 1|2)$ generated by

$$P^a\,, \qquad L^{ab}\,, \qquad D\,, \qquad K^a\,. \tag{2.2}$$

When $a, b = 1 \ldots d$ these generators coincide respecively with the usual translations, rotations, dilations and special conformal transformations. However we have now extra generators for $a, b = \theta, \bar{\theta}$.[2] Every time an index takes a fermionic value, the correspondent tensor changes grading. The R-symmetry is $Sp(2)$ and is generated by $L_{ab}$ with both indices in fermionic directions, namely $L_{\theta\bar{\theta}}, L_{\theta\theta}, L_{\bar{\theta}\bar{\theta}}$.[3] The remaining $2(2+d)$ generators have a single fermionic index and thus are fermionic.

Generators with fermionic indices are simple to understand, e.g. $P_\theta$ is just a translation in direction $\theta$ which is implemented by the derivative $\partial_\theta$, similarly $L_{\theta\mu}$ rotates a bosonic direction into a fermionic one.

We also notice that all fermionic generators lie either in the scalar (like $P^\theta$, $P^{\bar{\theta}}$, $K^\theta$, $K^{\bar{\theta}}$) or in the vector (like $L^{\theta\mu}$, $L^{\bar{\theta}\mu}$) representations of $SO(d)$. There are no spinorial supercharges, which are required for a conventional type of supersymmetry. As a result, PS CFTs violate spin-statics theorem and therefore are non-unitary.

We can classify the operators of the PS CFT. Superprimaries are annihilated by $K^a$ and have definite dimensions $\Delta$ and $OSp(d|2)$-spin defined respectively by the action of $D$ and $L^{ab}$. Superdescendants are obtained by acting with $P^a$ on the superprimaries.

By exploiting the symmetries, one can constrain the form of correlators of superprimaries, revealing that they follow the same structure as those in a standard CFT, except now the operator positions lie in superspace. For example two and three point functions of superspace operators $\mathcal{O}_i$ with dimension $\Delta_i$ take the form

$$\langle \mathcal{O}_i(y_1)\mathcal{O}_j(y_2)\rangle = \frac{\delta_{\Delta_i\Delta_j}}{(y_{12}^2)^{\Delta}}\,, \quad \langle \mathcal{O}_1(y_1)\mathcal{O}_2(y_2)\mathcal{O}_3(y_3)\rangle = \frac{\lambda_{123}}{(y_{12}^2)^{\frac{\Delta_1+\Delta_2-\Delta_3}{2}}(y_{23}^2)^{\frac{\Delta_2+\Delta_3-\Delta_1}{2}}(y_{13}^2)^{\frac{\Delta_1+\Delta_3-\Delta_2}{2}}}\,,$$

where $\lambda_{123}$ are the OPE coefficients and $y_{ij}^2 = x_{ij}^2 - 2\theta_{ij}\bar{\theta}_{ij}$ (with e.g. $y_{ij}^a \equiv y_i^a - y_j^a$, $\theta_{ij} \equiv \theta_i - \theta_j$). Similarly a four point function of scalar superprimaries takes the form

$$\langle \mathcal{O}_1(y_1)\mathcal{O}_2(y_2)\mathcal{O}_3(y_3)\mathcal{O}_4(y_4)\rangle = \frac{\left(\frac{y_{14}^2}{y_{24}^2}\right)^{-\frac{\Delta_{12}}{2}}\left(\frac{y_{14}^2}{y_{13}^2}\right)^{\frac{\Delta_{34}}{2}}}{(y_{12}^2)^{\frac{1}{2}(\Delta_1+\Delta_2)}(y_{34}^2)^{\frac{1}{2}(\Delta_3+\Delta_4)}}\, f(U,V)\,, \tag{2.3}$$

which is fully fixed in terms of a function of two super cross ratios defined as $U \equiv \frac{y_{12}^2 y_{34}^2}{y_{13}^2 y_{24}^2}$, $V \equiv \frac{y_{14}^2 y_{23}^2}{y_{13}^2 y_{24}^2}$, where we defined $\Delta_{ij} \equiv \Delta_i - \Delta_j$.

All the correlators above are identical to those of a standard CFT upon the substitution $y \to x$. In [29], additional properties of PS CFTs were studied, such as the OPE, superconformal blocks, superembedding space formalism, and more. We do not review this material here, but we emphasize that most features of these theories match those of typical CFTs, with the distinction that they live in superspace. Indeed, as we explain below, PS CFTs can (for the most part) be captured by a standard CFT living in two fewer dimensions.

## 2.2 PS reduction

Let us first focus on the QFT defined by the action (2.1). For this model PS reduction implies that (2.1) is captured by the $(d-2)$-dimensional action

$$\hat{S} = \int d^{d-2}\hat{x}\left[\frac{1}{2}\partial\phi(\hat{x})\partial\phi(\hat{x}) + V(\phi(\hat{x}))\right]\,, \tag{2.4}$$

[2] To show the isomorphism with $OSp(d+1,1|2)$ one can repackage the generators into a single graded antisymmetric generator $J^{AB}$ with indices $A, B = 0, \ldots, d+1, \theta, \bar{\theta}$ as follows $J^{0d+1} = D$, $J^{0a} = \frac{P^a - K^a}{2}$, $J^{d+1a} = \frac{P^a + K^a}{2}$, $J^{ab} = L^{ab}$.

[3] $L_{ab}$ is a graded-antisymmetric tensor, which means $L_{ab} = -(-1)^{[a][b]}L_{ba}$ where $[a]$ measures the grading of the index $a$, which is equal to 0 for bosonic indices, and 1 for fermionic indices. This implies that $L_{\theta\bar{\theta}} = L_{\bar{\theta}\theta}$.

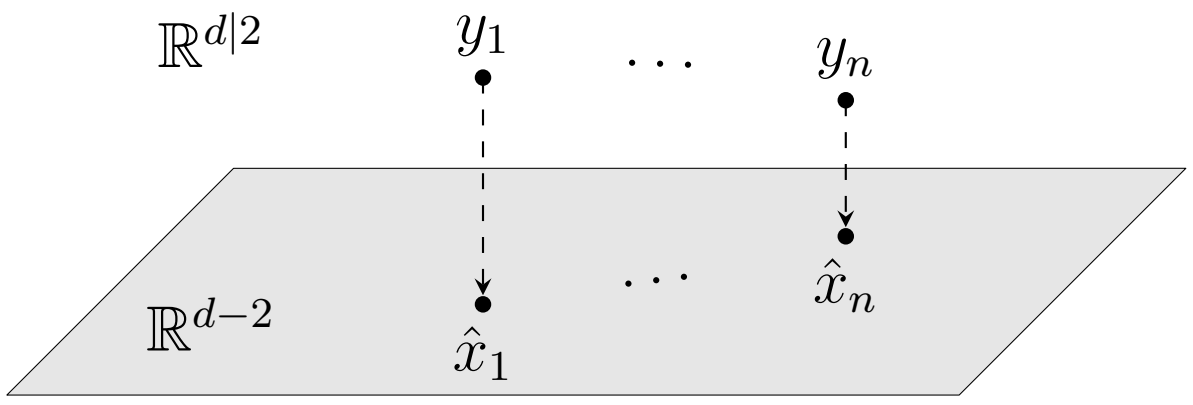

Figure 2: Dimensional reduction: all insertions $y_i \in \mathbb{R}^{d|2}$ are projected to $\mathbb{R}^{d-2}$.

where $\hat{x} \in \mathbb{R}^{d-2}$ and the derivatives also live in $d-2$ dimensions. The two actions compute the same observables when restricted to the same $\mathbb{R}^{d-2}$ as follows,

$$\langle \Phi(y_1) \dots \Phi(y_n) \rangle_S \, \big|_{\mathbb{R}^{d-2}} \, , = \langle \phi(\hat{x}_1) \dots \phi(\hat{x}_n) \rangle_{\hat{S}} \, , \tag{2.5}$$

where we define the restriction to $\mathbb{R}^{d-2}$ by setting the last four components of the superspace insertions to zero, namely $x_i^{d-1}, x_i^d, \theta_i, \bar{\theta}_i = 0$. We stress that (2.5) is very non-trivial since the correlator in the left-hand side is computed in the $d$-dimensional supersymmetric theory (2.1), while the right-hand side is computed in the $(d-2)$-dimensional theory (2.4). Nevertheless this formula for the dimensional reduction of Lagrangian theories was proven in various ways, perturbatively [1] and non-perturbatively [25–29]. See also [31,32] for arguments based on SUSY localization.

Now let us consider a generic PS CFT. Here we do not assume the existence of a Lagrangian, but we define the model as an infinite set of correlation functions of superprimaries which are closed under OPE and respect crossing symmetry. The reduction of this PS CFT is defined by taking the full set of correlators and simply setting all superprimaries to $\mathbb{R}^{d-2}$ as in figure 2, namely

$$\langle \mathcal{O}_1(y_1) \dots \mathcal{O}_n(y_n) \rangle |_{\mathbb{R}^{d-2}} \, . \tag{2.6}$$

The statement that the set of correlators (2.6) defines a good $\text{CFT}_{d-2}$ is already very non trivial. In fact, when imposing this type of restriction — sometimes referred to as a "trivial defect" — one must also account for infinitely many additional observables that arise from considering derivatives in the transverse direction $\mathbb{R}^{2|2}$, which are not included in (2.6). Without including these operators one expects that the set of correlators does not to close under OPE. However, because of PS SUSY, one finds that all such operators decouple and that (2.6) defines a consistent $\text{CFT}_{d-2}$, as shown in [29].

Let us describe this decoupling in more detail. The theory living on the trivial defect respects a smaller symmetry $SO(d-1,1) \times OSp(2|2)$ (since we selected a special $\mathbb{R}^{d-2}$), where $SO(d-1,1)$ plays the role of the conformal symmetry of the reduced theory while $OSp(2|2)$ is a global symmetry. All the operators defined through the prescription (2.6) are singlets under $OSp(2|2)$. It is well known that the singlet sector of a CFT defines itself a closed subsector, however there are many $OSp(2|2)$-singlets that we did not consider in (2.6). E.g. given an operator $\mathcal{O}(y)$ one could also consider the infinite tower of primary operators $(\partial_\perp^2)^n \mathcal{O}(\hat{x})$, where $\partial_\perp^a$ are derivatives in the transverse space $\mathbb{R}^{2|2}$. The statement is that this infinite tower decouples when inserted in any correlator of the type (2.6) when $n > 0$. Therefore we are left with a one-to-one map between the operator $\mathcal{O}(y)$ and $\mathcal{O}(\hat{x})$ instead of a map one-to-infinity. The reason of the decoupling is simple: the idea is that one cannot construct any $OSp(2|2)$-invariant out of the transverse $OSp(2|2)$-metric (which is used to define $\partial_\perp^2$), since it is supertraceless (see [29] for more details).

A similar argument demonstrates that, given a conserved superstress tensor $\mathcal{T}^{ab}(y)$, there exists a single operator obtained by restricting both $y$ and the indices to $\mathbb{R}^{d-2}$,[4] which remains conserved and plays the role of the lower dimensional stress tensor. The conservation in lower dimensions implies that energy cannot leak orthogonally to the defect, which is quite counterintuitive. This happens because the orthogonal leak is measured by an operator that decouples.

Another similar phenomenon happens at the level of the OPE. When we restrict the OPE of two superprimaries of a PS CFT and we plug it in (2.6) we find that each exchanged operator is also mapped to (at most) one operator in lower dimensions, instead of infinitely many. This can be also explained in terms of conformal blocks: a PS superspace superconformal block is equal to a single conformal block in two less dimensions (while a normal conformal block is expanded in a infinite sum of blocks in lower dimensions).

All these facts point to the conclusion that the PS $\text{CFT}_{d|2}$ and the $\text{CFT}_{d-2}$ are in one-to-one correspondence at a kinematic level, which suggest that instead of dimensionally reducing the $\text{CFT}_{d|2}$, one could try to dimensionally uplift a $\text{CFT}_{d-2}$. This means that given any $\text{CFT}_{d-2}$ one would want to embed this into a $\text{CFT}_{d|2}$. This is what we want to achieve in this note.

Unfortunately, the uplift is not such a straightforward task because the map between operators of the PS $\text{CFT}_{d|2}$ and the $\text{CFT}_{d-2}$ is not really one-to-one. Specifically, the $\text{CFT}_{d|2}$ is actually a larger theory than the $\text{CFT}_{d-2}$. In fact, there are superprimaries in the PS $\text{CFT}_{d|2}$ that do not have any counterpart in the $\text{CFT}_{d-2}$; in other words, they are projected to zero by the dimensional reduction (2.6). These operators are relatively rare, as they belong to sufficiently complicated spin representations. Let us explain this in more detail. A superprimary transforms in a specific representation of $OSp(d|2)$, which is labeled by a Young tableau $\ell$. When we restrict its indices to $\mathbb{R}^{d-2}$, this representation is mapped to a representation of $SO(d-2)$, characterized by the same Young tableau $\ell$. The key observation is that in some cases, the dimension of the representation $\ell$ of $SO(d-2)$ is zero. When this happens, the operator is projected to zero due to dimensional reduction. Let us give a simple example. A spin-two (traceless and graded-symmetric) super tensor $\mathcal{O}^{ab}$ has components $\mathcal{O}^{\mu\nu}$, $\mathcal{O}^{\mu\bar{\theta}}$, $\mathcal{O}^{\mu\bar{\theta}}$, $\mathcal{O}^{\mu\theta}$ (and is traceless) so its dimension is $\frac{d(d+5)}{2}$, meaning that it exists in any positive dimension. Conversely, the dimension of the spin two representation of $SO(d-2)$ is $\frac{d(d-3)}{2}$, which vanishes for $d=3$ (reflecting the fact that there is no spin in one dimension). Thus, in this case, the operator $\mathcal{O}^{ab}$ is projected to zero in $d=3$. When $d>3$ one can show that all spin $\ell$ superprimaries reduce to non-vanishing spin $\ell$ primaries of the $\text{CFT}_{d-2}$, however other more complicated (mixed-symmetric) spin representations may reduce to zero when $d$ is low enough. Such operators do not exist in the lower dimensional theory so it is hard to reconstruct them in the uplifted theory.

This will not pose an obstruction for the rest of the note, as we will focus on the infinitely many cases where this subtlety does not arise, allowing us to simply uplift the correlation functions.

[4] The operator is also made into a traceless and symmetric representation of $SO(d-2)$. The restriction of the indices can be done simply using polarization vectors. Supertensors are contracted with a set of graded-symmetric superspace polarization vectors that square to zero. The restriction is obtained by replacing them with usual $\mathbb{R}^{d-2}$ polarization vectors, which also square to zero. As explained below, this reduced operator exists only for $d>3$.

### 2.3 PS uplift

Let us now discuss a simple, yet very rich case, where we can fully reconstruct PS CFT correlators from the knowledge of the lower-dimensional correlators. To do so, we consider a four-point function of scalar operators $O_i(\hat{x}_i)$ (with dimensions $\Delta_i$) in a given CFT$_{d-2}$. We focus on $d \geq 4$ such that the correlator is defined at least on a plane.

Now, by knowing the four-point function on the plane, we can simply reconstruct it on $\mathbb{R}^{d|2}$ using symmetries, namely:

$$\langle O_1(\hat{x}_1)\cdots O_4(\hat{x}_4)\rangle_{\mathrm{CFT}_{d-2}} = \langle \mathcal{O}_1(y_1)\cdots \mathcal{O}_4(y_4)\rangle\Big|_{\mathbb{R}^{d-2}} \underset{\text{symmetries}}{\longrightarrow} \langle \mathcal{O}_1(y_1)\cdots \mathcal{O}_4(y_4)\rangle\,. \tag{2.7}$$

In practice, to achieve this uplift, one can simply take the standard function $f(u,v)$ of the cross-ratios of the CFT$_{d-2}$ and replace $u, v$ with $U, V$. The full four-point function in superspace is then defined by $f(U,V)$. This contains a lot of information, as each operator $\mathcal{O}_i(y_i)$ can be expanded into components. In fact, there are generally 43 different components, which we label by $s = 1, \ldots, 43$, obtained by expanding $f(U,V)$ in terms of $\theta_i, \bar{\theta}_i$.[5] Each term corresponds to a well-defined four-point function of a CFT$_d$, characterized by a function $f_s(u,v)$. Since the $f_s(u,v)$ are derived from the Taylor expansion of $f(U,V)$, these functions can be written as the action of a differential operator $\mathcal{D}_s$ on $f(u,v)$. This provides a prescription for the dimensional uplift: given any scalar four-point function defined by $f$, the full set of uplifted four-point functions $f_s$ is determined by

$$f(u,v) \underset{\text{uplift}}{\longrightarrow} f_s(u,v) = \mathcal{D}_s f(u,v)\,, \qquad (s = 1, \ldots, 43)\,, \tag{2.8}$$

where $\mathcal{D}_s$ are 43 differential operator in $u$ and $v$ which can be function of $\Delta_i$ and $d$. These are explicitly computed in [30]. The simplest differential operator is the one associated to the lowest component which is just equal to 1. Less trivial cases arise e.g. by considering the component proportional to $\theta_1\bar{\theta}_2$ and $\theta_1\theta_2\bar{\theta}_3\bar{\theta}_4$ which respectively give the differential operators

$$(-\Delta_1 - \Delta_2) + 2u\partial_u\,, \tag{2.9}$$
$$u\left[4u^2\partial_u^2 + 4(v-1)v\partial_v^2 + 8uv\partial_u\partial_v - \Delta_{12}\Delta_{34} - 2\left(\Delta_{12} - \Delta_{34} - 2\right)\left(u\partial_u + (v-1)\partial_v\right)\right]\,. \tag{2.10}$$

Now that we have a definition for $f_s(u,v)$, we can check that these define a consistent CFT$_d$. In particular imposing that $f(u,v)$ is crossing covariant and can be expanded in $(d-2)$-dimensional conformal blocks, we must check that

- $f_s(u,v)$ are crossing covariant,
- $f_s(u,v)$ are decomposed in $d$-dimensional conformal blocks.

In [30], it was shown that both properties are always satisfied. Crossing is quite involved, as it can relate different components, but it occurs automatically due to the form of the differential operators. The fact that the conformal block decomposition is also automatic is a very non-trivial result, leading to 43 unexpected formulae that relate conformal blocks in different dimensions, namely[6]

$$\mathcal{D}_s g^{(d-2)}_{\Delta,\ell} = \sum_{\substack{i,j \\ \leq 5 \text{ terms}}} c^{(s)}_{i,j}\, g^{(d)}_{\Delta+i,\ell+j}\,, \tag{2.11}$$

[5] To be precise, we need to expand in primary components. In particular, the highest component of a superprimary $\mathcal{O}(y)$ is defined by $\mathcal{O}_{\theta\bar{\theta}}(x) \equiv \partial_\theta\partial_{\bar{\theta}}\mathcal{O}(x)$, which is not a primary. The highest primary component is simply obtained by improving the operator $\mathcal{O}_{\theta\bar{\theta}}$ as follows: $[\partial_x^2 + (2-d+2\Delta)\partial_\theta\partial_{\bar{\theta}}]\mathcal{O}(x)$.

[6] The formula is schematic. In particular the dependence on $\Delta_i$ of the conformal blocks in the right hand side should be shifted by some units to take into account that the external operators might be higher components with dimensions $\Delta_i$, $\Delta_i+1$ or $\Delta_i+2$. This is defined more precisely in [30].

where $g^{(d)}_{\Delta,\ell}$ is a $d$-dimensional block for the exchange of an operator with dimension $\Delta$ and spin $\ell$. The explicit set of summed values of $i,j$ depends on $s$. For concreteness, depending on $s$, the sum over $(i,j)$ runs over one of the following three sets: $\{(0,0),(0,-2),(1,-1),(2,0),(2,-2)\}$, $\{(0,-1),(1,0),(1,-2),(2,-1)\}$, or $\{(1,-1)\}$. The full set of kinematical coefficients $c^{(s)}_{i,j}$ is known in a closed form. See [30] for the precise definitions of all ingredients appearing in (2.11).

Let us exemplify (2.11) for the case of the component proportional to $\theta_1\theta_2\bar\theta_3\bar\theta_4$, where the sum over $i,j$ contains only one term,

$$\mathcal{D}\, g^{(d-2)}_{\Delta,\ell} = 2(\Delta-1)\ell\, g^{(d)}_{\Delta+1,\ell-1}\,. \tag{2.12}$$

Equation (2.12) is very useful on its own: by knowing the blocks in $d-2$ dimensions, it allows the computation of blocks in $d$ dimensions simply by applying the differential operator $\mathcal{D}$ defined in (2.10). Equation (2.12) was already known in the literature (see equation (4.37) of [33]), but its physical meaning was unclear. We now understand that it arises due to PS supersymmetry.

# 3 Applications

Now that we have the prescription (2.8), we can select any scalar four point function of any $\mathrm{CFT}_{d-2}$ (for $d \geq 4$) and obtain a set of 43 uplifted correlators which are compatible with the bootstrap axioms and live in two higher dimensions. This procedure can be iterated any time and for each iteration we obtain a theory with larger and larger SUSY which lives in higher and higher dimensions.

Let us stress that (2.11) not only implies that $f_s$ are decomposed in the blocks $g^{(d)}_{\Delta,\ell}$, but also it can be used to determine the precise spectrum of the operators and OPE coefficients exchanged in the uplifted correlators. Namely

$$f(u,v) = \sum_{\Delta,\ell\in\mathcal{S}} a_{\Delta,\ell}\, g^{(d-2)}_{\Delta,\ell} \quad \underset{\text{uplift}}{\longrightarrow} \quad f_s(u,v) = \sum_{\Delta,\ell\in\mathcal{S}} \sum_{\substack{i,j \\ \leq 5 \text{ terms}}} a_{\Delta,\ell}\, c^{(s)}_{i,j}\, g^{(d)}_{\Delta+i,\ell+j}\,, \tag{3.1}$$

where the set $\mathcal{S}$ of exchanged operators is the same in both sums, however on the right hand side they have dimensions and spins opportunely shifted by $i$ and $j$. The knowledge of the OPE coefficients $a_{\Delta,\ell}$ in lower dimensions fully specifies the OPE coefficients in the uplift which are the same dressed by the known kinematic factors $c^{(s)}_{i,j}$.

By examining (3.1), we can identify signs of the non-unitarity in the uplifted theory. Specifically, we observe that the uplifted spectrum may include operators that fall below the unitarity bounds. For instance, a scalar operator with dimension $\Delta = \frac{d-4}{2} + x$ lies above the unitarity bounds of the $\mathrm{CFT}_{d-2}$ for any $x \geq 0$. This operator can be exchanged in $\mathcal{S}$ and thus appear also in the uplifted exchanges (notice that for some components $s$, the sum in $i,j$ contains also the term $i,j = 0$). From this, we can already deduce that for $0 \leq x < 2$ the operator lies below the unitarity bounds of the uplifted theory. Therefore, a unitary operator of the $\mathrm{CFT}_{d-2}$ becomes an operator below unitarity in the PS CFT. A similar situation arises for OPE coefficients: the coefficients $c^{(s)}_{i,j}$ are not always positive even when the associated correlators contains equal operators. So the uplifted OPE coefficients are not always real.

In the following we focus on two simple examples of uplifts —of GFF and minimal models— where the lower dimensional correlators are known in a closed form and thus we can explicitly apply (2.8).

### 3.1 Bootstrapping GFF using SUSY

The first application is to GFF, where we will see that PS SUSY can be used as a tool to "bootstrap" an infinite set of OPE coefficients.

Let us consider GFF in $d-2$ dimensions. GFF is defined by the non-local Lagrangian $\phi\Box^\kappa\phi$ for real $\kappa$ and reduces to the usual local free theory when $\kappa = 1$. Another way to define it is by specifying the two-point function of $\phi$ with scaling dimension $\Delta_\phi$ and impose that correlators with an arbitrary number of $\phi$ are obtained by Wick contractions. It is easy to define the full uplift of this theory, since we can uplift its Lagrangian as $\Phi(y)(\partial_y^a\partial_{ya})^\kappa\Phi(y)$ in the superspace $\mathbb{R}^{d|2}$ where we keep the same conventions as in (2.1). Similarly we can define the uplift also at the level of Wick contractions, schematically

$$\text{GFF}_{d-2}\left\{\begin{array}{c}\langle\phi(x_1)\phi(x_2)\rangle = (x_{12}^2)^{-\Delta_\phi}\\ +\\ \text{Wick contractions}\end{array}\right. \underset{\text{uplift}}{\longrightarrow} \text{PS GFF}_{d|2}\left\{\begin{array}{c}\langle\Phi(y_1)\Phi(y_2)\rangle = (y_{12}^2)^{-\Delta_\phi}\\ +\\ \text{Wick contractions}\end{array}\right. .$$

We can now consider any scalar four-point and apply (2.8).[7] Let us first consider the simplest: the four-point function of $\phi$. By following the prescription (2.8) we find a somewhat unexpected feature of the PS GFF: some components have vanishing four-point function, namely

$$f(u,v) = 1 + u^{\Delta_\phi} + \left(\frac{u}{v}\right)^{\Delta_\phi} \underset{\text{uplift}}{\longrightarrow} f_s(u,v) = \mathcal{D}_s f(u,v) = 0\,, \quad \text{for } s \in N\,, \tag{3.2}$$

where the precise set $N$ of vanishing components is detailed in [30] and $f(u,v)$ is the usual GFF four-point function of $\phi$. It is actually very easy to predict when a component of $\langle\Phi\Phi\Phi\Phi\rangle$ is zero. It is due to the factorization provided by Wick contractions together with the simple fact that two-point functions of different primaries vanish. For instance let us consider the case of 3 lowest components $\varphi$ with 1 highest component $\hat{\omega}$.[8] By Wick contractions, the four point function is always proportional to terms of the form $\langle\varphi\hat{\omega}\rangle\langle\varphi\varphi\rangle$ which vanish because of $\langle\varphi\hat{\omega}\rangle = 0$.

From (3.2) we learn that the function $f(u,v)$ must satisfy the differential equation $\mathcal{D}_s f(u,v) = 0$. This might look trivial since we know the explicit form of $f(u,v)$. However what is non-trivial is that $f(u,v)$ was computed in the usual GFF, which should not know about SUSY. Nevertheless this function secretely satisfies properties dictated by the supersymmetry of the uplift.

A much more powerful application of this philosophy allows us to determine the conformal block decomposition $f(u,v)$. The idea is to use (3.1) when $s \in N$. In this case we notice that

$$0 = \sum_{\Delta,\ell\in\mathcal{S}} \sum_{\substack{i,j\\ \leq 5 \text{ terms}}} a_{\Delta,\ell}\, c_{i,j}^{(s)}\, g^{(d)}_{\Delta+i,\ell+j} = \sum_{\Delta,\ell\in\mathcal{S}'} \underbrace{b^{(s)}_{\Delta,\ell}}_{=0}\, g^{(d)}_{\Delta,\ell}\,, \quad s \in N\,. \tag{3.3}$$

In the last equality we only reorganized the sum in a way that all conformal blocks appear with different dimensions and spin, so that they are linearly independent (thus implying that they all must be multiplied by zero, for the sum to be zero). The coefficients $b^{(s)}_{\Delta,\ell}$ are just linear combinations of $a_{\Delta,\ell}$ and the kinematical coefficients $c^{(s)}_{i,j}$. Therefore $b^{(s)}_{\Delta,\ell} = 0$ is a constraint on the OPE coefficients $a_{\Delta,\ell}$ of the original GFF! In [30] it was further inputted that the exchanged operators in GFF contains double twist operators and it was shown that the constraint takes the form of a recurrence relations for $a_{\Delta,\ell}$ which could be solved exactly. This is quite nice since the same logic can be applied to correlators of the form $\langle\phi\phi^n\phi^m\phi^k\rangle$ and when $m = k$ it was found a closed form expression for the exchanged operators in the OPE of this doubly-infinite family of correlators. Besides the explicit result, it is also interesting to see that all the operators of the double twist families of (non-SUSY) GFF are tied together by the supersymmetry of the uplift. This is a case where the existence of the uplift helps in bootstrapping the original theory. In [30] it is also discussed how this philosophy can be extended to perturbation theory around GFF, giving rise to sets of differential equation that GFF Feynman integrals must satisfy.

[7] Note that in this case we have an independent definition of the uplifted theory so we could also compute by Wick contractions all possible observables, even correlators of spinning operators and higher point functions.

[8] To be precise, according to footnote 5, we consider the highest *primary* component of $\Phi$, which we call $\hat{\omega}$.

## 3.2 Defining solvable CFT$_{d>2}$

Let us now focus on minimal models (MMs), the most famous class of interacting theories where four-point functions are known in closed form.

For any such scalar four-point function, we can easily apply equation (2.8) to define the associated uplifted correlators. However, for infinite classes of MMs, we can take a step further: we also have an RG argument that guarantees the existence of a complete uplifted theory. This stems from the fact that many MMs arise as the IR fixed point of a scalar action, and we know that any scalar action (2.4) can be uplifted to a superspace action (2.1). Thus, not only can we uplift the IR fixed point, but we can uplift the entire flow. The simplest example is given by the diagonal MMs which are obtained by starting with a UV free scalar and perturbing it with $\phi^n$ (one might focus on even powers of $\phi$ to keep unitarity), where to reach the (multicritical) fixed point we further need to tune the couplings of lower powers of $\phi$. This RG can be fully uplifted by considering free theory in superspace and perturb it by $\Phi^n$ as shown in figure 3.[9]

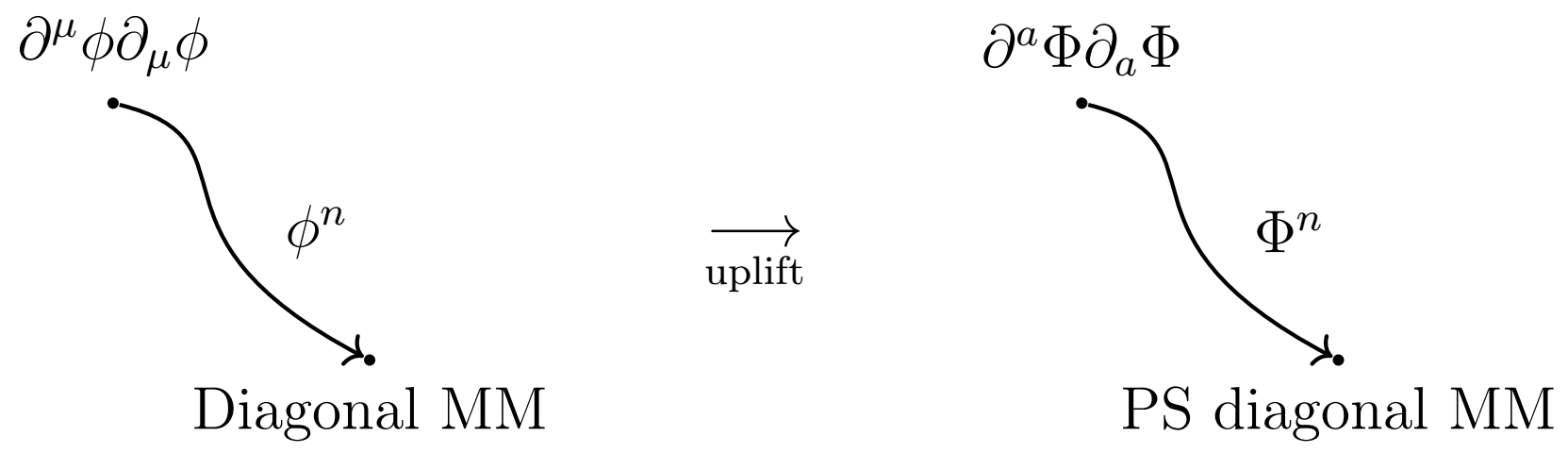

Figure 3: Uplift of diagonal minimal models.

Now that we know that these models have a well defined uplift, we can study their properties. To actually study MMs and their uplift we give up the Lagrangian description above since the correlation functions can be bootstrapped exactly [35]. One of the features of MMs is that their CFT data is rational. Applying (3.1), we notice that dimensions and OPE coefficients must also be rational numbers in the uplifted model. However, as expected, there are operators below the unitarity bounds and squared OPE coefficients are sometimes negative.

MMs have infinitely many conserved global primary currents, which arise from the holomorphic (or antiholomorphic) part of the Virasoro module of the identity. Notice that the number of higher spin conserved currents grows exponentially with the spin (while for a usual free boson in $d > 2$ there is a single one per spin). Each spin-$\ell$ conserved currents is uplifted to a conserved superprimary current $\mathcal{J}^{a_1 \cdots a_\ell}$ whose supermultiplet contains various conserved primaries currents. Thus, the uplifted MMs are interacting $4d$ theories with an infinite number of currents (growing exponentially with the spin). Notice that Maldacena and Zhiboedov [36] proved that higher spin conserved currents lead to non-interacting theories in $d > 2$. Their theorem however required unitarity as an assumption and does not apply to our case.

[9]Other scalar Ginzburg-Landau Lagrangians are expected to flow to MMs, see e.g. [34] for some recent results and a review of the state of the art of the subject. One could also uplift all these RG flows to define solvable 4$d$ models.

Of course from each conserved current one can build several topological operators, which can be thought as the conserved charges of the theory. We notice that the number of such charges is much larger for the PS MMs, just because the dimension of traceless and symmetric representations is larger in higher dimensions. E.g. we have two translations in $\mathbb{R}^2$, while there are four bosonic and two fermionic translations in $\mathbb{R}^{4|2}$. In this sense we can say that the symmetry of the PS MMs must be larger than the one of MMs.

Let us exemplify the case of the Ising MM, which contains three Virasoro primaries: the identity $\mathbb{I}$, $\sigma$, and $\epsilon$, where $\sigma$ and $\epsilon$ are scalars with dimensions $\Delta = 1/8$ and 1, respectively. The four-point functions of these primaries are known and can be uplifted using (2.8). By applying Virasoro generators, one can obtain the four-point functions for any global primary, and if the global primaries are scalars, they can also be uplifted using (2.8). Thus, in principle, we can express the uplifted four-point functions of all scalar (global primary) operators in the PS Ising MM.

The PS Ising MM contains primaries below the $4d$ unitarity bound ($\Delta \geq 1$): for instance the lowest component of the uplift of $\sigma$ has dimension $1/8$.

In the PS Ising MM a special phenomenon occurs: the model contains null states that cannot be modded out. This is because of the presence of $\epsilon$ as well as all the global primaries in the holomorphic (or antiholomorphic) part of its Virasoro multiplet. Let us focus on the superprimary associated with the uplift of $\epsilon$. Its lowest and highest primary components $\mathcal{E}_L$ and $\mathcal{E}_H$, are $4d$ scalars with dimensions $\Delta = 1$ and 3, respectively. It happens that $\mathcal{E}_H = \partial_x^2 \mathcal{E}_L$, thus $\mathcal{E}_H$ is both a primary and a descendant, so it has zero norm, as do all states in its multiplet. Despite this, correlators involving $\mathcal{E}_H$ can be computed and generally do not vanish. Similarly, $\mathcal{E}_L$ has the quantum numbers of a free scalar in $4d$, but it is not annihilated by $\partial_x^2$, so its multiplet includes the multiplet of $\mathcal{E}_H$, which contains infinitely many null states. The presence of zero norm states may seem puzzling, as their exchange in the OPE would typically lead to divergences. However, SUSY ensures that any exchange of $\mathcal{E}_L$ is precisely compensated an exchange of $\mathcal{E}_H$, which cancels the contributions from all null states. In terms of conformal blocks, the divergent block $g_{\mathcal{E}_L}$ by SUSY always appears in a linear combination $g_{\mathcal{E}_L} + c g_{\mathcal{E}_H}$, where $g_{\mathcal{E}_H}$ is finite, and the divergent coefficient $c$ is fine tuned to regularizes the exchange. From the representation theory point of view, the regularized block defines the exchange of a conformal multiplet which is reducible but not decomposable into irreducible ones, and it is referred to as an "extended" or "staggered" module (see e.g. [37–39]).[10]

To conclude we showed that the uplift of MMs exist and that all scalar four-point functions in these theories can be computed. The PS MMs have many interesting features that would be interesting to further investigate, like their infinite dimensional symmetry. Also one could in principle iterate the uplift and generate non-unitary solvable models in every even dimension.

[10] Following section 3.1 of [38], the highest component $\mathcal{E}_{\theta\bar{\theta}}$ of $\mathcal{E}$ is neither a primary nor a descendant. To be precise the highest component $\mathcal{O}_{\theta\bar{\theta}}(x) = \partial_\theta \partial_{\bar{\theta}} \mathcal{O}(y)$ of a generic scalar superprimary $\mathcal{O}$ with dimension $\Delta$ is never a primary operator since $[K^\mu, P_\theta P_{\bar{\theta}} \mathcal{O}(0)] = -2P^\mu \mathcal{O}(0)$. For this reason we define a *primary* highest component $\mathcal{O}_H(x) = \partial_x^2 \mathcal{O}_L + (2 - d + 2\Delta)\mathcal{O}_{\theta\bar{\theta}}$, and $\mathcal{O}_{\theta\bar{\theta}}$ is a linear combination of the primary $\mathcal{O}_H$ with the descendant $\partial_x^2 \mathcal{O}_L$. In our case $d = 4$ and $\Delta = 1$ we notice that $\mathcal{E}_H = \partial_x^2 \mathcal{E}_L$, thus $\mathcal{E}_{\theta\bar{\theta}}$ is still a well defined operator but it cannot be written as linear combination of primaries and descendants. By acting with $K^\mu$ on any state of the form $\mathcal{E}_{cycl} = \mathcal{E}_H + \alpha \mathcal{E}_{\theta\bar{\theta}}$ for $\alpha \neq 0$, we land on the multiplet of $\mathcal{E}_L$. Thus $\mathcal{E}_{cycl}$ is a cyclic vector from which we can recover the full staggered module by the action of the conformal algebra. PS SUSY thus provides natural examples of staggered modules in interacting higher-dimensional theories.

# 4 Conclusions

In this note we reviewed PS supersymmetry and how this can be used to define a dimensional uplift of a given model.

We demonstrated that the uplift is automatic at the level of scalar four-point functions. Specifically any such correlators of a $\mathrm{CFT}_{d-2}$ can be uplifted following (2.8), giving rise to a set of 43 four-point functions related by PS SUSY in the PS $\mathrm{CFT}_d$. The uplifted correlators are inherently crossing covariant and decomposable into $d$-dimensional conformal blocks. This decomposition is completely determined by that of the reduced theory, as shown in (3.1).

As a byproduct, we found that PS SUSY provides 43 relations between conformal blocks in different dimensions, shown schematically in (2.11). The simplest of these, in equation (2.12), allows for the computation of blocks in higher dimensions from those in lower dimensions.

We then exemplified the uplift in two cases. First we considered GFF and showed that the uplifted PS GFF has some special features which allow for a straightforward computation of the OPE coefficients in the conformal block decomposition of an infinite set of correlators. This, in turn, implies that usual GFF is strongly constrained by SUSY of the uplifted PS GFF.

Next, we turned to minimal models. We explained why the complete uplifted theories must exist, at least in certain infinite classes of examples. We also highlighted some of their features, such as the presence of infinitely many higher-spin conserved currents and associated charges. Additionally, we described how the uplifted Ising model contains zero-norm states in its spectrum that do not decouple. Despite this, we explained that the resulting theory remains well-defined.

There are several open directions for further exploration. First, one could extend the uplift (i.e., formula (2.8) and its consequences, such as (2.11) and (3.1)) to more general types of observables. For instance, one might investigate four-point functions of spinning operators or higher-point functions. Another promising avenue is to consider the uplift of correlators involving extended observables, such as defects [40]. In all cases, we expect the uplift to be automatic, yielding new relations between more general types of conformal blocks across dimensions.

It would also be interesting to study these more general observables in specific examples to explore whether they offer novel approaches to bootstrapping a model. For instance, one could examine the uplifted four-point function of spinning operators, focusing on the scalar components of their supermultiplets. By bootstrapping the resulting scalar four-point function in $d$-dimensions, one could gain valuable insights into the spinning bootstrap problem in $(d-2)$-dimensions, which is technically much more challenging.

Another intriguing direction is to better understand the uplift of minimal models, their symmetries, and whether it exists a prescription (such as an uplifted version of the BPZ equations [35]) to generate all possible correlators directly in four dimensions. Similarly, one could apply the uplift to virtually any CFT and seek to understand the resulting model. For example one could study the uplift of the $2d$ free boson or the $2d$ Liouville theory, which both have a Lagrangian description. The resulting models have $4d$ Lagrangian descriptions which completely define the full theory. A compelling avenue is to define these uplifted models through a probabilistic construction (see e.g. [41]).

It would also be interesting to consider deformations of the uplifted models. If these deformations are chosen to break SUSY, the resulting flow would not have any lower-dimensional counterpart and might lead to the discovery of new higher dimensional CFTs.

Finally, it would be highly desirable to establish a proof of the following conjecture: given any $\mathrm{CFT}_{d-2}$, there exists a unique uplifted PS $\mathrm{CFT}_d$. While we believe this to be true and have provided several arguments in support of it, we hope that a complete proof will be found in the future.

# Acknowledgements

We would like to thank the participants of the conference "Analytic results in CFT" for valuable discussions and Sylvain Ribault for comments on the draft. This research was partially supported by the European Research Council (ERC) under the European Union's Horizon 2020 research and innovation programme (grant agreement number 949077).

# Symmetry TFT, chiral TFT, and 2dCFT

**Ingo Runkel**

*Fachbereich Mathematik, Universität Hamburg,*
*Bundesstraße 55, 20146 Hamburg, Germany*

*E-mail:* `ingo.runkel@uni-hamburg.de`

Abstract: A brief exposition of symmetry topological field theory will be given in low dimensions. For suitably finite symmetries, the symmetry TFT has a state sum description which is presented in detail in the case of 1d and 2d quantum field theories. In the example of 2d conformal field theory, there is a second relation to 3d TFT, namely to the so-called chiral TFT which is related to symmetry TFT via folding.

# Contents

# 1 Introduction

SymTFT, or symmetry topological field theory, refers to the idea that topological defects in a $d$-dimensional quantum field theory (QFT) can be used to define a coupled system where the $d$-dim. QFT becomes a non-topological boundary condition of a $(d+1)$-dim. topological field theory (TFT).

For 1d QFTs one can be very explicit, and we will go through the construction of the corresponding 2d state sum TFT slowly and in detail in Section 2. Even though the 1d setting is very simple, the key ideas are already visible there. For 2d, the construction is more involved, but not much so, and in Section 3 we present it again in some detail as this is the first place where higher categorical data enters in the description of topological symmetries. Analogous constructions to those presented in Sections 2 and 3 should work in any dimension $d$, but the description of the state sum TFT becomes much more involved and is explicitly available only in the next dimension up, i.e. for 4d TFT. Higher dimensional examples can still be found, for instance when the TFT arises naturally in an action formulation without the need to go through a state sum construction.

In the case of 2d CFT, topological symmetry can be related to the separation into holomorphic and anti-holomorphic degrees of freedom, resulting in the description of 2d CFT via the chiral 3d TFT. The corresponding SymTFT is related to chiral TFT via folding. The relation between 2d CFT and 3d TFT makes it possible to transfer many questions about CFT to questions about TFT, and the latter can often be addressed by mathematical techniques in braided tensor categories. This allows one to prove analytic results about 2d CFTs, such as the classification of CFTs with given rational chiral symmetry. This is briefly reviewed in Section 4.

All references and pointers to further reading are collected in the bibliography in Section 5.

# 2 One-dimensional QFT and SymTFT

## 2.1 1d QFT and 0d topological defects

A one-dimensional quantum field theory $Q$ is just a quantum mechanical system, determined by a Hilbert space $\mathcal{H}$ and a Hamiltonian $H : \mathcal{H} \to \mathcal{H}$, i.e. $Q = (\mathcal{H}, H)$. If one thinks of the one dimension as time, $H$ determines the unitary time evolution via $U_t = \exp(iHt)$. However, in these notes will will work in euclidean signature. In that case, $H$ determines the bounded linear map assigned to an interval of length $\ell$,

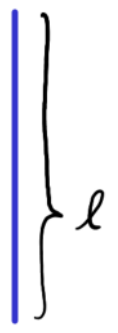

$$\rightsquigarrow \quad e^{-\ell H} : \mathcal{H} \to \mathcal{H} \ . \tag{2.1}$$

A *zero-dimensional topological defect* is a point observable $D$ whose expectation value is independent of its position, as long as one does not move it past other observables $\mathcal{O}$,

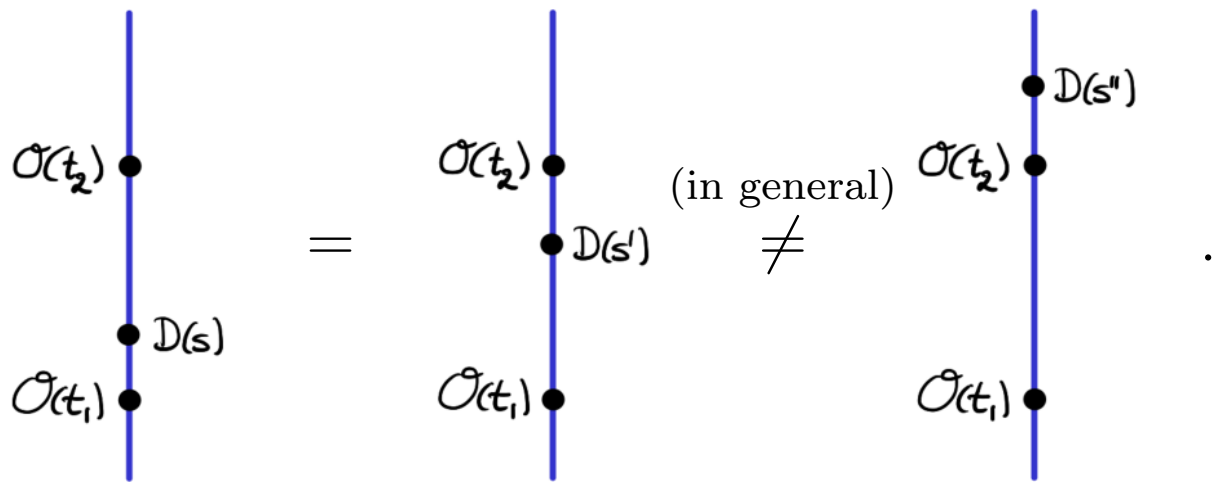

(2.2)

In the present setting, this amounts to a bounded operator $D : \mathcal{H} \to \mathcal{H}$ which commutes with the Hamiltonian, $[H, D] = 0$.

**Example:** Consider two distinguishable particles in two identical boxes (or any two identical copies of a given quantum mechanical system). The state space is $\mathcal{H}_{\text{tot}} = \mathcal{H} \otimes \mathcal{H}$ and the Hamiltonian is $H_{\text{tot}} = H \otimes \text{id} + \text{id} \otimes H$. Write

$$\tau : \mathcal{H} \otimes \mathcal{H} \to \mathcal{H} \otimes \mathcal{H} \quad , \quad a \otimes b \mapsto b \otimes a \ , \tag{2.3}$$

for the linear map that interchanges the two copies of $\mathcal{H}$. Then $D = \tau$ is an example of a 0d topological defect as clearly $[H, \tau] = 0$.

The *fusion of topological defects* is the process of moving two topological defects close to each other so that they become a new topological defect. Here this is just the composition of linear maps,

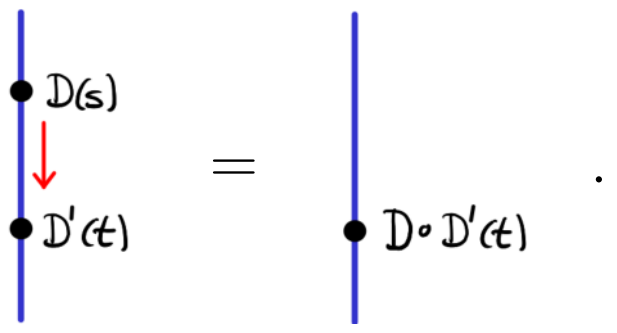

(2.4)

Suppose now we have a finite list of 0d topological defects $D_1, \dots, D_r$ which closes under composition,

$$D_i \circ D_j \ = \ \sum_{k=1}^{r} \mu_{ij}^{\,k} \, D_k \qquad (\mu_{ij}^{\,k} \in \mathbb{C}) \ . \tag{2.5}$$

We will assume that $D_1 = \text{id}$ and that the $D_i$ are linearly independent (we will see later how this assumption can be dropped). Then the $\mu_{ij}^{\,k}$ are the structure constants of a unital associative algebra over $\mathbb{C}$. Explicitly, let $F$ be the $\mathbb{C}$-vector space with basis $e_1, \ldots, e_r$, and with product and unit given by

$$e_i \, e_j \;=\; \sum_{k=1}^{r} \mu_{ij}^{\,k} \, e_k \quad , \quad 1 = e_1 \; . \tag{2.6}$$

We want to think of $F$ as a *finite topological symmetry* of the 1d QFT $Q$. It generalises finite group symmetries as the $D_i$ need not be invertible. Of course one may also consider infinite topological symmetries, but we will restrict ourselves to the simplest case in these notes.

To prepare the description of the 2d SymTFT, we use string diagrams to write product and unit,

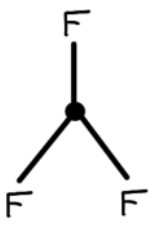 $= \mu : F \otimes F \to F$ , 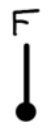 $= \eta : \mathbb{C} \to F$ , (2.7)

where $\eta(\lambda) = \lambda \cdot 1$ with $\lambda \in \mathbb{C}$ and $1 \in F$ is the unit element. In terms of string diagrams, the associativity and unit condition are

= 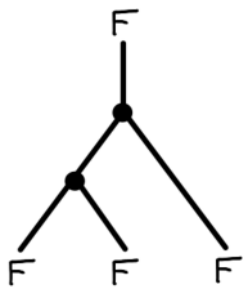 and 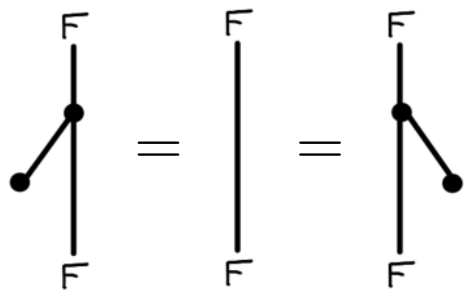 . (2.8)

To apply the two-dimensional state sum construction for topological field theories, we need to make a further assumption on $F$, namely $F$ should be semisimple (isomorphic to a direct sum of matrix algebras). Just as the assumption of finiteness above, this is a restriction on the kind of topological symmetry we consider, but it still generalises the case of a finite group symmetry. Equivalently, we assume that $F$ is a $\Delta$-*separable symmetric Frobenius algebra*. This means that $F$ is in addition equipped with a coproduct $\Delta$ and counit $\epsilon$,

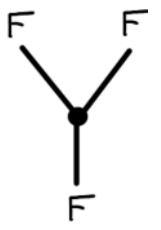 $= \Delta : F \to F \otimes F$ , 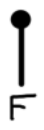 $= \epsilon : F \to \mathbb{C}$ . (2.9)

These have to satisfy counitality, as well as the *fusion move* and *bubble move*:

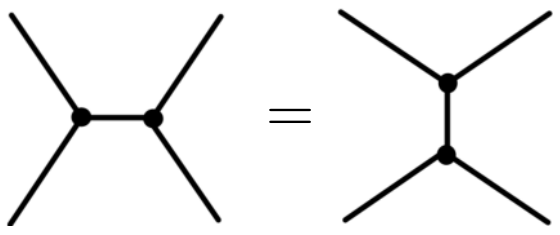 and 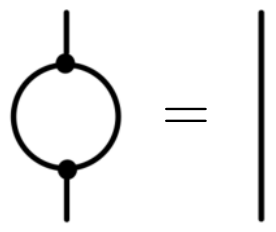 . (2.10)

These two identities have to hold for all allowed orientations on the edges, that is, all orientations so that each vertex has two incoming and one outgoing edge or vice versa.[1]

**Example:** Continuing with the product of two identical quantum systems, $\mathcal{H}_{\text{tot}} = \mathcal{H} \otimes \mathcal{H}$, we take $D_1 = \text{id}$ and $D_2 = \tau$. We have $\tau \circ \tau = \text{id}$, so that $F = \mathbb{C}[\mathbb{Z}_2]$, the group algebra of $\mathbb{Z}_2$. Write $\{1, t\}$ for the basis of $F$ with $t^2 = 1$. Then

$$\Delta(1) = \tfrac{1}{2}(1 \otimes 1 + t \otimes t) \quad , \quad \epsilon(1) = 2 \quad , \quad \epsilon(t) = 0 \; . \tag{2.11}$$

**Exercise:** *Check this. What fails without the factor of $\frac{1}{2}$ for $\Delta$ and 2 for $\epsilon$?*

[1]The bubble move amounts to the qualifier "$\Delta$-separable" (algebraically $\mu \circ \Delta = \text{id}_F$). For an algebra $F$ to be $\Delta$-separable symmetric Frobenius is actually a property, that is, the coproduct and counit are in this case uniquely determined by the product and unit. The coproduct will be convenient in the state sum construction below.

This condition of being "$\Delta$-separable" should (and can) be weakened to just "separable". $F$ being separable is equivalent (over $\mathbb{C}$) to it being semisimple. In this case the separability idempotent is an extra choice which enters the state sum construction and which, in our quest to stick to the simplest setting, we will avoid. Some pointers are given in the bibliography in Section 5.

## 2.2 2d state sum TFT

The state sum construction of a two-dimensional topological field theory takes as an input a $\Delta$-separable symmetric Frobenius algebra $F$ over $\mathbb{C}$. The output is a topological field theory described as a functor from two-dimensional oriented bordisms to vector spaces. We will again restrict to the simplest case and only describe the closed invariants,[2] i.e. the numbers the TFT assigns to (compact oriented) 2-manifolds.

The construction is as follows. Let $M$ be a 2-manifold. For now we take $M$ to have empty boundary and no line defects, but we will add these later. Pick a cell-decomposition of $M$ with contractible 2-cells and only 3-valent vertices, and with oriented edges. The orientation of the edges has to be admissible in the sense that each vertex has two incoming and one outgoing edge or vice versa. Then decorate each edge by $F$, and the vertices by $\mu$ or $\Delta$ according to the orientations and evaluate the corresponding string diagram. The resulting number is independent of the choice of cell decomposition as any two such decomposition are related by a finite sequence of fusion and bubble moves. Here is an illustration in the case of $M$ being a 2-torus,

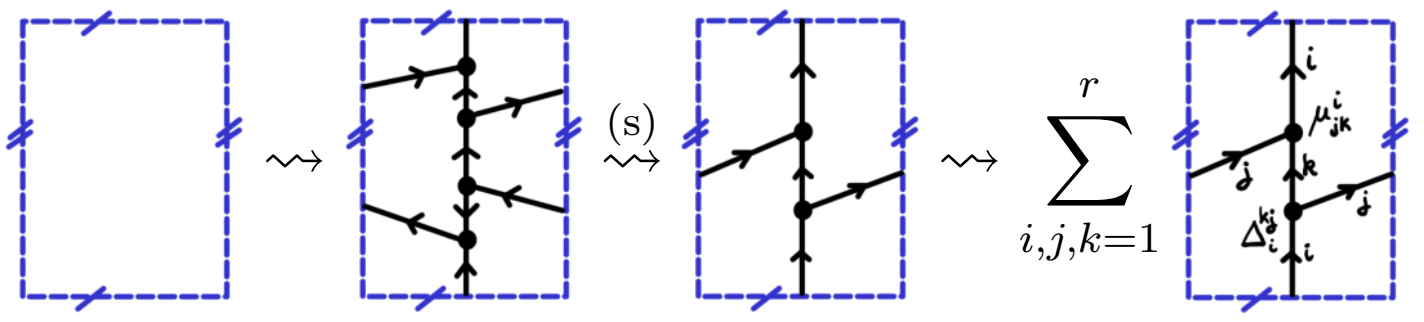

$$, \quad \mathcal{T}_F(T^2) = \sum_{i,j,k=1}^{r} \mu_{jk}^{\,i}\, \Delta_i^{kj} \quad . \qquad (2.12)$$

**Exercise:** *Do the simplification (s) using the fusion and bubble moves. Compute the torus invariant for $F = \mathbb{C}[\mathbb{Z}_2]$. (*) Show that in general the torus invariant is equal to the dimension of the centre of $F$.*

Let us refer to the resulting 2d TFT as $\mathcal{T}_F$. This is the *symmetry topological field theory (SymTFT) for the topological symmetry $F$.*

We will need to be able to evaluate $\mathcal{T}_F$ also on surfaces with boundaries. *Topological boundaries* are boundaries that do not require a metric on the boundary. In the state sum construction, these are described by finite-dimensional left $F$-modules $\Lambda$. An $F$-module $\Lambda$ is a vector space $\Lambda$ and a linear map $\rho_\Lambda : F \otimes \Lambda \to \Lambda$ compatible with the product of $F$,

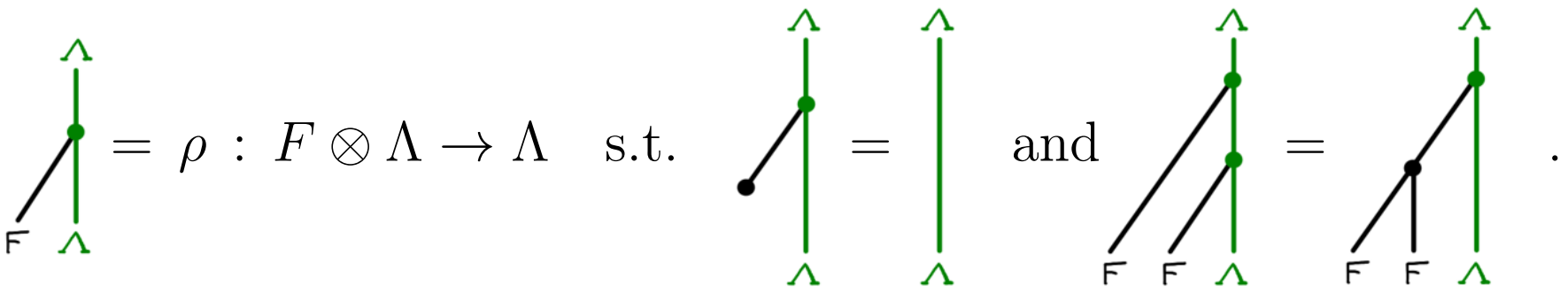

(2.13)

It will be useful to note that because $F$ is a Frobenius algebra, an $F$-module has a canonical co-action which is given by

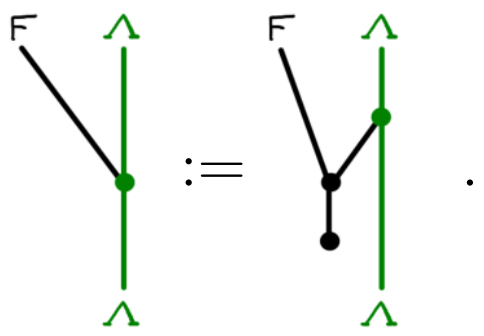

(2.14)

[2]There is a potentially confusing terminology with the use of "closed invariants", in that manifolds – or bordisms – in TFT have different types of boundaries. One type is used when formulating the composition in the bordism category and describes where bordisms are glued along a common boundary to form a new bordism. These are sometimes called *gluing boundaries*. In the type of TFT we consider, the bordisms can also have so-called *free boundaries*. For example, in the bordism category, an annulus with two free boundaries can be obtained by composing two rectangles (think of them as horseshoe shaped), with alternating free and gluing boundaries. The term "closed invariants" refers to the absence of gluing boundaries, i.e. to bordisms from the empty set to the empty set. These bordisms are, however, allowed to have free boundaries.

**Example:** $F = \mathbb{C}[\mathbb{Z}_2]$ has two simple modules, namely the one-dimensional modules $\mathbb{C}_\pm$ with $t$ acting by $\pm 1$. Any algebra is a module over itself with action given by the product, called the *regular module* ${}_F F$.

The description of boundary conditions as $F$-modules is simply the answer to the question "What data and conditions do I need for the state sum construction to be defined in the presence of a boundary such that it remains independent of the choice of cell-decomposition?" The same principle will later give us topological defects in $\mathcal{T}_F$ as $F$-$F$-bimodules.

At this point we are able to evaluate $\mathcal{T}_F$ on surfaces with boundaries labelled by finite-dimensional $F$-modules.

**Exercise:** *Consider the torus with a disc cut out and the resulting boundary labelled* $\Lambda$*:*

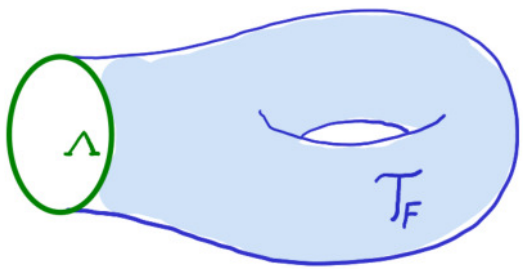

. (2.15)

*In the example of* $F = \mathbb{C}[\mathbb{Z}_2]$*, what is the value of* $\mathcal{T}_F$ *on that surface for the three different boundary conditions* $\Lambda = \mathbb{C}_+, \mathbb{C}_-, {}_F F$*?*

## 2.3 The non-topological boundary condition $\widetilde{Q}$ for $\mathcal{T}_F$

Recall that $Q = (\mathcal{H}, H)$ denoted a 1d QFT, $D_1, \dots, D_r$ a collection of 0d topological defects that closed under fusion, and $e_1, \dots, e_r$ the corresponding basis of the algebra $F$. Write $e_i^* : F \to \mathbb{C}$, $i = 1, \dots, r$ for the dual basis.

We want to turn $Q$ into a non-topological boundary condition $\widetilde{Q}$ for the SymTFT $\mathcal{T}_F$. To this end we define the topological junction

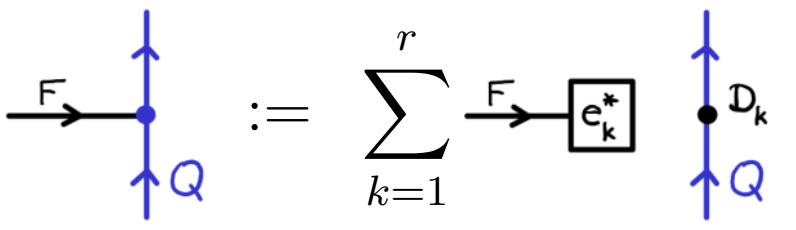

. (2.16)

In the present setting this is just a linear map $F \otimes \mathcal{H} \to \mathcal{H}$. The construction of $F$ is set up in such a way that the following identity is automatically satisfied:

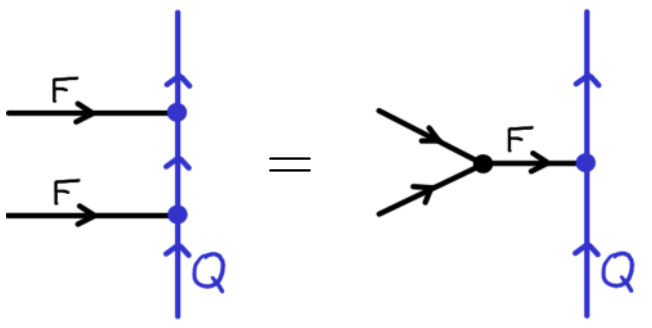

. (2.17)

Indeed, substituting the definition on the left hand side gives

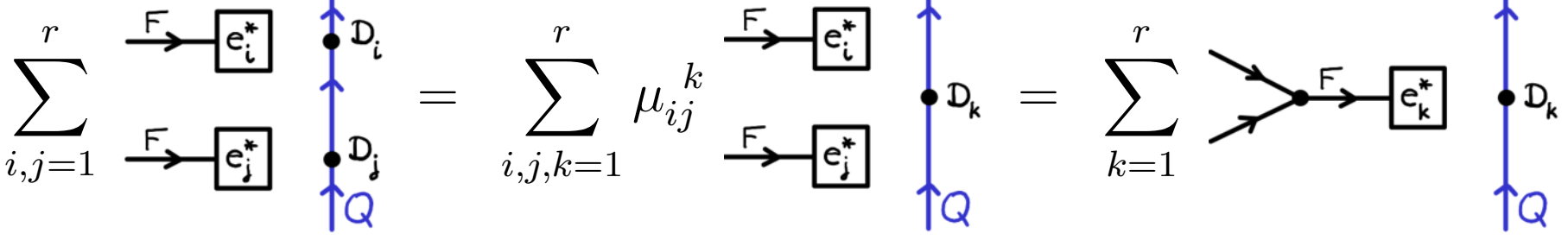

, (2.18)

which is precisely the right hand side.

More abstractly this is not surprising as the above defines a left $F$-module structure on $\mathcal{H}$, which is the same as an algebra homomorphism $F \to \mathrm{End}(\mathcal{H})$, and that is how we defined $F$ in the first place.

Altogether, we have now turned $Q$ into a non-topological boundary condition $\widetilde{Q}$ for the state sum TFT $\mathcal{T}_F$. We can now evaluate $\mathcal{T}_F$ ("compute the partition function") on surfaces like

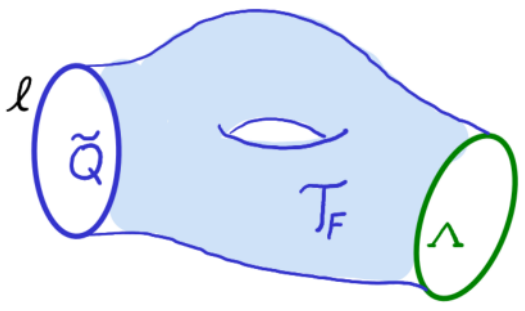

. (2.19)

The result will depend on the length $\ell$ of the non-topological boundary $\widetilde{Q}$.[3]

**Exercise:** *(*) The result of evaluating $\mathcal{T}_F$ on the above surface can be written as*

$$\sum_{k=1}^{r} \lambda_k \, \mathrm{tr}_{\mathcal{H}}\big( D_k \, e^{-\ell H} \big) \tag{2.20}$$

*for some constants $\lambda_k \in \mathbb{C}$. Take $\Lambda$ to be the regular module ${}_F F$ and express the $\lambda_k$ in terms of the structure constants of $F$.*

After all this preparation, we can now state the main property of the SymTFT, namely that we can recover $Q$ from the 2d TFT $\mathcal{T}_F$ with a specific choice of non-topological (or "physical") and topological (or "symmetry") boundary condition:

> Let $\Sigma$ be a 1d manifold with metric, and set $M = \Sigma \times [0,1]$. The boundary of $M$ consists of $\Sigma \times \{0\}$, to which we assign the topological boundary condition ${}_F F$, and $\Sigma \times \{1\}$, to which we assign the non-topological boundary condition $\widetilde{Q}$. Then evaluating $\mathcal{T}_F$ on $M$ is equal to evaluating $Q$ on $\Sigma$.

Pictorially,

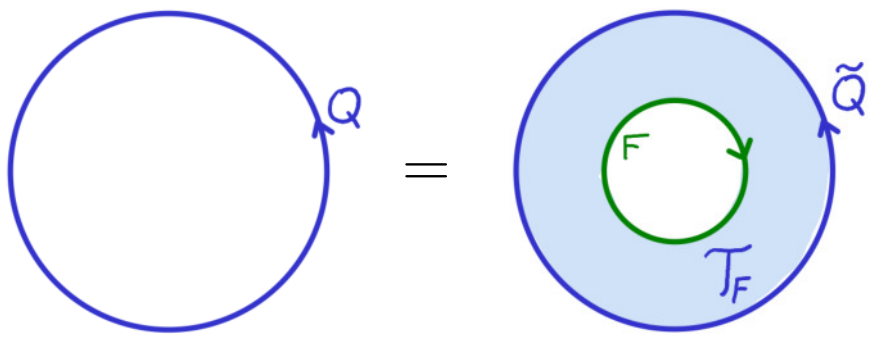

. (2.21)

The first few steps in the calculation to shows this are (only a segment of $\Sigma \times [0,1]$ is shown)

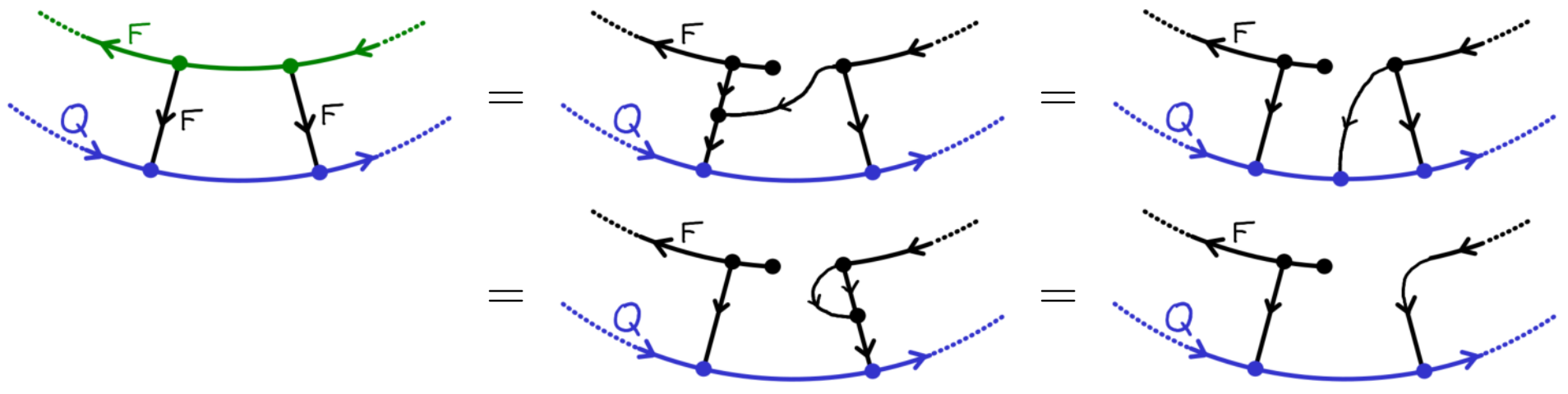

.

More abstractly, the above computation amounts to the statement that $F^* \otimes_F \mathcal{H} \cong \mathcal{H}$.

**Exercise:** *Complete the proof.*

In the computation to recover $Q$, we used the specific topological boundary condition ${}_F F$. Using a general topological boundary condition in its place leads to the second important property of the SymTFT:

[3] More generally, the result of evaluating the $(d+1)$-dimensional TFT with a $d$-dimensional non-topological boundary $\widetilde{Q}$ depends on the $d$-dimensional metric on that boundary, up to $(d+1)$-dimensional diffeomorphisms which restrict to $d$-dimensional isometries. In the present case where $d = 1$, the only invariant is the overall length of the topological boundary.

Choosing other topological boundary conditions $\Lambda$ on $\Sigma \times \{0\}$ (while keeping $\widetilde{Q}$ on $\Sigma \times \{1\}$) leads to new 1d QFTs $Q_\Lambda$.

**Example:** Let $Q$ be the product system $\mathcal{H}_{\mathrm{tot}} = \mathcal{H} \otimes \mathcal{H}$ with 0d topological defects id and $\tau$, resulting in $F = \mathbb{C}[\mathbb{Z}_2]$ as above. For the topological boundary condition $\Lambda = \mathbb{C}_+$, the new QFT $Q_\Lambda$ is given by $Q$ restricted to the symmetric product $S^2(\mathcal{H})$, i.e. to the subspace of $\mathcal{H} \otimes \mathcal{H}$ spanned by vectors of the form $a \otimes b + b \otimes a$.

**Exercise:** *(*) Show this. What about* $\mathbb{C}_-$*?*

## 2.4 Observables in $\mathcal{T}_F$

By a (zero-dimensional) observable of $Q$ we simply mean a bounded operator[4] $\mathcal{O} : \mathcal{H} \to \mathcal{H}$. As opposed to a topological defect, $\mathcal{O}$ need not commute with $H$. A given choice $D_1, \dots, D_r$ of topological symmetry in $Q$ separates the observables into two classes: those which commute with the $D_i$, and those which do not.

Suppose $[D_i, \mathcal{O}] = 0$ for $i = 1, \dots, r$. In terms of the state sum construction of the SymTFT, this means that the topological junction attaching $F$ to $Q$ commutes with an insertion of $\mathcal{O}$. Thus $\mathcal{O}$ is a *local operator* on the non-topological boundary $\widetilde{Q}$:

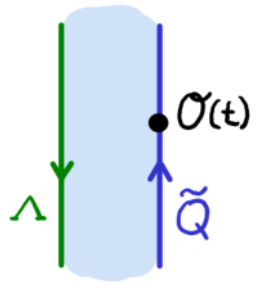

. (2.22)

Observables $\mathcal{O}$ of $Q$ which do not commute with the chosen topological symmetry are described as endpoints of topological line defects in $\mathcal{T}_F$:

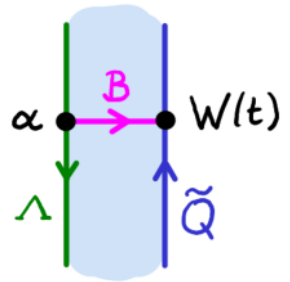

. (2.23)

More precisely, to obtain the observables in $Q$ we need to choose $\Lambda = {}_F F$, for a general $\Lambda$ this describes the observables of $Q_\Lambda$. Note also that local operators on $\widetilde{Q}$ correspond the special case that the line defect $B$ is the trivial (or invisible, or unit) line defect.

Algebraically, the line defect $B$ is described by an $F$-$F$-bimodule $B$, the non-topological junction $W$ by an $F$-module intertwiner $W \in \mathrm{Hom}_F(B \otimes_F \mathcal{H}, \mathcal{H})$, and the topological junction $\alpha$ by an $F$-module intertwiner $\alpha \in \mathrm{Hom}_F(\Lambda, B \otimes_F \Lambda)$. The trivial line defect is given by the regular bimodule ${}_F F_F$.

**Exercise:** *(*) Show that for* $\Lambda = {}_F F$*, every* $\mathcal{O} : \mathcal{H} \to \mathcal{H}$ *can be written in terms of a topological line defect as above. Hint: Consider the diagram*

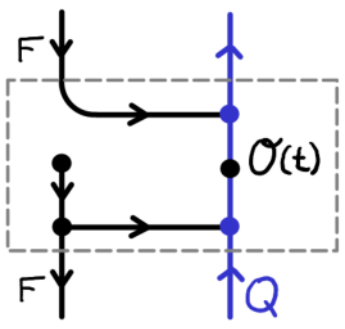

. (2.24)

[4]More generally one should work with families of bounded operators $\mathcal{O}_{s,t} : \mathcal{H} \to \mathcal{H}$, $s, t \in \mathbb{R}_{>0}$, which correspond to evaluating the 1d QFT on the interval $[-t, s]$ with an insertion of $\mathcal{O}$ at zero. The limit $s, t \to 0$ need not exist, and $\mathcal{O}_{s,t}\mathcal{O}_{t,s'}$ can be singular for $t \to 0$ as one would expect as two fields approach each other.

*What can you say about the space of non-topological junctions* $\mathrm{Hom}_F(B \otimes_F \mathcal{H}, \mathcal{H})$ *as a bimodule over the local observables on* $\widetilde{Q}$*? What are the observables of* $Q_\Lambda$ *in the example of the product system with* $\Lambda = \mathbb{C}_+$*?*

The logic of the construction so far was to start from a 1d QFT $Q$, pick a suitable topological symmetry $D_1, \dots, D_r$, construct the algebra $F$ and the associated 2d TFT $\mathcal{T}_F$, couple $Q$ to $\mathcal{T}_F$ as a non-topological boundary condition $\widetilde{Q}$, and finally to recover $Q$ for a particular choice of topological boundary condition $\Lambda = {}_F F$.

Conceptually, it is better to start from $F$ and to *realise* $F$ *via topological symmetries of* $Q$. The realisation then becomes part of the data, concretely here a choice of algebra homomorphism $\rho : F \to \mathrm{End}(\mathcal{H})$. From this point of view, a given symmetry $F$ can be realised in different QFTs $Q$, and it can be realised in different ways in a given $Q$. This is akin to separating the study of abstract groups from the study of their representation theory. Note that there is no need to fix a basis for $F$ (i.e. a specific choice of $D_i$) or to demand that $\rho$ is injective (linear independence of the $D_i$).

# 3 Two-dimensional QFT and SymTFT

## 3.1 Topological line defects and fusion categories

While in the 1d QFT case we had the advantage of a precise mathematical framework, this is no longer the case for 2d QFTs. Instead, we suppose that we are given, in one way or another, a 2d QFT $Q$ (in euclidean signature) which has observables localised on points ("fields") and on lines ("defects").

By a *topological line defect* we mean a defect such that correlators of $Q$ do not depend on the precise location of the defect, that is, the defect can be deformed as long as it is not taken past other operators,

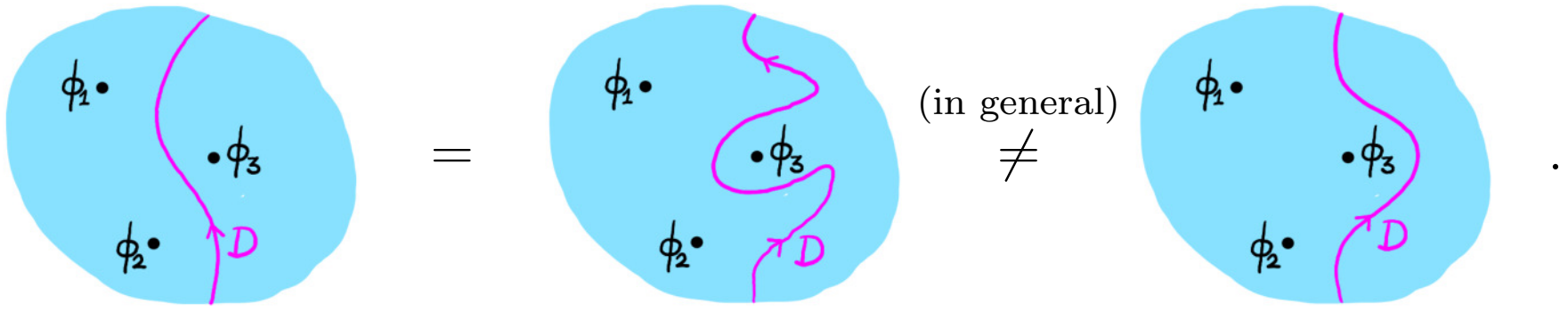

(3.1)

These and the following pictures are to be understood as showing a local patch of the surface for which a 2d QFT correlator is evaluated. The equal signs are meant as "these equalities hold inside correlation functions, independent of what surface and operators are considered outside the patch shown, as long as they are the same on both sides of the equality".

Topological line defects allow for a fusion operation where two parallel topological line defects $D$ and $D'$ are moved close to each other to form a single new line defect $D * D'$,

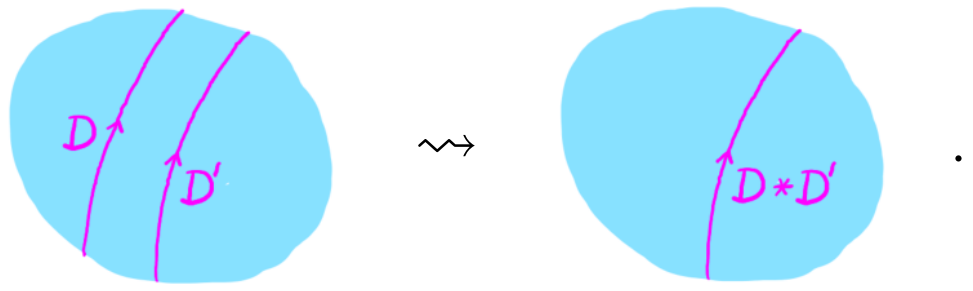

(3.2)

Actually, this picture is not quite accurate in that it is not a modification localised in a surface patch. Indeed, what would the fusion procedure even mean when the defects cannot be made parallel globally:

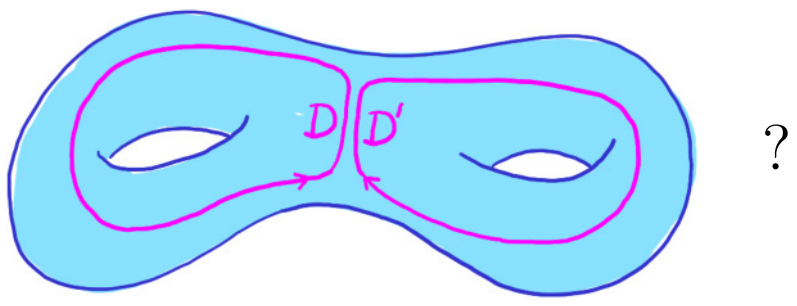

(3.3)

To make the fusion of defects a local operation, we are forced to introduce another ingredient: 0d topological defects, in this case topological junction fields,

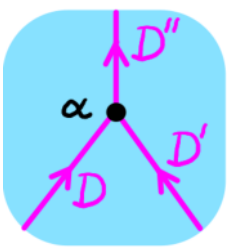

. (3.4)

Just as other fields of the QFT $Q$, these topological junction fields also form a $\mathbb{C}$-vector space.

After these initial considerations, let us say more carefully what we mean by a finite (and semisimple) topological symmetry of a 2d QFT $Q$. Let $D$ be a topological line defect in $Q$. We say that $D$ is *elementary* if the space of topological defect fields on $D$ is one-dimensional, and we say two elementary defects $D, D'$ are *distinct* if there are no non-zero topological junctions $D \leftrightarrow D'$. Let $D_1, \dots, D_r$ be a collection of mutually distinct elementary topological line defects of $Q$. We will assume[5] that the spaces of topological 3-fold junctions $\lambda$, $\kappa$ in

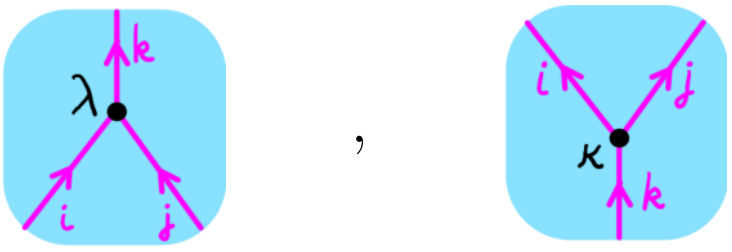

. (3.5)

are one-dimensional (here and below we abbreviate $D_k = k$, etc.). We make, once and for all, a choice of non-zero junction field $i * j \to k$ (when it exists), and we choose a dual field $k \to i * j$ such that

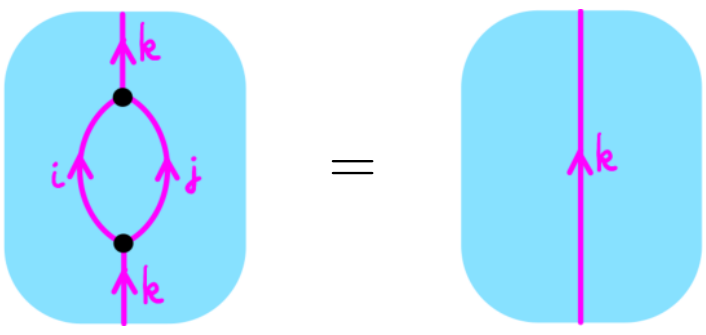

. (3.6)

We can now finally say what it means for $D_1, \dots, D_r$ to *close under fusion*:

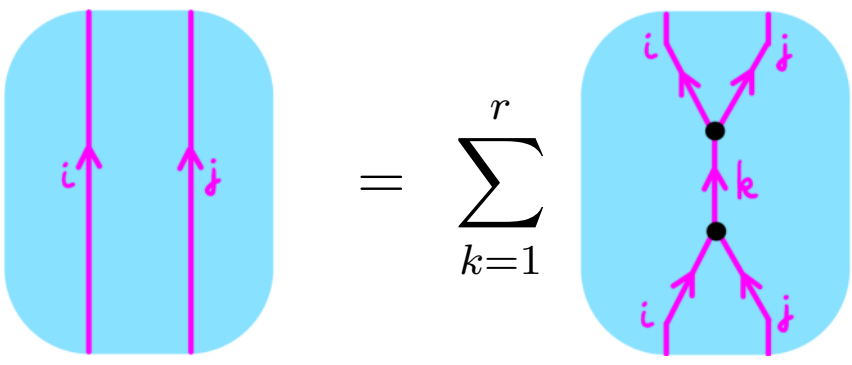

. (3.7)

Note that some terms in this sum may be zero, namely if the corresponding space of junctions $i * j \to k$ is zero-dimensional. We write $D_i * D_j = \sum_{k=1}^r N_{ij}^{\ \ k} D_k$, where $N_{ij}^{\ \ k} \in \{0, 1\}$ is the multiplicity of $D_k$ in the fusion of $D_i$ with $D_j$ (also called *fusion rule*).

We assume that $D_1$ is the trivial defect $\mathbf{1}$ which simply amounts to not inserting a line defect on the surface.

We need our collection $D_1, \dots, D_r$ to be *closed under orientation reversal.* That means that for each $k \in \{1, \dots, r\}$ there is a $\bar{k}$ in the same range, such that

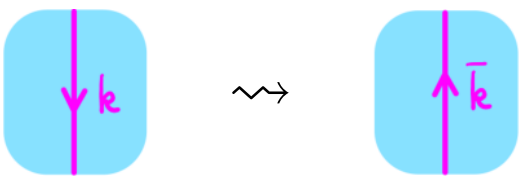

. (3.8)

**Exercise:** *Again this is not a local identity in a patch as the two configurations differ on the boundary of the patch. Hence we cannot write this as an equality. Can you change the above definition so that it makes sense as an equality? Hint: "There exists topological junctions* $k * \bar{k} \to \mathbf{1}$, $\mathbf{1} \to \bar{k} * k$ *such that..."*

[5]This is done for simplicity in order to avoid a proliferation of multiplicity labels. In a comprehensive treatment of finitely semisimple topological symmetries one needs to drop this assumption as fusion categories can have higher fusion rule multiplicities.

As one final assumption we will need that small defect loops labelled by $D_i$ can be shrunk to a non-zero constant time the identity field of the QFT $Q$. These constants are called *quantum dimensions*,

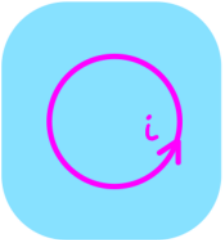

$=$
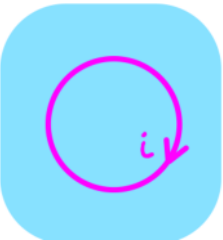

$= \dim(D_i)$

. (3.9)

The $\dim(D_i)$ are typically not integers (but one can show that they are always real numbers).[6]

Under the fairly elaborate assumptions on $D_1, \dots, D_r$ we made above, the QFT $Q$ with these topological defects defines for us a *spherical fusion category* $\mathcal{F}$. It has

- simple objects $d_1, \dots, d_r$, all other objects are direct sums of these,
- tensor product $d_i \otimes d_j = \bigoplus_{k=1}^{r} N_{ij}^{\ k}\, d_k$ and unit object $\mathbf{1} = d_1$,
- quantum dimensions $\dim(d_i) = \dim(D_i)$,

and one more ingredient, the F-symbols, which describe the associator for the tensor product, and which we introduce in the next section.

## 3.2 F-symbols and group cohomology

Given a 2d QFT $Q$ and a collection of defects $D_1, \dots, D_r$ as in the previous section, consider the topological three-to-one junctions $i * j * k \to \ell$. The space of these junctions can be of higher dimension even under our assumption that $N_{ij}^{\ k} \in \{0, 1\}$. In fact, one can build two different bases for this space out of three-fold junctions, and the change of basis is precisely given by the F-symbols,

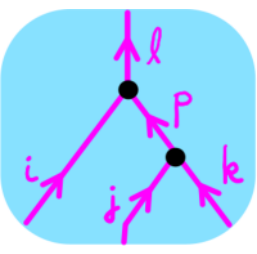

$= \sum_{q=1}^{r} \mathrm{F}^{(ijk)\ell}_{pq}$
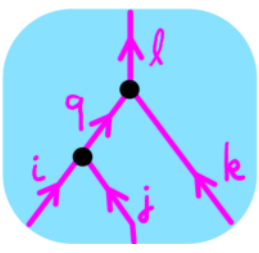

. (3.10)

**Exercise:** *The F-symbols satisfy the pentagon identity, which is a polynomial equation of the form "*FF $= \sum$ FFF*". Derive this with all the indices in place.*

Note that the F-symbols are not extra input, but they are computed from correlation functions of $Q$. Algebraically they enter the definition of the monoidal structure on $\mathcal{F}$, but we will not go through these details here.

Consider the special case where the defect fusion is given by a finite group $G$. That is, our selection of elementary topological line defects is $D_g$, $g \in G$, with fusion $D_g * D_h \cong D_{gh}$ and $D_e$ the trivial defect. In particular, the defects are invertible under fusion. In that case, there is no summation in the change-of-basis relation above, and the two sides just differ by a constant,

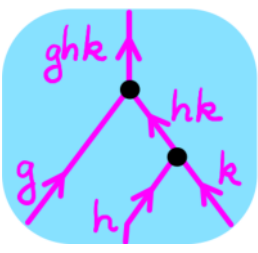

$= \omega(g, h, k)$
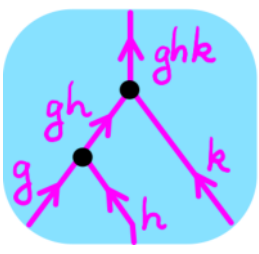

, where $\omega(g, h, k) \in \mathbb{C}^\times$ . (3.11)

[6] It follows from the assumptions so far that each of the two differently oriented circles equals to a constant (as they can be interpreted as topological junctions on the identity defect $D_1$). The additional assumption here is that the two constants coincide. This is again a simplification which in general needs to be dropped. In fusion category language this means that one should consider pivotal fusion categories, rather than just spherical fusion categories. On the other hand, unitary fusion categories are automatically spherical.

The pentagon relation becomes the cocycle condition in group cohomology, and changing the choice of topological three-fold junctions by constants modifies $\omega$ by a coboundary. The invariant information therefore is a class in the third group cohomology,

$$[\omega] \in H^3(G, \mathbb{C}^\times) \ . \tag{3.12}$$

**Exercise:** *Look up the differential for group cohomology and verify this.*

This provides another important lesson from studying topological symmetries. Even in the simple case of a finite group symmetry, if the symmetry is realised by invertible topological line defects, the QFT $Q$ provides us with an extra piece of data, namely a class $[\omega] \in H^3(G, \mathbb{C}^\times)$. One physical interpretation of this class is as an obstruction to being able to gauge $G$.

Altogether, in two dimensions a finitely semisimple topological symmetry is described by a spherical fusion category $\mathcal{F}$. It is again better to take the opposite point of view, where one starts from $\mathcal{F}$ and asks if it can be realised by topological defects in a 2d QFT $Q$.

## 3.3 3d state sum TFT

Starting from the input of a spherical fusion category $\mathcal{F}$ and a realisation of $\mathcal{F}$ by topological defects in a 2d QFT $Q$, we would like to produce a 3d TFT $\mathcal{T}_\mathcal{F}$ and turn $Q$ into a non-topological boundary condition $\widetilde{Q}$ for $\mathcal{T}_\mathcal{F}$.

Let us briefly look back at the 2d state sum TFT. We can think of that construction as building a 2d network out of 1d TFTs and 0d junctions ("gauging a topological symmetry of the trivial 2d TFT"). The corresponding construction in 3d is to build a foam in 3d out of 2d TFTs, together with 1d and 0d junctions.

With this analogy in mind, let us go through the ingredients for the 3d state sum TFT in more detail. Recall that the algebraic input is the spherical fusion category $\mathcal{F}$ with (representatives of isomorphism classes of) simple objects $d_1, \dots, d_r$. This defines:

- A 2d state sum TFT specified by a $\Delta$-separable symmetric Frobenius algebra $A$. We label a surface on which this 2d TFT is to be evaluted by $A$, as in the patch below:

  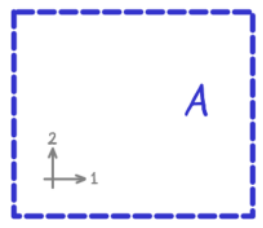
  
  . (3.13)

  The little coordinate system indicates the orientation of the surface.

  Concretely, $A = \mathbb{C}^r$, that is, $r$ copies of $\mathbb{C}$ with element-wise multiplication. Equivalently, $A = \mathrm{End}_\mathcal{F}(G)$ with $G = d_1 \oplus \cdots \oplus d_r$ a minimal generator. The algebra $A$ encodes the simple objects of $\mathcal{F}$.

- A 1d topological junction $T$ from the product of two 2d TFTs as above to one:

  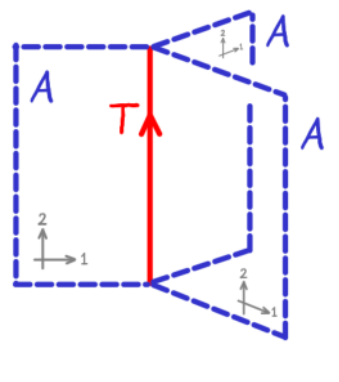
  
  . (3.14)

  Algebraically, $T$ is an $A$-$(A \otimes A)$-bimodule. Explicitly,

$$T = \bigoplus_{i,j,k} T_{ij}^{\ \ k} \quad \text{where} \quad T_{ij}^{\ \ k} = \mathrm{Hom}_\mathcal{F}(i \otimes j, k) \ . \tag{3.15}$$

Write $1_i$ for the unit of the $i$'th copy of $\mathbb{C}$ in $A$. Then acting with $1_k$ from the left and $1_i \otimes 1_j$ from the right projects $T$ to $T_{ij}{}^k$. Note that $\dim T_{ij}{}^k = N_{ij}{}^k$, so that the 1d topological junction encodes the fusion rules.

- Two 0d topological junctions $\alpha$ and $\overline{\alpha}$ which describe the crossing of two $T$-lines,

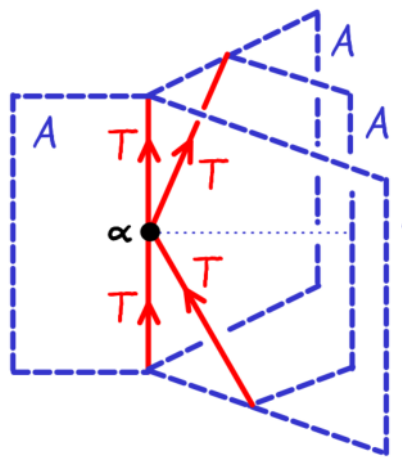

, (3.16)

with $\overline{\alpha}$ labelling the mirrored crossing. Its components are given by F-symbols and their inverses, that is, the junctions $\alpha$ and $\bar{\alpha}$ encode the associator and its inverse (up to dimension factors).

This data encodes the spherical fusion category $\mathcal{F}$ as follows. As a $\mathbb{C}$-linear category it is equivalent to the category $A$-mod of finite-dimensional $A$-modules over $\mathbb{C}$. The bimodule $T$ defines a tensor product functor on $A$-mod via $(M, N) \mapsto T \otimes_{A\otimes A} M \otimes N$ which is equivalent to the tensor product on $\mathcal{F}$. The junction $\alpha$ is an $A$-$A^{\otimes 3}$-bimodule map from $T \otimes_{A\otimes A} (A \otimes T)$ to $T \otimes_{A\otimes A} (T \otimes A)$ and gives the associator.

Given a closed oriented 3-manifold, pick a cell-decomposition such that all 2- and 3-cells are contractible (i.e. they are topologically a disc and a ball, respectively), and all 2- and 1-cells can be compatibly oriented. Label the 2-cells by $A$, the 1-cells by $T$, and the 0-cells by $\alpha$ or $\bar{\alpha}$, depending on orientations.[7] Evaluate the resulting diagram as a network of 2d TFTs with defects and junctions to get a number. This number is a 3-manifold invariant of $M$ and does not depend on the specific choice of cell decomposition.

We denote the resulting 3d state sum TFT by $\mathcal{T}_{\mathcal{F}}$. This is the SymTFT for the finitely semisimple topological symmetry $\mathcal{F}$.

The TFT $\mathcal{T}_{\mathcal{F}}$ can also be evaluated on 3-manifolds with boundary. The boundary components then carry a topological boundary condition which algebraically is given by an $\mathcal{F}$-module category $\mathcal{B}$ (with module trace). One choice of module category is $\mathcal{F}$ itself. In terms of the state sum, this boundary condition simply amounts to "ending the foam", that is, the boundary of $M$ is part of the cell-decomposition and decorated by the same rules as the interiour of $M$.

## 3.4 The non-topological boundary condition $\widetilde{Q}$ for $\mathcal{T}_{\mathcal{F}}$

We can turn the 2d QFT $Q$ into a non-topological boundary $\widetilde{Q}$ for the 3d TFT $\mathcal{T}_{\mathcal{F}}$ by attaching the cell-decomposition of the 3-manifold to line defects on the boundary where we evaluate $Q$. The procedure is similar in spirit to the 1d case treated in Section 2.3, and requires a bit of preparation.

We fix a collection of boundary conditions $E_i$, $i = 1, \ldots, r$ for the 2d state sum TFT defined by $A$. As we saw in Section 2.2, these are described by $A$-modules, and we let $E_i$ be the left $A$-module with underlying vector-space $\mathbb{C}$ (write $e_i = 1 \in E_i$) and action of $1_j \in A$ given by $1_j.e_i = \delta_{i,j}\, e_i$. We also need a point junction which can sit at the end of

[7]One has to include weight factors for 2-cells and 3-cells which are determined by the quantum dimensions of $\mathcal{F}$ – the bibliography lists references with the details.

a $T$-line defect,

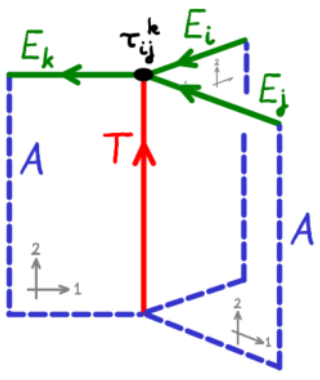

$$. \tag{3.17}$$

Algebraically, $\tau_{ij}^{\ k}$ is a left $A$-module morphism $T \otimes_{A\otimes A} E_i \otimes E_j \to E_k$.

We can now couple the 2d TFT defined by $A$ and the junction lines $T$ to the 2d QFT $Q$ by setting

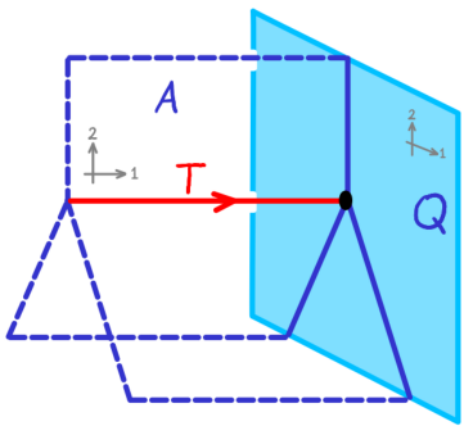

$$:= \sum_{i,j,k=1}^{r}$$

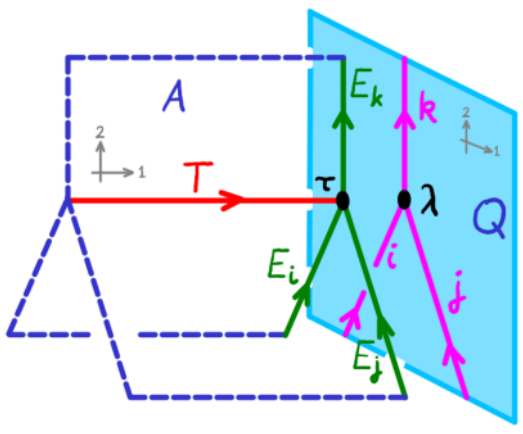

$$. \tag{3.18}$$

Note that this again just shows a patch of the surface on which we evaluate $Q$.[8]

Altogether, this allows us to define a non-topological boundary condition $\widetilde{Q}$ for the 3d state sum TFT $\mathcal{T}_{\mathcal{F}}$. Namely, we can connect the foam defining the 3d state sum to the 2d QFT $Q$ via the junctions defined above (and the dual of $\tau$, which we did not introduce),

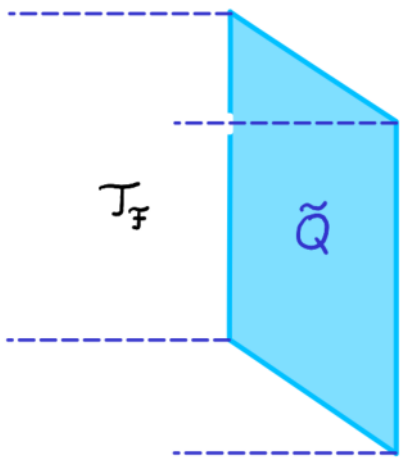

$$\rightsquigarrow$$

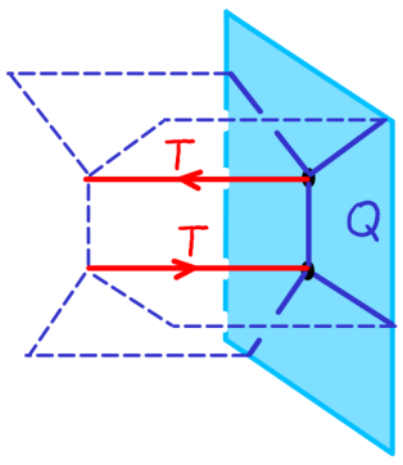

$$. \tag{3.19}$$

We will not go through the full proof that this is independent of the cell decomposition, but let us at least check this in one example. The key identity to use is the following:[9]

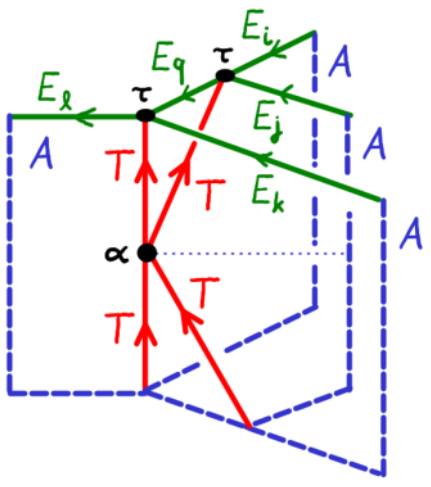

$$= \sum_{q=1}^{r} F^{(ijk)\ell}_{pq}$$

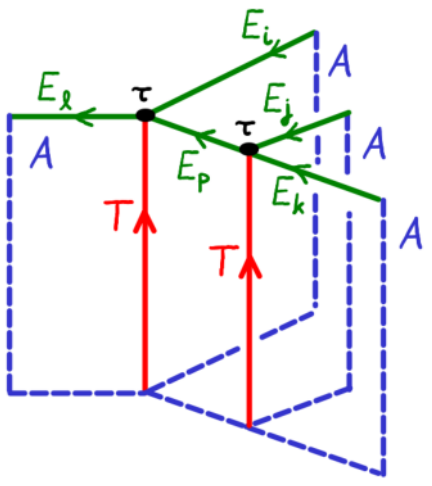

$$. \tag{3.20}$$

Using this, one gets

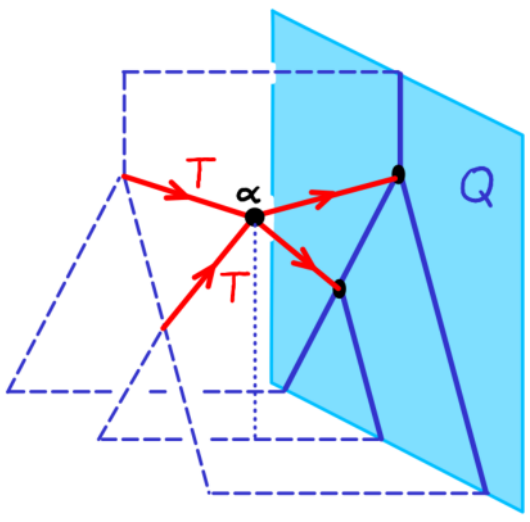

$$=$$

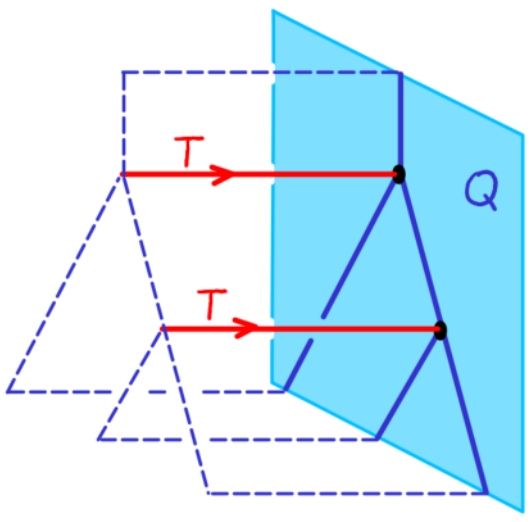

$$. \tag{3.21}$$

---

[8] We cannot write "=" here as we did in Section 3.1 as the configuration will not look the same on both sides outside the patch shown. Instead the replacement of a topological junction line attaching the 2d TFT $A$ to $Q$ by a sum over elementary defects $D_k$ and boundary conditions $E_k$ has to be done everywhere at the same time.

[9] There is a technical detail that was omitted here: in the triangular surface bounded by $(T, T, E_q)$ one has to place a point insertion of $\psi^2$ (which we did not introduce). This compensates a factor of a quantum dimension, see [1, Sec. 4] for the detailed conventions.

**Exercise:** *Show this by substituting the definition of the junctions between the 2d TFT for A and Q, and using the key identity above. (*) Using the conventions in [1, Sec. 4.2] for $\alpha$, check the key identity.*

An equivalent construction of $\widetilde{Q}$ is to note that the topological boundary condition $\mathcal{F}$ of $\mathcal{T}_\mathcal{F}$ has $\mathcal{F}$ as category of line defects, and that one can use these to couple $\mathcal{T}_\mathcal{F}$ to $Q$. Schematically,

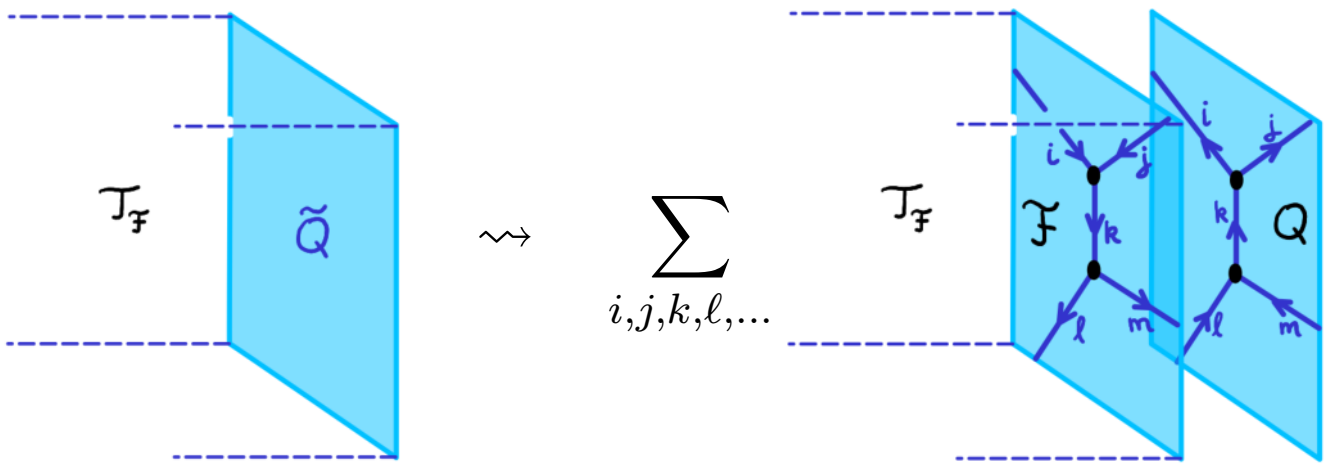

, (3.22)

where the line defects are placed on the 1-cells of a cell decomposition of the surface $\Sigma$ with contractible 2-cells.

As in the 1d case one can recover $Q$ from the SymTFT $\mathcal{T}_\mathcal{F}$ with appropriate boundary conditions. Namely, let $\Sigma$ be a surface with metric, and let $M = \Sigma \times [0,1]$. On the boundary $\Sigma \times \{0\}$ we place the topological boundary condition $\mathcal{F}$, and to the boundary $\Sigma \times \{1\}$ we assign the non-topological boundary condition $\widetilde{Q}$. Then evaluating $\mathcal{T}_F$ on $M$ is equal to evaluating $Q$ on $\Sigma$:

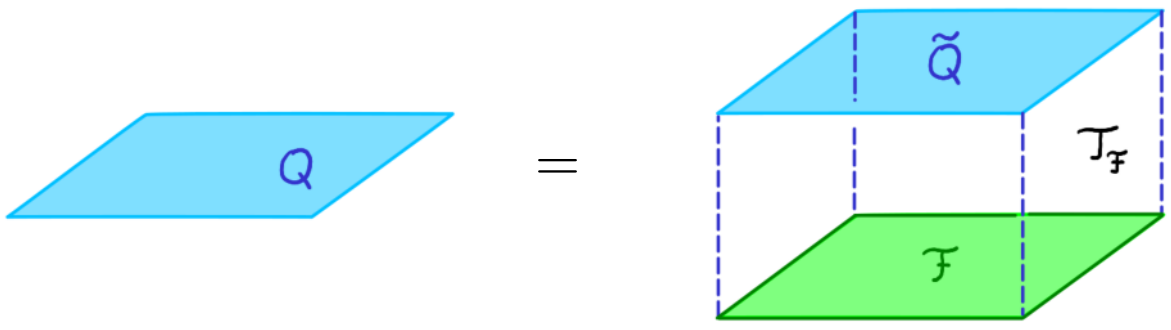

. (3.23)

Choosing other topological boundary conditions $\mathcal{B}$ instead of $\mathcal{F}$ defines new 2d QFTs $Q_\mathcal{B}$.

The (non-topological) fields $\phi(x)$ of $Q$ contain as a subspace those that commute with all the line defects $D_i$, $i = 1, \dots, r$ as in

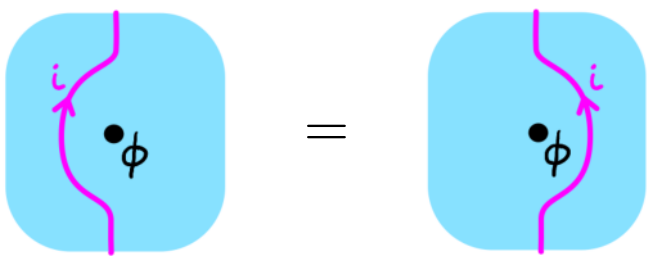

. (3.24)

These fields are local fields on the non-topological boundary $\widetilde{Q}$. General fields of $Q$ that do not necessarily commute with the $D_i$ are described in the SymTFT $\mathcal{T}_\mathcal{F}$ via line defects $L$ with a non-topological endpoint $W(x)$ on $\widetilde{Q}$ and a topological endpoint $\zeta$ on $\mathcal{F}$,

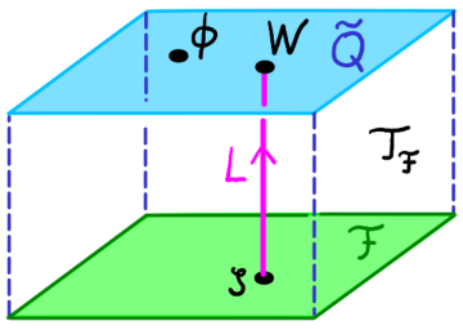

. (3.25)

Note that local fields on $\widetilde{Q}$ arise as a special case of $W$ when one chooses $L$ to be trivial.

**Example:** Let $Q$ be the two-dimensional Ising conformal field theory, a Virasoro minimal model of central charge $c = \frac{1}{2}$. It has three primary fields, the identity $\mathbf{1}$ with conformal weights $h_\mathbf{1} = \bar{h}_\mathbf{1} = 0$, the spin field $\sigma(x)$ with weights $h_\sigma = \bar{h}_\sigma = \frac{1}{16}$, and the energy field $\epsilon(x)$ with weights $h_\epsilon = \bar{h}_\epsilon = \frac{1}{2}$. All other fields are descendants of these, obtained by acting with modes of the holomorphic and anti-holomorphic copy of the Virasoro algebra.

There are three elementary topological line defects, the trivial defect id, the spin flip defect $S$, and the duality defect $D$. The composition rules are $S * S = \mathrm{id}$, $S * D = D$, $D * D = \mathrm{id} \oplus S$. There are thus three choices for a finitely semisimple topological symmetry in the Ising CFT,

$$\mathcal{F}_1 = \langle \mathrm{id} \rangle \quad , \quad \mathcal{F}_2 = \langle \mathrm{id}, S \rangle \quad , \quad \mathcal{F}_3 = \langle \mathrm{id}, S, D \rangle \ , \tag{3.26}$$

where we list the choice of elementary defects, and the pointed brackets mean the fusion category obtained by taking direct sums of these.

For each of these three choices $\mathcal{F}_i$ of topological symmetry, we can now ask which fields of the Ising CFT commute with the corresponding line defects. Recall that these are precisely the local fields on the topological boundary condition $\widetilde{Q}$ for the SymTFT $\mathcal{T}_{\mathcal{F}_i}$. One finds

| top. sym. | local fields on $\widetilde{Q}$ |
|---|---|
| $\mathcal{F}_1$ | $\mathbf{1}$, $\sigma$, $\epsilon$ |
| $\mathcal{F}_2$ | $\mathbf{1}$, $\epsilon$ |
| $\mathcal{F}_3$ | $\mathbf{1}$ |

Here only the primary fields are listed, the local fields on $\widetilde{Q}$ consist of these primary fields plus all their descendants.

Of particular interest is the last entry, which is the largest of the three choices of topological symmetry. In this case, the local fields on $\widetilde{Q}$ are given by $V \otimes_{\mathbb{C}} \overline{V}$, where $V$ denotes the holomorphic fields in the Ising CFT, and $\overline{V}$ the antiholomorphic fields. In fact, also the SymTFT factorises into a product of two 3d TFTs, and this is the setting we consider in the next section.

# 4 Two-dimensional CFT and chiral TFT

The discussion of SymTFT for 1d and 2d QFTs in the previous two sections continues analogously for higher dimensional QFTs. In this section, we want to give a formulation that is specific to conformal field theory in two dimensions, and which builds on the Ising example just discussed.

Namely, let $V$ be a rational vertex operator algebra ("the chiral algebra of the 2d CFT"). Let $Q$ be a 2d CFT whose fields contain $V \otimes_{\mathbb{C}} \overline{V}$, that is, a holomorphic and an antiholomorphic copy of $V$. Then there exists a choice of topological symmetry $\mathcal{F}$ such that the fields that commute with $\mathcal{F}$ are precisely $V \otimes_{\mathbb{C}} \overline{V}$. In this case, the SymTFT $\mathcal{T}_{\mathcal{F}}$ is a product

$$\mathcal{T}_{\mathcal{F}} \;\cong\; \mathcal{S}_{\mathcal{C}} \otimes \mathcal{S}_{\overline{\mathcal{C}}} \ . \tag{4.1}$$

Here, $\mathcal{S}_{\mathcal{C}}$ is a surgery 3d TFT obtained from a modular fusion category $\mathcal{C}$ (a fusion category which is in addition non-degenerately braided and has a ribbon twist). $\overline{\mathcal{C}}$ is $\mathcal{C}$ equipped with opposite braiding and twist. The category $\mathcal{C}$ in turn is obtained as representations of $V$,

$$\mathcal{C} \;=\; \mathrm{Rep}(V) \ . \tag{4.2}$$

Since the 3d TFT $\mathcal{S}_{\mathcal{C}}$ is fully determined by the chiral symmetry of the 2d CFT, as formalised by the vertex operator algebra $V$, we refer to $\mathcal{S}_{\mathcal{C}}$ as the *chiral TFT*.

One consequence of $\mathcal{T}_{\mathcal{F}}$ being a product is that the elementary line defects in the 3d TFT $\mathcal{T}_{\mathcal{F}}$ are products $x \boxtimes \bar{y}$ of elementary line defects $x$ in $\mathcal{S}_{\mathcal{C}}$ and $\bar{y}$ in $\mathcal{S}_{\overline{\mathcal{C}}}$. Algebraically, the line defects in $\mathcal{T}_{\mathcal{F}}$ are parametrised by the Drinfeld centre $\mathcal{Z}(\mathcal{F})$ of $\mathcal{F}$, and those in $\mathcal{S}_{\mathcal{C}}$ are labelled by $\mathcal{C}$ itself, i.e. by $V$-modules. We have

$$\mathcal{Z}(\mathcal{F}) \;\cong\; \mathcal{C} \boxtimes \overline{\mathcal{C}} \tag{4.3}$$

as $\mathbb{C}$-linear ribbon categories.

Geometrically, the SymTFT $\mathcal{T}_{\mathcal{F}}$ is obtained by folding from the more fundamental chiral TFT $\mathcal{S}_{\mathcal{C}}$. Namely, a CFT correlator on a conformal surface $\Sigma$ can be represented in SymTFT on the 3-manifold $\Sigma \times [0, 1]$ with a non-topological and a topological boundary condition as before. But it can also be presented in chiral TFT on $\Sigma \times [-1, 1]$ with two non-topological boundary conditions (one supports the holomorphic and the other the anti-holomorphic fields), together with a topological surface defect $\mathcal{A}$ placed at $\Sigma \times \{0\}$,

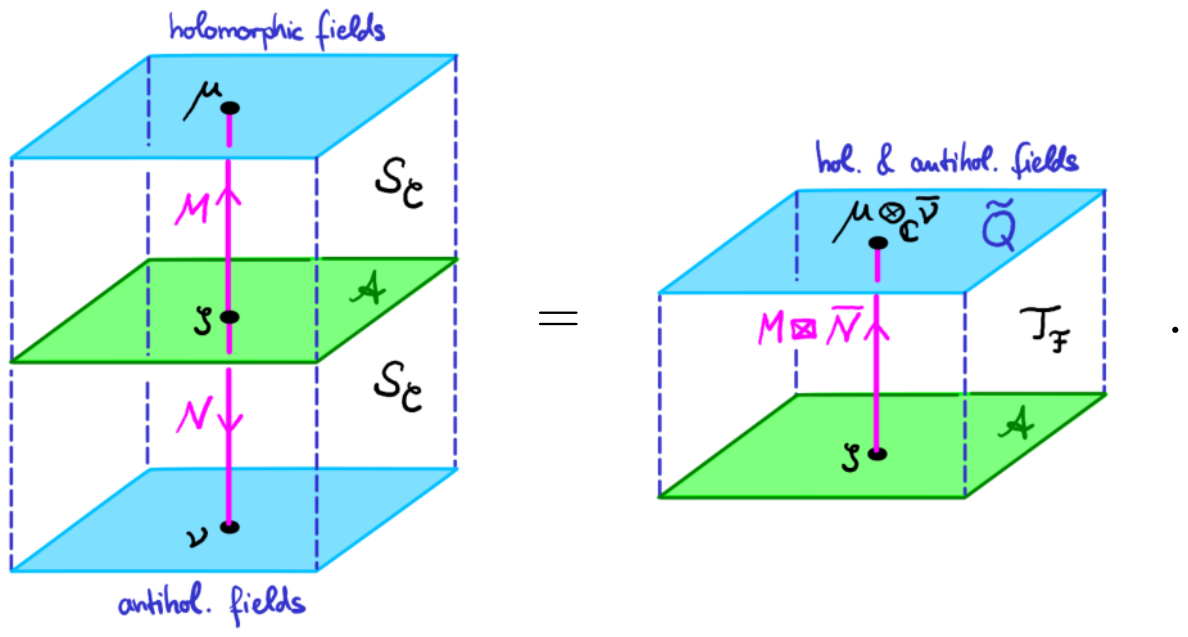

(4.4)

The folding happens at the surface defect $\mathcal{A}$, which becomes the topological boundary for $\mathcal{T}_{\mathcal{F}}$. In terms of chiral TFT $\mathcal{S}_{\mathcal{C}}$, a bulk field of the CFT is described by:

- two representations $M, N$ of $V$, which label line defects that attach to the holomorphic and antiholomorphic boundary for $\mathcal{S}_{\mathcal{C}}$,
- two vectors $\mu \in M$, $\nu \in N$ which label the endpoint of the corresponding line defect on the (anti)holomorphic boundary,
- a topological point junction $\zeta$ on the surface defect $\mathcal{A}$ which parametrises the ways in which the line defects $M, N$ can start on $\mathcal{A}$.

The chiral TFT pictures is well suited for working with 2d CFTs. For example, it encodes the separation into holomorphic and antiholomorphic fields already into the geometry. As an example, consider a local patch showing two bulk fields, one inserted at 0 and one inserted at $z$ as in the middle picture below. For the latter field, the endpoint of the line defect on the holomorphic boundary is at $z$ and on the anti-holomorphic boundary at the complex conjugate $\bar{z}$. We carry out two different deformations of this situation:

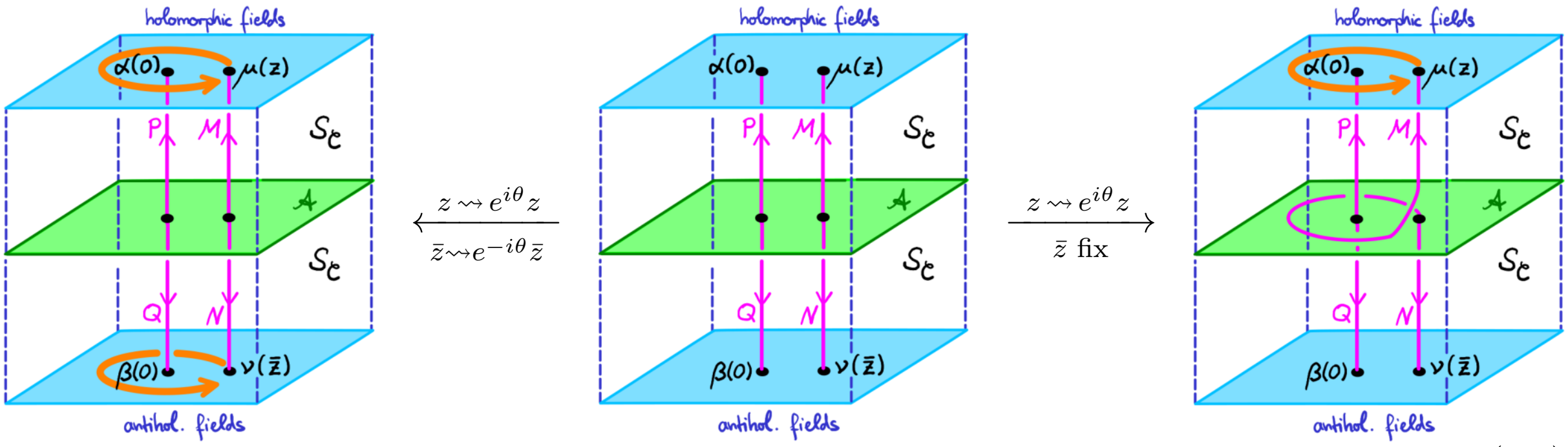

(4.5)

On the left hand side, we move $z$ around 0 and move $\bar{z}$ accordingly. In this way one returns to the initial situation, showing that correlators are single valued in the positions of the bulk fields. On the right hand side, we just change $z \rightsquigarrow e^{i\theta} z$ while keeping $\bar{z}$ fixed at its original point – we can do this because of the geometric separation of the endpoints on the holomorphic and antiholomorphic boundaries. A priori this seems to not make sense from the point of view of the 2d CFT, but the chiral TFT picture gives a natural interpretation to this process: The bulk field is split into a pair of disorder fields, one purely holomorphic, one purely antiholomorphic, and the deformation generates a circular topological defect surrounding the bulk field at zero, and it changes the bulk field at $z$ into a defect field on that defect. Note that even though $z$ returns to its original position, the correlator is in general not single valued under this deformation. This is visible from the TFT as the defect configuration inside the three-manifold has changed.

There is one surface defect which one always gets for free, namely the trivial one. This describes the CFT where $M, N$ in (4.4) are conjugate, $M \cong N^*$, for $M, N$ simple – also called the charge-conjugation CFT. For general surface defects, one can show:

1. All surface defects (which can end on a line defect with non-zero dimension) are obtained by gauging a line defect, that is, from a Δ-separable symmetric Frobenius algebra in $\mathcal{C}$.
2. All 2d CFTs which contain $V \otimes_{\mathbb{C}} \overline{V}$ in their bulk fields ($V$ a rational VOA), have a unique vacuum, and non-degenerate two-point correlators, are obtained from the chiral TFT in this way, i.e. are in 1-1 correspondence to elementary surface defects as above.

In this sense, the chiral TFT picture allows for a complete understanding of rational CFTs, given the "local data", i.e. the chiral algebra $V$ and its representation theory.

# 5 Bibliography

**Section 1:** Some references and lecture notes on SymTFT are [2–10]. The name "SymTFT" was first used in [4]. I am not aware of a detailed treatment of the 1d case as given in these notes. The 2d case is briefly mentioned in [3]. State-sum models in 4d have been defined in terms of suitable fusion-2-categories in [11]. The description of 2d CFT via chiral 3d TFT was developed earlier in the series of papers [12–20].

**Section 2:** One way to include continuous symmetries is to couple to area-dependent field theories such as two-dimensional Yang-Mills theory, see [21] for details on state sum constructions of such theories including defects with length parameters, as well as for further references. How to weaken the condition of Δ-separability to just separable is contained in [22]. That being Δ-separable symmetric Frobenius is a property of an algebra is explained e.g. in [13, Thm. 3.6]. Lecture notes on TFT with and without defects are e.g. [23, 24]. 2d state sums were defined in [25, 26], and the description of boundary conditions can be obtained as a special case of [13] (take the ambient modular fusion category to be that of vector spaces) and is directly given in [27]. Line defects in 2d state sums are described in [28].

**Section 3:** Topological line defects and the monoidal category they form were described in 2d CFT in [13, 18], and in more general 2d QFTs in [28–32].

Supersymmetric Landau-Ginzburg models provide examples where the left and right quantum dimension do not agree (and are in general not real-valued) [33, 34]. That quantum dimensions in a spherical fusion category over the complex numbers are necessarily real is shown in [35, Cor. 2.10].

The pentagon identity for the F-symbols is written out in the present conventions e.g. in [15, App. C.2]. The connection to group cohomology is already given in [36, App. E] (and e.g. in [15, App. A] in the present conventions). That the class $[\omega]$ in the third group-comology is an obstruction to gauging is stated in the present setting in [15, Sec. 3.3] and [37], see also [29, 30].

The 3d state sum TFTs are also called Turaev-Viro-Barrett-Westbury theories [38–40]. The version used here is actually a special case of the generalised orbifold construction for 3d TFTs from [1, 41] (this includes the dimension factors that were skipped in the main text), and the coupling of these orbifolds to boundaries is described in [42] (for topological boundaries, but the conditions for non-topological boundaries are analogous). Boundary conditions for 3d TFTs in terms of module categories are described in [43, 44].

That different topological boundary conditions for the $(n+1)$-dimensional TFT define different $n$-dimensional QFTs is discussed the literature on SymTFT cited above. The topological line defects of the Ising CFT and the corresponding 3d TFT description are given in [45–47].

**Section 4:** The description of 2d CFT via chiral 3d TFT was developed before SymTFT, and references were already given above. The foundational example is 3d Chern-Simons theory and 2d Wess-Zumino-Witten models [48]. The relation between chiral TFT and SymTFT is briefly discussed in [10]. The relation between the chiral TFT construction in [13] and surface defects was found in [43, 49], and surgery TFTs with surface defects were developed in [50, 51]. That surface defects which can end on a line are obtained by gauging line defects is shown in [43, 49].

The classification of 2d CFTs in terms of surface defects, or, more precisely, in terms of (Morita classes of) $\Delta$-separable symmetric Frobenius algebras, is given in [19, 52, 53].

That for any rational CFT $Q$ there is a choice of topological symmetry $\mathcal{F}$ such that only $V \otimes_{\mathbb{C}} \overline{V}$ commutes with $\mathcal{F}$ can be seen as follows: by the classification results in Section 4, $Q$ corresponds to a certain Frobenius algebra $A$ in $\mathrm{Rep}(V)$. Take $\mathcal{F}$ to be all $A$-$A$-bimodules in $\mathrm{Rep}(V)$. The claim follows from the non-degeneracy of a generalised $S$-matrix pairing between bulk fields and line defects [18, Thm. 4.2]. A closely related result is the Galois correspondence between sub-VOAs and fusion sub-categories described in [54, Sec. 7].

### Acknowledgements

I would like to thank the organisers of the "28$^e$ rencontre Itzykson: Analytic results in Conformal Field Theory" (Sept. 10–12, 2024) at the IPhT Saclay for the opportunity to present this overview and for a very inspiring meeting. Many thanks to Aaron Hofer and Sylvain Ribault for helpful comments on a draft of these proceedings. I also thank the Deutsche Forschungsgemeinschaft (DFG, German Research Foundation) under Germany's Excellence Strategy - EXC 2121 "Quantum Universe" - 390833306, and the Collaborative Research Center - SFB 1624 "Higher structures, moduli spaces and integrability" - 506632645, for support.

# Quantum group symmetric conformal field theories

**Bernardo Zan**

*Dipartimento di Fisica, Università di Genova and INFN, Sezione di Genova, Via Dodecaneso 33, 16146, Genoa, Italy*

*Department of Applied Mathematics and Theoretical Physics, University of Cambridge, CB3 0WA, UK*

ABSTRACT: Quantum groups are algebras which have been known to describe the symmetry of certain lattice systems. In the continuum, instead, they seem to play a different and more subtle role, appearing in two-dimensional CFTs in a more indirect manner; in the literature, they are sometimes referred to as "hidden symmetries". In this contribution, we clarify the role of quantum groups as global symmetries in the continuum by considering CFTs which have the quantum group as a bona fide global symmetry in the case of $U_q(sl_2)$. Finally, we discuss how these theories are related to minimal models, where quantum groups are known to appear indirectly. The following is based on [1, 2].

# Contents

# 1 Introduction

Symmetries have long played a fundamental role in the study of QFTs and their renormalization group flows. Historically, symmetries have been described by groups, but in the last decade or so, signficant progress has shown that such a view of global symmetries is too narrow; this progress goes broadly under the name of generalized symmetries. This progress includes higher form symmetries [3], which can act on extended operators rather than local ones; non-invertible symmetries, which do not form a group but rather a fusion category (see [4] for a complete set of references); and higher group symmetries, which combine symmetries of different forms [5,6]. Given these generalizations, it is only natural to ask if this is all there is, or if further generalization of the concept of global symmetry can play a role in QFT.

A hint that this is not the full story can be seen by considering some lattice models which have different symmetry structures. The XXZ model with imaginary surface terms [7] is an example of a spin chain whose global symmetry is the quantum group $U_q(sl_2)$. Given that this model is critical, we expect the long distance physics of the spin chain to be described by a CFT, which should inherit the same global symmetry structure.

With the larger question in mind of understanding what is the most general symmetry that a QFT can have, in this contribution we will study CFTs with quantum group symmetry. For simplicity, we focus on the quantum group $U_q(sl_2)$. The following is based on [1,2].

# 2 A lightning-quick review of $U_q(sl_2)$

**Commutation relations.** The first thing to know about the quantum group $U_q(sl_2)$ is that it's not a group, but rather an algebra. To be precise, it is a quasitriangular Hopf algebra, which can be obtained as a deformation of the enveloping algebra of $sl_2$, $U(sl_2)$. Similarly to $sl_2$, we have a raising generator $E$, a lowering generator $F$, and the Cartan element $H$. The commutation relations are the $q$-deformed commutation relations of the $sl_2$ algebra

$$q^H E = q^2 E q^H\,, \qquad q^H F = q^{-2} F q^H\,, \qquad [E,F] = [H]_q \equiv \frac{q^H - q^{-H}}{q - q^{-1}}\,. \tag{2.1}$$

The parameter $q \in \mathbb{C}$ is arbitrary, and we recover the $sl_2$ commutation relations in the limit $q \to 1$.

**Coproduct.** A very important property of Hopf algebras is that they are endowed with a coproduct $\Delta : U_q(sl_2) \to U_q(sl_2) \otimes U_q(sl_2)$. Physically speaking, if we know how the algebra acts on one operator, the coproduct tells us how it acts on two or more operators, and it will play a role in constraining the correlation functions through Ward identities. In the case of $sl_2$, this is the very simple rule $\Delta(X) = X \otimes \mathbb{1} + \mathbb{1} \otimes X$, for $X \in sl_2$. The situation is slighlty trickier for $U_q(sl_2)$, because the coproduct needs to be compatible with the commutation relations, meaning that

$$\Delta([X, Y]) = [\Delta(X), \Delta(Y)] , \qquad \forall\, X, Y \in U_q(sl_2) . \tag{2.2}$$

Given that the commmutation relations have been deformed, we need to deform the coproduct as well. The choice is not unique; we will work with

$$\begin{aligned} \Delta(E) &= E \otimes \mathbb{1} + q^{-H} \otimes E , \\ \Delta(F) &= F \otimes q^{H} + \mathbb{1} \otimes F , \\ \Delta(q^H) &= q^H \otimes q^H . \end{aligned} \tag{2.3}$$

**R-matrix.** The coproduct $\Delta$ treats the two copies of $U_q(sl_2)$ differently. We could have defined $\tilde{\Delta}$, a coproduct where 'left' and 'right' copies of $U_q(sl_2)$ are swapped, e.g. $\tilde{\Delta}(E) = E \otimes q^{-H} + 1 \otimes E$. The R-matrix is an element of $U_q(sl_2) \otimes U_q(sl_2)$ which allows us to go from one coproduct to the other

$$\mathcal{R}\Delta(X)\mathcal{R}^{-1} = \tilde{\Delta}(X) . \tag{2.4}$$

An explicit formula for the R-matrix can be found in [8].

**Representation theory.** In the CFT, we will want operators to transform in representations of $U_q(sl_2)$. The representation theory of $U_q(sl_2)$ depends on whether $q$ is a root of unity.

For generic $q$, i.e. not root of unity, the finite-dimensional representations of $U_q(sl_2)$ are the same of those of $su(2)$ or $sl_2$; they are labelled by a half-integer spin $\ell$ and are $2\ell+1$ dimensional. Every element of the representation is therefore labelled by two quantum number $|\ell, m\rangle$, with $m = -\ell, \dots, \ell$. The generators $E$ and $F$ act by raising and lowering $m$, while $H$ reads off $m$

$$\begin{aligned} H\,|\ell, m\rangle &= 2m\,|\ell, m\rangle \\ E\,|\ell, m\rangle &= e_{\ell,m}\,|\ell, m+1\rangle \qquad & e_{\ell,m} &= q^{-m}\sqrt{[\ell - m]_q[\ell + m + 1]_q} \\ F\,|\ell, m\rangle &= f_{\ell,m}\,|\ell, m-1\rangle \qquad & f_{\ell,m} &= q^{m-1}\sqrt{[\ell + m]_q[\ell - m + 1]_q} \end{aligned} \tag{2.5}$$

The state $|\ell, \ell\rangle$ is a highest weight state and $|\ell, -\ell\rangle$ is a lowest weight one , $E\,|\ell, \ell\rangle = F\,|\ell, -\ell\rangle = 0$.

The situation is more complicated when $q$ is a root of unity, because then for some integer $p$, $E^p = F^p = 0$. Then we might not be able to build the representations by starting with a highest weight state and acting with $F$ several times until we reach a lowest weight state. In this contribution we take $q$ to be a phase, but we focus on the generic case. See [9] for more details about the $U_q(sl_2)$ representations at $q$ root of unity.

**$3j$ and $6j$ symbols.** Finally, $U_q(sl_2)$ has $3j$-symbol and $6j$-symbols. Their value depends non-trivially on $q$, but the role they play is the same we are familiar with from the $su(2)$ algebra. Later on, we will need the quantum Clebsch Gordan coefficients, which are closely related to $3j$-symbols, and which tells us how to decompose the product of two representations of $U_q(sl_2)$

$$|\ell_1, m_1\rangle \otimes |\ell_2, m_2\rangle = \sum_{\ell_3=|\ell_1-\ell_2|}^{\ell_1+\ell_2} \sum_{m_3=-\ell_3}^{m_3=\ell_3} \begin{bmatrix} \ell_1 & \ell_2 & \ell_3 \\ m_1 & m_2 & m_3 \end{bmatrix}_q |\ell_3, m_3\rangle \, . \tag{2.6}$$

The explicit form of the quantum Clebsch-Gordan coefficients can be found in [10].

# 3 CFTs with quantum group symmetry

## 3.1 General properties

At this point we are ready to say what a CFT with quantum group symmetry looks like. We will work in $d = 2$, for reasons that will be clear soon, and in Euclidean space, given that these theories will not be unitary.

We will say that a CFT has an internal $U_q(sl_2)$ global symmetry if

1. Operators transform linearly under the elements of the algebra

$$\mathcal{O} \to X \cdot \mathcal{O} \, , \qquad X = E, F, H \, ; \tag{3.1}$$

2. The quantum group commmutes with the spacetime symmetries

$$[U_q(sl_2), \text{Virasoro}] = 0 \, . \tag{3.2}$$

This means that operators of the theories will be labeled by the usual Virasoro lables $(h, \bar{h})$, but also by the quantum group labels $\ell, m$, specifying the $U_q(sl_2)$ representation $|\ell, m\rangle$ in which they transform.

A consequence is that correlation functions satisfy Ward identities

$$\langle X \cdot (\mathcal{O}_1 \dots \mathcal{O}_n) \rangle = 0 \, , \qquad X = E, F, H \tag{3.3}$$

where $X$ has to be intended as its coproduct the appropriate number of times, $\Delta^{n-1}(X)$, because it's acting on the tensor product of $n$ operators transforming under $U_q(sl_2)$. We are also assuming that the vacuum is a singlet under $U_q(sl_2)$, which is therefore not spontaneously broken.

This already has important consequences: the Ward identities imply that operators are not mutually local. Two operators in an Euclidean QFT are mutually local if their correlation functions are permutation invariant (up to minus signs in the case of fermions), e.g.

$$\langle \mathcal{O}_1(x) \mathcal{O}_2(y) \rangle = \pm \langle \mathcal{O}_2(y) \mathcal{O}_1(x) \rangle \tag{3.4}$$

Let us see what the consequences of the Ward identities are in the simplest setting of an operator transforming in the spin $1/2$ representation of $U_q(sl_2)$. The multiplet consists of two operators $\mathcal{O}_\pm$ with magnetization $m = \pm\frac{1}{2}$.

The Ward identity for the two point function of these operators is

$$0 = \langle F \cdot (\mathcal{O}_+(x)\mathcal{O}_+(y)) \rangle = f_{\frac{1}{2},\frac{1}{2}} \left( q \langle \mathcal{O}_-(x)\mathcal{O}_+(y) \rangle + \langle \mathcal{O}_+(x)\mathcal{O}_-(y) \rangle \right) \tag{3.5}$$

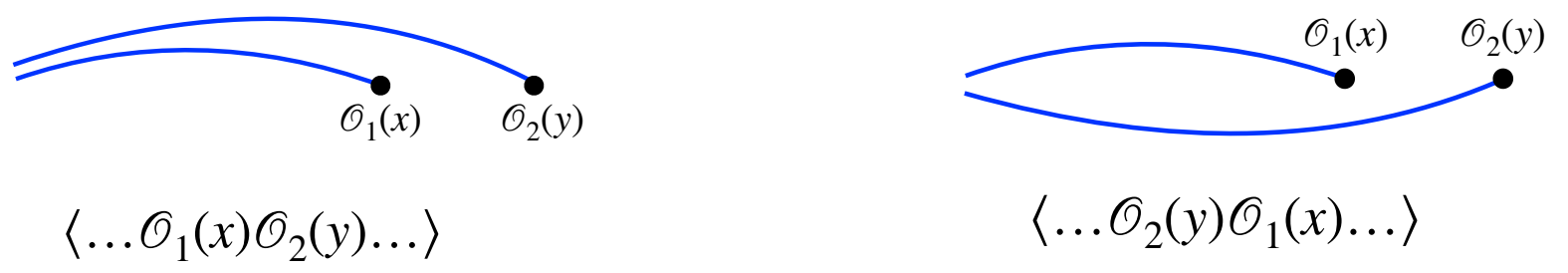

Figure 1: Example of different correlation functions related to different permutations of topological lines. We do not discuss here the other endpoint of the topological lines, and defer to [1] for more discussion.

where we have used the coproduct (2.3) and the matrix elements (2.5). Now we can use the Ward identities for rotations, which state that

$$\langle \mathcal{O}_1(x)\mathcal{O}_2(y)\rangle = (-1)^{2\mathfrak{s}} \langle \mathcal{O}_1(y)\mathcal{O}_2(x)\rangle \tag{3.6}$$

with $\mathfrak{s}$ the spacetime spin. With this, we can finally rewrite the Ward identities as

$$\langle \mathcal{O}_+(x)\mathcal{O}_-(y)\rangle = -q(-1)^{2\mathfrak{s}} \langle \mathcal{O}_-(y)\mathcal{O}_+(x)\rangle \tag{3.7}$$

For generic $q$ we have two possibilities: either $\mathfrak{s}$ is not a half-integer, or $\langle \mathcal{O}_+(x)\mathcal{O}_-(y)\rangle \neq \langle \mathcal{O}_-(y)\mathcal{O}_+(x)\rangle$. In any of case, we have lost mutual locality.[1]

What does this mean? We would like to think of operators $\mathcal{O}_i(x)$ living at a point $x$, but the lack of mutual locality means that this needs to be modified. The simplest modification to this picture is that operators live at points but are attached to a line. This tail only needs to be topological, i.e. it can be moved around freely as long as it does not cross other operators. Different permutations of operators correspond to different configurations of the topological lines, see Figure 1.

Lines can still cross operators, but not for free. The price to pay for this is the R-matrix of the quantum group, which allows us to go from the left to the right of Figure 1. This has a further consequence about $U_q(sl_2)$ symmetric CFTs. We can start from a correlation function $\langle \mathcal{O}_1(x)\mathcal{O}_2(y)\rangle$, and by moving operators around as well as swapping the topological lines with the R-matrix, we can get the correlation function $\langle \mathcal{O}_2(x)\mathcal{O}_1(y)\rangle$. The requirement that these two quantities match gives the non-trivial constraint between spacetime and $U_q(sl_2)$ spin

$$\mathfrak{s} = \pm \left( \frac{\ell(\ell+1)}{\pi i} \log q - \ell \right) + \mathbb{Z}\,, \tag{3.8}$$

where the $\pm$ is fixed in a given theory. As we can see, generically the spacetime spin will not be a half-integer.

## 3.2 XXZ$_q$ as a check

After discussing the general characteristics of a CFT with quantum group symmetry, we are ready to discuss a particular example. This will allow us to check the validity of our construction by checking that this CFT obeys the properties mentioned above.

We will consider a quantum group symmetric spin chain which is critical, and gives rise to a CFT in the continuum limit. In principle we would like to work with periodic boundary conditions, however we do not know of any system with $U_q(sl_2)$ symmetry and periodic boundary conditions. The closest thing we can do is to take the spin chain introduced in [11], which we dub $XXZ_q$. The Hamiltonian is built out of the following objects

$$R_{i=1,\dots,N-1} = \frac{1}{2}\left(\sigma_i^x\sigma_{i+1}^x + \sigma_i^y\sigma_{i+1}^y + \frac{q+q^{-1}}{2}\left(\sigma_i^z\sigma_{i+1}^z + 1\right) - \frac{q-q^{-1}}{2}\left(\sigma_i^z - \sigma_{i+1}^z - 2\right)\right) \tag{3.9}$$

where $N$ is the number of lattice sites. Notice that the index $i$ only runs up to $N-1$. These objects have the nice properties that they commute with the generators of the quantum group on the lattice

$$\begin{aligned} E &= q^{1/2}\sum_{i=1}^{N} q^{-\sum_{j<i}\sigma_j^z}\sigma_i^+ \\ F &= q^{-1/2}\sum_{i=1}^{N} \sigma_i^- q^{\sum_{j>i}\sigma_j^z} \\ q^H &= \prod_{i=1}^{N} q^{\sigma_i^z} \end{aligned} \tag{3.10}$$

with $\sigma^\pm = \frac{1}{2}(\sigma^x \pm i\sigma^y)$. Furthermore, they are generators of a Hecke algebra [12]. Using these properties, we can define an operator $G \equiv R_1 R_2 \dots R_{N-1}$ which acts as a translation operator on the $R_i$'s,

$$GR_iG^{-1} = R_{i+1}\,, \tag{3.11}$$

which allows us to define the missing $R_N$

$$R_N \equiv GR_{N-1}G^{-1}\,. \tag{3.12}$$

[1] The fact that non-integer spacetime spin implies the loss of mutual locality can be seen by taking two point functions of operators and rotating one a full circle around the other. The correlation function is unchanged only if $\mathfrak{s} \in \mathbb{Z}/2$.

An expression of $R_N$ in terms of Pauli matrices is very complicated, so we do not write it here. It is important to notice, however, that $R_N$ is as non-local as it gets, with $N$-sites interaction. Then the $XXZ_q$ model is defined by the Hamiltonian

$$\mathcal{H}_{XXZ_q} = -\sum_{i=1}^{N} R_i\,, \tag{3.13}$$

which is indeed $U_q(sl_2)$ invariant. It is not quite translation invariant, but it commutes with $G$.[2]

This Hamiltonian is not unitary, but it is PT symmetric and has real eigenvalues [13]. It is critical, and we will study the CFT arising in the thermodynamic limit. Its spectrum has been studied in [14], and we report it here. It is convenient to parametrize $q = \exp(i\pi\frac{\mu}{\mu+1})$ with $\mu > 1$. Then the central charge of the CFT is given by

$$c = 1 - \frac{6}{\mu(\mu+1)}\,. \tag{3.14}$$

This is the same central charge of generalized minimal models $\mathcal{M}_{\mu,\mu+1}$.

We denote primary operators of the theory as $W^m_{r,s}$ where $r$ is a positive integer, $s$ an odd positive integer, and $m$ an integer between $-\frac{s-1}{2} \le m \le \frac{s-1}{2}$. The weights of these operators under the Virasoro algebra are

$$L_0 W^m_{r,s} = h_{r,s} W^m_{r,s}\,, \qquad \bar{L}_0 W^m_{r,s} = h_{r,1} W^m_{r,s} \tag{3.15}$$

[2]Notice that if we only considered the sum from $i = 1$ to $N-1$, we would still obtain a quantum group symmetric system. Specifically, we would obtain a system very closely related to that of [7].

with

$$h_{r,s} = \frac{(r(\mu+1) - s\mu)^2 - 1}{4\mu(\mu+1)} \tag{3.16}$$

the usual dimension of degenerate operators in $c < 1$ theories.[3]

For what concerns the quantum group global symmetry, the operators $W^m_{r,s}$ come in $s$ dimensional degenerate multiplets, which means that they transform in $\ell = \frac{s-1}{2}$ spin representation for $U_q(sl_2)$, while $m$ represent (half of) their $H$ eigenvalue.

The first thing we notice is that the spacetime spin $\mathfrak{s} = h - \bar{h}$ is generically not a half-integer, which we expect from these operators being not mutually local. The construction of Section 3.1 predicts the very constraining relation (3.8) between the spacetime and the $U_q(sl_2)$ spin. A little work shows that this relation is satisfied in $\mathrm{XXZ}_q$, providing a first check that everything is working as intended.

A further check of the consistency of our construction is that the theory is crossing symmetric. While the spectrum of the CFT has been known for some time [14], OPE coefficients have only recently been computed [1, 2] following two independent mathods:

- we can use the fact that operators $W_{r,s}$ are degenerate, and have a null descendant at level $rs$ in the holomorphic sector (and $r$ in the antiholomorphic one). This means that we can study differential equations *à la* BPZ [15] to obtain the Virasoro blocks, and then impose crossing symmetry to fix the OPE coefficients;

- we can develop a Coulomb gas method, building from [16], in which correlation functions of a free theory, after integration over some contours, give correlation functions of the theory we're interested in. This allows us to get the OPE coefficients.

We refer the reader to [1, 2] for the details of these constructions and the explicit results. The important outcome is that both method give the same results and we have the non-trivial result that the theory admits a solution of OPE coefficients that make it crossing symmetric. This provides another important check that our construction of a CFT is consistent with the critical limit of quantum group symmetric theories.

# 4 Connection with minimal models

$\mathrm{XXZ}_q$ has some connections with other theories which have no quantum group symmetries, i.e. generalized minimal models. Here we briefly explore one.[4]

We remind the reader that when expanding a four point function, one can do so in different channels

$$\begin{aligned}\langle \mathcal{O}_1(0)\mathcal{O}_2(z,\bar{z})\mathcal{O}_3(1)\mathcal{O}_4(\infty)\rangle &= \sum_j C_{12j}C_{j34}\mathcal{F}^{(s)}_{h_j}(z)\bar{\mathcal{F}}^{(s)}_{\bar{h}_j}(\bar{z}) \\ &= \sum_k C_{23k}C_{1k4}\mathcal{F}^{(t)}_{h_k}(z)\bar{\mathcal{F}}^{(t)}_{\bar{h}_k}(\bar{z})\end{aligned} \tag{4.1}$$

where $C_{ijk}$ are OPE coefficients, and $\mathcal{F}_h(z)$ are Virasoro blocks. It has long been known that, in minimal models, the crossing kernel of Virasoro blocks $\mathbb{F}_{k,j}$, defined as

$$\mathcal{F}^{(t)}_{h_k}(z) = \sum_j \mathbb{F}_{k,j}\mathcal{F}^{(s)}_{h_j}(z) \tag{4.2}$$

[3] The CFT is not symmetric between left and right movers because the lattice Hamiltonian (3.13) is not invariant under spacetime parity.

[4] Another interesting connection, which we do not discuss here, is related to $\mathrm{XXZ}_q$ developing a subsector of the theory which coincides with minimal models for integer $\mu$. See [1] for more details.

contains $6j$ symbols of $U_q(sl_2)$ [17]. This is puzzling: we know minimal models do not have quantum group symmetry. Then why do these $6j$ symbols appear? We will see that the existence and crossing symmetry of the XXZ$_q$ theory provide an explanation of this.

The OPE of operators $W_{r,s}$ in XXZ$_q$ has the following form

$$W^{m_1}_{r_1,s_1}(z_1,\bar{z}_1)W^{m_2}_{r_2,s_2}(z_2,\bar{z}_2) = \sum_k \frac{C_{12k}}{z_{21}^{h_{12k}}\,\bar{z}_{21}^{\bar{h}_{12k}}} \begin{bmatrix} \frac{s_1-1}{2} & \frac{s_2-1}{2} & \frac{s_k-1}{2} \\ m_1 & m_2 & m_k \end{bmatrix}_q \left(W^{m_k}_{r_k,s_k}(z_1,\bar{z}_1) + \ldots\right). \tag{4.3}$$

Using this, let us consider the two different $s$ and $t$-channel expansion of a correlation function. For simplicity, consider operators $W_{1,s}$, which have scaling dimensions $(h_{1,s}, h_{1,1} = 0)$, and therefore are purely chiral. We will consider the following correlation function and its two different expansions

$$\begin{aligned} &\langle W^{m_1}_{1,s_1}(0)W^{m_2}_{1,s_2}(z)W^{m_3}_{1,s_1}(1)W^{m_4}_{1,s_2}(\infty)\rangle = \\ &= \sum_j C^2_{12j} \begin{bmatrix} \frac{s_1-1}{2} & \frac{s_2-1}{2} & \frac{s_j-1}{2} \\ m_1 & m_2 & m_j \end{bmatrix}_q \begin{bmatrix} \frac{s_j-1}{2} & \frac{s_1-1}{2} & \frac{s_2-1}{2} \\ m_j & m_3 & -m_4 \end{bmatrix}_q \begin{bmatrix} \frac{s_2-1}{2} & \frac{s_2-1}{2} & 0 \\ -m_4 & m_4 & 0 \end{bmatrix}_q \mathcal{F}^{(s)}_{1,s_j}(z) = \\ &= \sum_k C^2_{12k} \begin{bmatrix} \frac{s_2-1}{2} & \frac{s_1-1}{2} & \frac{s_k-1}{2} \\ m_2 & m_3 & m_k \end{bmatrix}_q \begin{bmatrix} \frac{s_1-1}{2} & \frac{s_k-1}{2} & \frac{s_2-1}{2} \\ m_1 & m_k & -m_4 \end{bmatrix}_q \begin{bmatrix} \frac{s_2-1}{2} & \frac{s_2-1}{2} & 0 \\ -m_4 & m_4 & 0 \end{bmatrix}_q \mathcal{F}^{(t)}_{1,s_k}(z) \end{aligned} \tag{4.4}$$

where we've used the cyclicity property of $C_{ijk}$ for chiral operators $W_{1,s}$ [1].

Using orthogonality properties of quantum Clebsch Gordan coefficients [18], and the definition of $6j$-symbol as the product of four quantum Clebsch Gordan coefficients [19], we can rewrite (4.4) as

$$\mathcal{F}^{(t)}_{1,s_k}(z) = \sum_j \frac{C^2_{12j}}{C^2_{12k}} \begin{Bmatrix} \frac{s_1-1}{2} & \frac{s_2-1}{2} & \frac{s_j-1}{2} \\ \frac{s_1-1}{2} & \frac{s_2-1}{2} & \frac{s_k-1}{2} \end{Bmatrix}_q \mathcal{F}^{(s)}_{1,s_j}(z) \tag{4.5}$$

We see therefore that $6j$-symbols naturally appear in the crossing kernel of Virasoro blocks for the XXZ$_q$ theory. However, Virasoro blocks only care about the central charge of the theory and the dimensions $h_i$ of operators. This means that, given that the XXZ$_q$ theory exists for continuous $c \leq 1$ and has degenerate operators, the property (4.5) applies to minimal models as well. Therefore, the existence and consistency (i.e. crossing symmetry) of XXZ$_q$ provides an explanation to the appearence of $6j$-symbols in the crossing kernel of minimal models.

# 5 Conclusions

In this contribution, we've answered the question of whether CFTs with quantum group symmetry exist in the affirmative. We've described their peculiarities, especially the appearence of defect ending operators, in the case of the quantum group $U_q(sl_2)$. We have provided an example of such theory, XXZ$_q$, and checked that it's well defined. This theory is closely related to generalized minimal models.

Let us finish by mentioning some open questions:

- How does this fit with the recent progress related to generalized symmetries? Specifically, what are the properties of the topological tails of defect-ending operators?
- What is the microscopic construction of these extended operators in the XXZ$_q$ lattice model?
- Can we find unitary quantum group symmetric CFTs?
- How does the picture generalize to higher dimensions?

# Logarithmic energy operators in percolation bulk CFT

**Yifei He**

*Laboratoire de Physique de l'École Normale Supérieure, ENS, Université PSL, CNRS, Sorbonne Université, Université Paris Cité, F-75005 Paris, France*

*E-mail:* `yifei.he@ens.fr`

Abstract: We study the energy and second-energy operators in the 2d percolation bulk CFT with central charge $c = 0$. Their normalizations at generic $c$ can be analyzed using the reality of the random cluster CFTs; at $c = 0$, they become zero-norm operators and mix into logarithmic multiplets. By taking the $c \to 0$ limit, we compute some of the three-point constants and find that the energy operator couples non-trivially to the rank-3 Jordan blocks of the second energy operator. Surprisingly, this coupling builds a non-vanishing four-point function of the energy operator in the percolation bulk CFT.

# Contents

# 1 Introduction

One of the most fascinating potential applications of 2d CFTs is to understand disordered systems or geometrical phase transitions with central charge $c = 0$, and a signature example is the critical percolation. Challenges arise not only because the $c = 0$ CFTs are non-unitary, but also because they are logarithmic [1]. In this case, one immediately encounters the "$c \to 0$ catastrophe" where the operator product expansions (OPE) become singular and logarithmic operators appear. While the logarithmic mixing involving the stress energy tensor $T$ have been analyzed over the years [2–6, 14], little progress on the $c = 0$ CFTs were made beyond that.

In this contribution, we consider the energy-type operators in the percolation bulk CFT that originate from Kac operators in the Potts cluster CFTs. This focus is motivated for two reasons: First, the energy-type operators have direct physical interpretations in critical percolation and thus their CFT correlation functions should encode interesting physical observables; second, Kac operators have played crucial roles in analytic approaches to 2d CFTs, so it is important to study their fates at $c = 0$ – hoping to gain an analytic handle on these fascinating CFTs.

We are able to make progress thanks to the recent numerical and analytic bootstrap studies of Potts cluster CFTs at generic $c$ [7–11]. In particular, it was understood recently that the cluster CFTs are logarithmic at generic $c$ [10, 12, 13]. In the limit $c \to 0$, this gives rise to the percolation CFT and has led to the discovery of a rank-3 Jordan block at $c = 0$ [14]. The existence of such a higher rank Jordan block turns out to have profoundly interesting effect on the percolation bulk CFT as we shall see.

The analysis we describe was originally done in [15] and here we give a brief summary and direct the interested readers to the original paper for details. Following [15], we parameterize the central charge as

$$c = 13 - \frac{6}{\beta^2} - 6\beta^2 \ , \tag{1.1}$$

where the critical cluster model CFTs correspond to

$$\frac{1}{2} \leq \beta^2 \leq 1 \,, \tag{1.2}$$

and the scaling dimensions are parametrized by the Kac indices as usual:

$$h_{r,s} = \frac{1}{4}\left(\left(r\beta^{-1} - s\beta\right)^2 - \left(\beta^{-1} - \beta\right)^2\right) \,. \tag{1.3}$$

At $c = 0$ or $\beta^2 = \frac{2}{3}$, the OPE of a generic primary operator $\mathcal{O}$ becomes singular:

$$\mathcal{O}(z,\bar{z})\mathcal{O}(0,0) = (z\bar{z})^{-2h_{\mathcal{O}}(c)} B_{\mathcal{O}}(c)\left(1 + \frac{h_{\mathcal{O}}(c)}{c/2}(z^2 T + \bar{z}^2\bar{T}) + \ldots\right) \,. \tag{1.4}$$

This is known as the "$c \to 0$ catastrophe". There are several resolutions to this [16, 17]: 1, the dimension $h_{\mathcal{O}}(c)$ could vanish at $c = 0$, cancelling the singular denominator of the OPE coefficients; 2, there could be other terms in the $\ldots$ from operators with dimension $(h, \bar{h}) = (2, 0)$ at $c = 0$ to cancel the divergent term of $T$. This has led to understanding the famous logarithmic pair $(t, T)$ [3]. Here, we are interested in the third resolution where $B_{\mathcal{O}} = 0$ at $c = 0$. This must be the case for all Kac operators $\mathcal{O} = \Phi_{r,s}$ since their OPEs are severely restricted by the Virasoro degeneracy and the previous two resolutions do not apply. The spectrum of the cluster model [18] contains the Kac operators $\Phi_{r,1}, r = 1, 2, \ldots$ . In particular, the $\Phi_{2,1}$ is known as the energy operator and the $\Phi_{3,1}$ is the second energy operator and we are interested in the fate of their correlation functions and OPEs at $c = 0$. The first step is to understand the proper normalization $B_{\Phi_{r,1}}$ of these operators.

# 2 Real CFT operators

To determine the normalization of Kac operators $\Phi_{r,1}$, we consider their amplitudes in the four-spin correlator in the cluster model at generic $c$, written in terms of the conformal data:

$$A^{\sigma}_{\Phi}(c) = \frac{C^2_{\sigma\sigma\Phi}(c)}{B_{\Phi}(c)} \tag{2.1}$$

where $C_{\sigma\sigma\Phi}$ is the three-point constant and $B_{\Phi}$ is the normalization. In unitary CFTs, operators can be conveniently normalized to 1, thanks to reflection positivity $B > 0$. In the case of the cluster model, we have a one-parameter family of non-unitary CFTs with $-2 < c < 1$ where reflection positivity is violated [19], so the normalization can be a non-trivial function of the central charge $c$. To analyze such a function, we resort to another physical principle – the cluster CFTs are real, since they arise from the long-distance description of random cluster ensembles with real and positive measures. Therefore, correlation functions should satisfy the reality condition [20]:

$$\langle \mathcal{O}_1(x_1)\mathcal{O}_2(x_2)\ldots\mathcal{O}_n(x_n)\rangle^* = \langle \mathcal{O}_1(x_1)\mathcal{O}_2(x_2)\ldots\mathcal{O}_n(x_n)\rangle \,, \tag{2.2}$$

where in particular for the three-point functions:

$$\langle \mathcal{O}_1(x_1)\mathcal{O}_2(x_2)\mathcal{O}_3(x_3)\rangle^* = \langle \mathcal{O}_1(x_1)\mathcal{O}_2(x_2)\mathcal{O}_3(x_3)\rangle \,. \tag{2.3}$$

This means that the three-point constants satisfy the reality condition

$$C^*_{\mathcal{O}_1\mathcal{O}_2\mathcal{O}_3} = C_{\mathcal{O}_1\mathcal{O}_2\mathcal{O}_3} \tag{2.4}$$

and their squares are non-negative for the whole range of central charge $c$ (1.2). The reality requirement puts constraints on the operator normalizations $B_{\Phi}(c)$ as a function of the continuous parameter $c$.

# 3 Kac operators in random cluster model

For Kac operator $\Phi_{r,1}, r = 1, 2, 3, \ldots$, we can extract their normalizations since their four-spin amplitudes (2.1) have been calculated using the interchiral symmetry and can be written in terms of the products of the amplitude ratios [9]

$$A^{\sigma}_{\Phi_{r,1}} = \prod_{i=1}^{r-1} R_{i,1}, \quad r = 2, 3, \ldots, \tag{3.1}$$

where

$$R_{i,1} = \frac{A^{\sigma}_{\Phi_{i+1,1}}}{A^{\sigma}_{\Phi_{i,1}}} = \frac{2^{4-\frac{4i+2}{\beta^2}}\Gamma\left(\frac{1}{2}-\frac{i}{2\beta^2}\right)\Gamma\left(\frac{3}{2}-\frac{i+1}{2\beta^2}\right)\Gamma\left(\frac{i}{2\beta^2}\right)\Gamma\left(\frac{i+1}{2\beta^2}\right)}{\Gamma\left(1-\frac{i}{2\beta^2}\right)\Gamma\left(\frac{i}{2\beta^2}+\frac{1}{2}\right)\Gamma\left(1-\frac{i+1}{2\beta^2}\right)\Gamma\left(\frac{i+1}{2\beta^2}-\frac{1}{2}\right)}, \quad i = 1, 2, \ldots . \tag{3.2}$$

We have taken the identity operator $\Phi_{1,1}$ to be unit normalized, consistent with the three-point connectivity [21]. Denoting the properly normalized Kac operators as $\hat{\Phi}_{r,1}$, to guarantee $C^2 \geq 0$ and finite for any $c$ (or $\beta$), we can extract from the expression (3.1) the normalization (up to a positive function of $c$):

$$B_{\hat{\Phi}_{r,1}} \sim \prod_{i=1}^{r-1} \left(\beta^2 - \beta^2_{\text{poles},i}\right)\left(\beta^2 - \beta^2_{\text{zeros},i}\right) \tag{3.3}$$

where $\beta$ are the poles and zeros of the amplitudes (3.1). See [15] for details. In particular, for the energy operator $\hat{\Phi}_{2,1}$ and the second energy operator $\hat{\Phi}_{3,1}$, we find

$$B_{\hat{\Phi}_{2,1}} \sim \left(\beta^2 - \frac{2}{3}\right), \tag{3.4a}$$

$$B_{\hat{\Phi}_{3,1}} \sim \left(\beta^2 - \frac{2}{3}\right)^2\left(\beta^2 - \frac{3}{5}\right)\left(\beta^2 - \frac{3}{4}\right). \tag{3.4b}$$

We see that indeed the norms of the Kac operators vanish at $c = 0$ ($\beta^2 = \frac{2}{3}$) and in particular, the norm of the second energy operator has a double zero. The norms (3.4) and the resulting non-negative finite $C^2$ are plotted in figs. 1 and 2. Similar analysis can be done for higher Kac operators with possibly even higher order zeros in the norm. Interestingly, as observed in [15], this agrees with the operator normalization in the $c < 1$ Liouville CFT. As we will see below, the higher order ($\geq 2$) zero in the norm results in higher logarithmic structures in the $c = 0$ percolation CFT and this has interesting consequence on the correlation functions.

# 4 Logarithmic mixing in percolation

At $c = 0$, the Kac operators correspond to zero-norm states which are at risk of being removed from the CFT state space. In this case, they necessarily acquire logarithmic partner(s) corresponding to states overlapping with the zero-norm states and this leads to logarithmic mixing in the percolation CFT.

## 4.1 Rank-2 Jordan block of energy operator $\varepsilon$

Consider first the energy operator $\hat{\Phi}_{2,1}$. At $c = 0$, its dimension coincides with the 2-hull operator $\hat{\Phi}_{0,2}$ that has a physical interpretation of joining two percolation clusters:

$$(h_{2,1}, h_{2,1}) = (h_{0,2}, h_{0,2}) = \left(\frac{5}{8}, \frac{5}{8}\right). \tag{4.1}$$

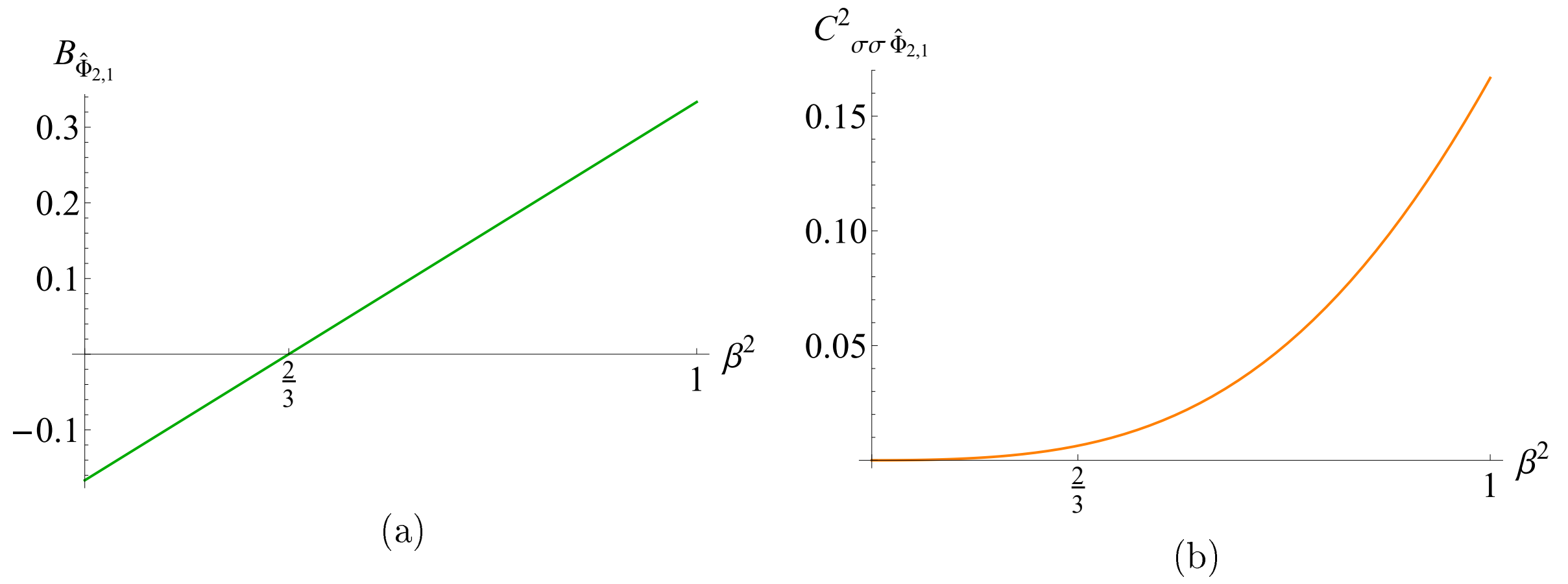

Figure 1: The normalization (3.4) and the square of the three-point constant $C_{\sigma\sigma\hat{\Phi}_{2,1}}$ for the energy operator $\hat{\Phi}_{2,1}$.

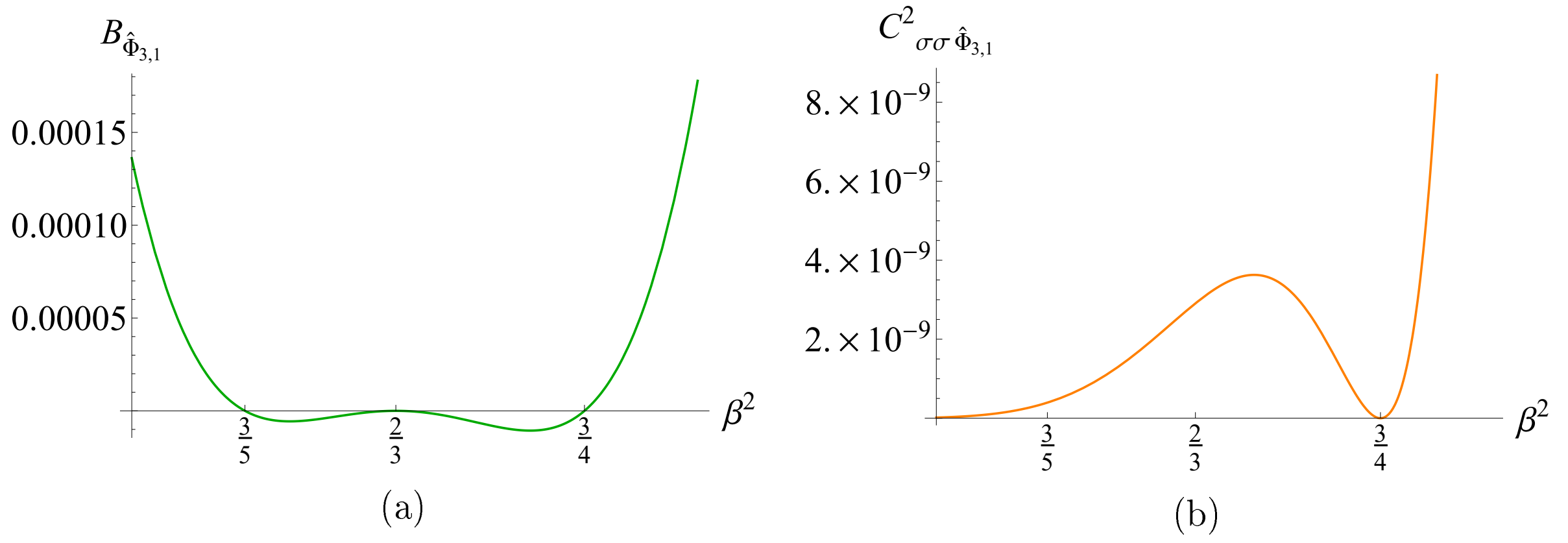

Figure 2: The normalization (3.4) and the square of the three-point constant $C_{\sigma\sigma\hat{\Phi}_{3,1}}$ for the second energy operator $\hat{\Phi}_{3,1}$.

Taking the two-point function of the energy operator with a zero at $c = 0$

$$\langle \hat{\Phi}_{2,1}(z,\bar{z})\hat{\Phi}_{2,1}(0,0)\rangle \overset{c\to 0}{\simeq} \frac{c/2}{(z\bar{z})^{2h_{2,1}(c)}} , \tag{4.2}$$

the 2-hull operator can be normalized as [15]

$$\langle \hat{\Phi}_{0,2}(z,\bar{z})\hat{\Phi}_{0,2}(0,0)\rangle \overset{c\to 0}{\simeq} -\frac{c/2}{(z\bar{z})^{2h_{0,2}(c)}} . \tag{4.3}$$

We can now define a logarithmic partner of the energy operator as

$$\tilde{\varepsilon} = \gamma\left(\frac{\hat{\Phi}_{2,1}}{B_{\hat{\Phi}_{2,1}}(c)} + \frac{\hat{\Phi}_{0,2}}{B_{\hat{\Phi}_{0,2}}(c)}\right) \tag{4.4}$$

which becomes the "top field" of a rank-2 Jordan block and the "bottom field" is defined symmetrically as

$$\varepsilon \equiv \hat{\Phi}_{2,1} \text{ or } \hat{\Phi}_{0,2} . \tag{4.5}$$

Their non-vanishing overlap $\langle\varepsilon|\tilde{\varepsilon}\rangle$ is characterized by the logarithmic coupling $\gamma$ that appears in the two-point function at $c = 0$:

$$\langle \tilde{\varepsilon}(z,\bar{z})\varepsilon(0,0)\rangle = \frac{\gamma}{(z\bar{z})^{2h_\varepsilon}} . \tag{4.6}$$

The definition (4.5) means that the energy operator and the 2-hull operator essentially become identical at $c = 0$. The value of $\gamma$ can be computed from the two-point function of $\tilde{\varepsilon}$ by taking the $c \to 0$ limit:

$$\langle \tilde{\varepsilon}(z,\bar{z})\tilde{\varepsilon}(0,0)\rangle = \frac{-2\gamma^2 \frac{h'_{2,1}-h'_{0,2}}{B'_{\hat{\Phi}_{2,1}}}\ln(z\bar{z}) + const.}{(z\bar{z})^{2h_\varepsilon}} = \frac{-2\gamma\ln(z\bar{z}) + const.}{(z\bar{z})^{2h_\varepsilon}} \, , \tag{4.7}$$

where $'$ denotes the derivative with respect to $c$ evaluated at $c = 0$ and the second equality in (4.7) is required by conformal Ward identities from (4.6). This gives

$$\gamma = \frac{B'_{\hat{\Phi}_{2,1}}}{h'_{2,1} - h'_{0,2}} = \frac{1}{2(h'_{2,1} - h'_{0,2})} = -\frac{5}{4} \, . \tag{4.8a}$$

The value is checked [15] to agree with the previous results analyzed using the $S_Q$ symmetry [22].

## 4.2 Rank-3 Jordan block of second energy operator $\varepsilon'$

The case of second energy operator $\hat{\Phi}_{3,1}$ is more complicated. As we have seen in (3.4), the norm of $\hat{\Phi}_{3,1}$ acquires a second order zero at $c = 0$ and it was found recently [14] that at $c = 0$, it mixes into a rank-3 Jordan block associated with $T\bar{T}$. Intuitively, the higher order zero in the norm means that it is more difficult to build a non-vanishing correlation function and not decouple the second energy operator from the percolation CFT. This is achieved by mixing into a rank-3 Jordan block involving more intricate cancellations and logarithmic structures. To see this, it is instructive to analyze this mixing in two steps, following [15].

At $c = 0$, the dimension of $\hat{\Phi}_{3,1}$ coincides with that of $T\bar{T}$, and a generic $c$ rank-2 Jordan block $(\hat{\Psi}, \bar{A}\hat{X})$ (with dimensions $(h_{1,-2}, h_{1,-2})$) whose logarithmic coupling vanishes at $c = 0$:

$$\langle \hat{\Psi}(z,\bar{z})\bar{A}\hat{X}(0,0)\rangle \overset{c\to 0}{\simeq} \frac{-b_{1,2}c/2}{(z\bar{z})^4} \, , \; b_{1,2} = \frac{5}{6} \, . \tag{4.9}$$

We can consider first a mixing between $\hat{\Phi}_{3,1}$ and $T\bar{T}$ to try to obtain some non-trivial correlation for the second energy operator. The generic $c$ two-point function of the second energy operator has a double zero at $c = 0$

$$\langle \hat{\Phi}_{3,1}(z,\bar{z})\hat{\Phi}_{3,1}(0,0)\rangle \overset{c\to 0}{\simeq} \frac{-c^2/4}{(z\bar{z})^4}, \tag{4.10}$$

where we have chosen a constant factor according to the normalization $c^2/4$ of $T\bar{T}$ fixed by the Virasoro symmetry. We can then try to construct a rank-2 Jordan block like before by defining

$$\Theta \equiv \frac{\theta_0}{c^2/4}\big(T\bar{T} - \hat{\Phi}_{3,1}\big), \tag{4.11}$$

with a logarithmic coupling $\theta_0$

$$\langle \Theta(z,\bar{z})\hat{\Phi}_{3,1}(0,0)\rangle \overset{c\to 0}{\simeq} \frac{\theta_0}{(z\bar{z})^4} \, . \tag{4.12}$$

The two-point function of $\Theta$ is

$$\langle \Theta(z,\bar{z})\Theta(0,0)\rangle \simeq \frac{-\frac{8\alpha^2 h'_{3,1}}{c}\ln(z\bar{z}) + \theta_1}{(z\bar{z})^4} = \frac{-2\theta_0\ln(z\bar{z}) + \theta_1}{(z\bar{z})^4} \tag{4.13}$$

so the conformal Ward identities give

$$\theta_0 = -\frac{c}{4h'_{3,1}} \sim O(c) \tag{4.14}$$

and $\theta_1 \sim O(c)$. We see that the logarithmic mixing with $T\bar{T}$ gives an order $c$ correlation for the second energy operator that still vanishes at $c = 0$. This is however not the end of the story as the two rank-2 Jordan blocks $(\Theta, \hat{\Phi}_{3,1})$ and $(\hat{\Psi}, \bar{A}\hat{X})$ can further mix into a rank-3 Jordan block. In particular, note the logarithmic couplings of the two rank-2 Jordan blocks satisfy

$$-\frac{b_{1,2}c}{2} = \frac{c}{4h'_{3,1}} \tag{4.15}$$

which will allow cancellation of singularities. Define the top field as

$$\Psi_2 \equiv \frac{a}{b_{1,2}c/2}(\Theta - \hat{\Psi}) = \frac{a}{c^2/4}(T\bar{T} - \hat{\Phi}_{3,1}) - \frac{a}{b_{1,2}c/2}\hat{\Psi} \tag{4.16}$$

where $a$ is the logarithmic coupling of a rank-3 Jordan block [14] and the bottom field is identified as equivalently

$$\Psi_0 = \hat{\Phi}_{3,1}, \text{ or } T\bar{T}, \text{ or } \bar{A}\hat{X}. \tag{4.17}$$

We can compute the two-point function of $\Psi_2$ by taking the $c \to 0$ limit and obtain

$$\langle \Psi_2(z, \bar{z})\Psi_2(0, 0)\rangle = \frac{-\frac{2a^2(b-2b_{1,2})}{bb_{1,2}^2}\ln^2(z\bar{z}) - 2a_1 \ln(z\bar{z}) + a_2}{(z\bar{z})^4}, \tag{4.18}$$

where

$$a = \frac{1}{4(2h'_{1,2} - h'_{3,1})h'_{3,1}} = -\frac{25}{48}. \tag{4.19}$$

The "middle field" can be obtained by acting $L_0 - 2$ or $\bar{L}_0 - 2$ on $\Psi_2$

$$\Psi_1 = \frac{2a}{b_{1,2}c}\hat{\Phi}_{3,1} - 2ah''_{3,1}\hat{\Phi}_{3,1} - \frac{2a}{b_{1,2}c}\bar{A}\hat{X} + \frac{a}{bb_{1,2}}\hat{\Psi} \tag{4.20}$$

and the two-point functions of the rank-3 Jordan block $(\Psi_2, \Psi_1, \Psi_0)$ take the standard form. See [15] for details. The second energy operator $\varepsilon' \sim \Psi_0 \equiv \hat{\Phi}_{3,1}$ that started with a double zero in its two-point function now has a non-trivial correlation with the top field:

$$\langle \Psi_0(z, \bar{z})\Psi_2(0, 0)\rangle = \frac{a}{(z\bar{z})^4} . \tag{4.21}$$

# 5 Three-point functions and OPE

The appearance of a rank-3 Jordan block has interesting consequences for higher point correlation functions in the percolation CFT. From the definition of the operators using the generic $c$ fields, we can compute the three-point functions involving two energy operator $\varepsilon$ by taking the limit $c \to 0$ from generic $c$ expressions. Here we are using the fact that $\hat{\Phi}_{2,1}$ is a Kac operator at generic $c$ so we know the generic $c$ expressions of the three-point functions. One arrives at [15]:

$$\langle \varepsilon(z_1, \bar{z}_1)\varepsilon(z_2, \bar{z}_2)\Psi_2(z_3, \bar{z}_3)\rangle = \frac{C_{\varepsilon\varepsilon\Psi_2} + C_{\varepsilon\varepsilon\Psi_1}\tau_3}{(z_{12}\bar{z}_{12})^{2h_\varepsilon - 2}(z_{13}\bar{z}_{13})^2(z_{23}\bar{z}_{23})^2} , \tag{5.1a}$$

$$\langle \varepsilon(z_1, \bar{z}_1)\varepsilon(z_2, \bar{z}_2)\Psi_1(z_3, \bar{z}_3)\rangle = \frac{C_{\varepsilon\varepsilon\Psi_1}}{(z_{12}\bar{z}_{12})^{2h_\varepsilon - 2}(z_{13}\bar{z}_{13})^2(z_{23}\bar{z}_{23})^2} , \tag{5.1b}$$

$$\langle \varepsilon(z_1, \bar{z}_1)\varepsilon(z_2, \bar{z}_2)\Psi_0(z_3, \bar{z}_3)\rangle = 0 . \tag{5.1c}$$

The last equation tells us that the three-point correlation of the energy operator $\varepsilon$ and the second energy operator $\varepsilon' = \Psi_0$ vanish at $c = 0$. The vanishing coupling makes physical sense: the energy-type operators describe bond occupations in the cluster model and at $c = 0$ in percolation, the bonds are random and have no long-range correlations. One can similarly see that for higher energy-type operators (i.e. properly normalized higher Kac operators), three-point constants also vanish [15]. However, the energy operator $\varepsilon$ couples non-trivially to the logarithmic partners $\Psi_2, \Psi_1$ of $\varepsilon'$, captured by the three-point constant $C_{\varepsilon\varepsilon\Psi_1}$. These operators arise as combinations of generic $c$ fields that give rise to 2-point logarithmic correlation functions at $c = 0$ consistent with conformal Ward identities. Although their physical interpretations are unclear, it is certainly fascinating that such combinations acquire non-vanishing three-point correlations with the energy operator which is supposed to describe random bonds.

It is worth stressing that the non-vanishing three-point coupling $C_{\varepsilon\varepsilon\Psi_1}$ is a result of the existence of higher (rank-3 in this case) logarithmic structure in the percolation CFT. As a comparison, we can calculate the three-point function of $\varepsilon$ with the intermediate-step rank-2 Jordan block $(\Theta, \hat{\Phi}_{3,1})$ from above and find that [15]

$$C_{\varepsilon\varepsilon\hat{\Phi}_{3,1}} = O(c), \quad C_{\varepsilon\varepsilon\Theta} = O(c) \,, \tag{5.2}$$

so they vanish at $c = 0$.

Using (5.1), and taking into account also the logarithmic pair $(t, T)$, we can write down the OPE of the energy operator $\varepsilon$ [15]:

$$\begin{aligned}\varepsilon(z,\bar{z})\varepsilon(0,0) =&(z\bar{z})^{-2h_\varepsilon}\Bigg\{ z^2\frac{C_{\varepsilon\varepsilon t}}{b}T + c.c. + \dots \\ &+ (z\bar{z})^2\left(\frac{C_{\varepsilon\varepsilon\Psi_1}}{a}\Psi_1 + \frac{C_{\varepsilon\varepsilon\Psi_1}}{a}\ln(z\bar{z})\Psi_0 + \frac{aC_{\varepsilon\varepsilon\Psi_2} - a_1 C_{\varepsilon\varepsilon\Psi_1}}{a^2}\Psi_0\right) + \dots \Bigg\} \,,\end{aligned} \tag{5.3}$$

with the percolation conformal data:

$$b = -5, \quad a = -\frac{25}{48}, \quad C_{\varepsilon\varepsilon t} = bh_\varepsilon = -\frac{25}{8}, \quad C_{\varepsilon\varepsilon\Psi_1} = \frac{ah_\varepsilon^2}{b_{1,2}} = -\frac{125}{512} \,. \tag{5.4}$$

# 6 Four-point function of $\varepsilon$

Consider now what this means for the four-point function. With the basic conformal data, we can construct the $s-$channel expansion of the four-point function of $\varepsilon$ using cluster decomposition. See appendix D of [15] for details. Interestingly, the non-vanishing coupling $C_{\varepsilon\varepsilon\Psi_1}$ gives rise to a non-vanishing leading term for the four-point function of $\varepsilon$ [15]:

$$\langle\varepsilon(\infty)\varepsilon(1)\varepsilon(z,\bar{z})\varepsilon(0)\rangle =(z\bar{z})^{-2h_\varepsilon}\left\{\frac{\left(C_{\varepsilon\varepsilon\Psi_1}\right)^2}{a}(z\bar{z})^2 + \dots\right\} \,. \tag{6.1}$$

This is a very curious result from the BPZ equation point of view. Since we have defined the energy operator – the bottom field of the rank-2 Jordan block as $\varepsilon \sim \hat{\Phi}_{2,1}$, one might expect that its four-point function at $c = 0$ could be obtained by taking the $c \to 0$ limit of the BPZ solution. Consider the properly normalized Kac operator $\hat{\Phi}_{2,1}$, the BPZ solution gives the following $s$-channel expansion [17]

$$\begin{aligned}\langle\hat{\Phi}_{2,1}\hat{\Phi}_{2,1}\hat{\Phi}_{2,1}\hat{\Phi}_{2,1}\rangle &= (z\bar{z})^{-2h_{2,1}(c)}\Bigg\{B^2_{\hat{\Phi}_{2,1}}(c)+\frac{B^2_{\hat{\Phi}_{2,1}}(c)h^2_{2,1}(c)}{c/2}(z^2+\bar{z}^2)+\frac{C^2_{\hat{\Phi}_{2,1}\hat{\Phi}_{2,1}T\bar{T}}(c)}{B_{T\bar{T}}(c)}(z\bar{z})^2\\ &\qquad+\frac{C^2_{\hat{\Phi}_{2,1}\hat{\Phi}_{2,1}\hat{\Phi}_{3,1}}(c)}{B_{\hat{\Phi}_{3,1}}(c)}(z\bar{z})^{h_{3,1}(c)}+\dots\Bigg\}\\ &\overset{c\to 0}{=}(z\bar{z})^{-2h_\varepsilon+2}\left(\frac{h^4_\varepsilon}{2b_{1,2}}\ln(z\bar{z})+const.\right)c+\dots\,.\end{aligned} \tag{6.2}$$

which vanishes as $c\to 0$. This vanishing four-point function at $c=0$ can also be understood from the cluster decomposition perspective. If we take the intermediate-step rank-2 Jordan block $(\Theta,\hat{\Phi}_{3,1})$ discussed above, and use their conformal data given in (5.2), we can write the expansion of the four-point function

$$\begin{aligned}&\langle\varepsilon(\infty)\varepsilon(1)\varepsilon(z,\bar{z})\varepsilon(0)\rangle\\ &\overset{?}{=}(z\bar{z})^{-2h_\varepsilon+2}\left(\frac{C_{\varepsilon\varepsilon T\bar{T}}(2\theta_0C_{\varepsilon\varepsilon\Theta}-C_{\varepsilon\varepsilon T\bar{T}}\theta_1)}{\theta_0^2}+\frac{C^2_{\varepsilon\varepsilon T\bar{T}}}{\theta_0}\ln(z\bar{z})\right)+\dots\end{aligned} \tag{6.3}$$

where the ? above the equal sign indicates that we have only taken into account the cluster decomposition involving the intermediate-step $(\Theta,\hat{\Phi}_{3,1})$ but not the full rank-3 Jordan block. The expression (6.3) agrees precisely with the limit of the BPZ solution (6.2), but contradicts the non-vanishing four-point function (6.1). Indeed, the BPZ solution contains only the conformal blocks of the identity operator (and thus $T\bar{T}$) and the second energy operator $\hat{\Phi}_{3,1}$, so its $c\to 0$ limit can only contain the information of the intermediate-step rank-2 Jordan block $(\Theta,\hat{\Phi}_{3,1})$ that results from mixing $\hat{\Phi}_{3,1}$ and $T\bar{T}$. The non-vanishing four-point function (6.1), however, gets contribution from the full rank-3 Jordan block that involves the other pair $(\hat{\Psi},\bar{A}\hat{X})$ which the Kac operator $\hat{\Phi}_{2,1}$ at generic $c$ has no knowledge of. On the other hand, the operator algebra of 2-hull operator $\hat{\Phi}_{0,2}$ at generic $c$ certainly contains the information of the $(\hat{\Psi},\bar{A}\hat{X})$ [15]. At $c=0$, the energy operator $\hat{\Phi}_{2,1}$ and the 2-hull operator $\hat{\Phi}_{0,2}$ become identical (4.4) and curiously, this identification allows a non-trivial couplings of the energy operator $\varepsilon$ to the rank-3 Jordan block (in particular the field $\Psi_1$) which builds a non-vanishing four-energy correlator (6.1). For future work, it would be important to understand precisely what happens to the Virasoro degeneracy of the $\hat{\Phi}_{2,1}$ upon this identification and what kind of analytic tools this could provide for the $c=0$ CFTs. Let us mention that the possibility of zero-norm operators having non-trivial higher point correlations have been previously discussed in [23].

# 7 Conclusions

In this contribution, following [15], we described the logarithmic mixing of the energy and second energy operators in the 2d percolation bulk CFT. We start by analyzing the Kac operator normalization at generic $c$ using that the Potts cluster CFTs are real. The Kac operators generically acquire zeros in their norms at $c=0$ and the energy and second energy operators mix into rank-2 and rank-3 Jordan blocks respectively at $c=0$. We compute some of the three-point constants in the percolation CFT and interestingly, although the energy-type operators (bottom fields of the Jordan blocks) themselves have vanishing three-point constants, there is a non-trivial coupling between the energy operator $\varepsilon$ and the logarithmic partners of the second energy operator $\varepsilon'$. Surprisingly, this coupling gives a non-trivial contribution to the four-point correlation function of the energy operator according to cluster decomposition. Similar results can be obtained in another $c = 0$ bulk CFT that describes self-avoiding polymer [15].

This work leaves interesting puzzles for future investigations. First, it would be important to understand the physical significance of the energy-type operators at $c = 0$ and test our results using alternative methods, for example on the lattice [24] or using the probabilistic approach [25, 26]. Naively, the energy-type operators describe bond occupations and in percolation, they become completely random and uncorrelated. However, the logarithmic mixing allows to identify them with the hull-type operators and it appears that it is this identification that gives them non-trivial operator algebra and correlation functions through the intricate logarithmic structure of the $c = 0$ bulk CFT. On the other hand, from the algebraic perspective, it would also be important to understand the fate of the BPZ equation at $c = 0$ and what kind of analytic handle we might gain to study the percolation and other $c = 0$ bulk CFTs.

## Acknowledgements

I would like to thank the organizers and participants of the 28th Rencontre Itzykson “Analytic results in Conformal Field Theory” for valuable feedback, and J. L. Jacobsen and H. Saleur for insightful discussions.

# Backbone exponent and annulus crossing probability for planar percolation

**Pierre Nolin**[1] **Wei Qian**[2] **Xin Sun**[3] **Zijie Zhuang**[4]

[1] *City University of Hong Kong*
*E-mail:* bpmnolin@cityu.edu.hk

[2] *City University of Hong Kong (on leave from CNRS, LMO, France)*
*E-mail:* weiqian@cityu.edu.hk

[3] *Peking University*
*E-mail:* xinsun@bicmr.pku.edu.cn

[4] *University of Pennsylvania*
*E-mail:* zijie123@wharton.upenn.edu

Abstract: We report the recent derivation of the backbone exponent for 2D percolation. In contrast to previously known exactly computed percolation exponents, the backbone exponent is a transcendental number, which is a root of an elementary equation. We also report an exact formula for the probability that there are two disjoint paths of the same color crossing an annulus. The backbone exponent captures the leading asymptotic, while the other roots of the elementary equation capture the asymptotic of the remaining terms. This suggests that the backbone exponent is part of a conformal field theory (CFT) whose bulk spectrum contains this set of roots. Our approach is based on the coupling between Schramm-Loewner evolution curves and Liouville quantum gravity (LQG) and the integrability of Liouville CFT that governs the LQG surfaces.

## Contents

# 1 Introduction

As a fundamental model of critical phenomena, percolation [BH57] has broad applications in physics and other natural sciences [Sab15]. The fractal geometry of critical percolation clusters is a classical topic in statistical physics. For two-dimensional Bernoulli percolation, the exact values of many fractal dimensions are explicitly known. These values were often first discovered in physics and later proved using probabilistic methods. For instance, the dimensions of the percolation cluster, cluster boundary, and pivotal points (also known as "red bonds") are $\frac{91}{48}$, $\frac{7}{4}$, and $\frac{3}{4}$, respectively. Despite these successful examples, the fractal dimension of the backbone of a percolation cluster remained elusive for a long time. The backbone is the part of a percolation cluster that remains after removing the dangling ends, where electrical current between distant vertices would flow. In this note, we report recent progress on the exact derivation of the backbone exponent and the related annulus crossing probability.

The backbone dimension $D_B$ can be characterized as follows. Consider critical Bernoulli percolation on a triangular lattice of small mesh size, where each site is colored black or white with equal probability. Let $A(r, R)$ be the annulus of inner radius $r$ and outer radius $R$. As the mesh size tends to 0, the probability that there are two disjoint black paths crossing an annulus $A(r, R)$ converges to a limiting probability $p_B(r, R)$. There exists an exponent $x_B$ such that $p_B(r, R)$ decays as $(\frac{r}{R})^{x_B+o(1)}$ as $r/R \to 0$ [BN11]. Then we have $D_B = 2-x_B$. The exponent $x_B$ is known as the backbone exponent or the monochromatic two-arm exponent.

In general, arm exponents of percolation describe the asymptotic behavior of the annulus crossing probability. The monochromatic $k$-arm exponent corresponds to $k$ disjoint black paths. For $k \geq 2$, the polychromatic $k$-arm exponents correspond to $k$ disjoint paths, not all of which are of the same color. The one-arm exponent and the polychromatic $k$-arm exponents are explicitly known to be $\frac{5}{48}$ and $\frac{k^2-1}{12}$, respectively [dN79, SD87, ADA99, SW01, LSW02b]. These, in particular, give the dimensions $\frac{91}{48}$, $\frac{7}{4}$, and $\frac{3}{4}$ mentioned above.

In our recent work [NQSZ23], the exact value of $x_B$ is shown to be the unique solution in $(\frac{1}{4}, \frac{2}{3})$ to

$$\frac{\sqrt{36x+3}}{4} + \sin\left(\frac{2\pi\sqrt{12x+1}}{3}\right) = 0. \tag{1.1}$$

The numerical value is 0.3566668..., which matches well with numerical simulations in [Gra99, DBN04, FKZD22]. Using this expression, we show that $x_B$ is a transcendental number, which is surprising since all other known exponents in Bernoulli percolation are rational.

Our derivation is based on the convergence of 2D percolation toward Schramm-Loewner evolution (SLE) [Sch00] with parameter 6. We focus on site percolation on the triangular lattice because the convergence to $\mathrm{SLE}_6$ is established only for this lattice [Smi01]. This convergence is believed to hold for a broad class of 2D Bernoulli percolation, such as bond percolation on the square lattice. Another key ingredient in our derivation is the coupling between SLE and Liouville quantum gravity (LQG), which describes the scaling limit of statistical physics models on random triangulations. The quantum gravity method for deriving fractal dimension has, for example, been applied to the dimension 4/3 of the Brownian frontier [Dup98] which was later proved using SLE in [LSW01, LSW02a]. This idea, known as the Knizhnik-Polyakov-Zamolodchikov relation [KPZ88], was put into a mathematical framework in [DS11].

Compared to [Dup98], the main novelty of our approach is to incorporate the integrability of Liouville conformal field theory, the field theory that governs the random surfaces in LQG [Pol81], into the study of SLE. This approach was developed in [AHS24] and has been successfully used in several problems. In fact, we use this method to derive in [SXZ24] that $p_B(r, R)$ exactly equals

$$\frac{q^{-\frac{1}{12}}}{\prod_{n=1}^{\infty}(1-q^{2n})} \sum_{s\in\mathcal{S}} \frac{-\sqrt{3}\sin(\frac{2\pi}{3}\sqrt{3s})\sin(\pi\sqrt{3s})}{\cos(\frac{4\pi}{3}\sqrt{3s}) + \frac{3\sqrt{3}}{8\pi}} q^s, \tag{1.2}$$

where $\mathcal{S} = \{s \in \mathbb{C} : \sin(4\pi\sqrt{\frac{s}{3}}) + \frac{3}{2}\sqrt{s} = 0\} \setminus \{0, \frac{1}{3}\}$ and $q = r/R$. The numerical values for $\mathcal{S}$ are $\{0.440,\ 2.194 \pm 0.601i,\ 5.522 \pm 1.269i, \ldots\}$. This equation is related to (1.1) by $s = x + \frac{1}{12}$. Hence (1.2) provides a physical interpretation of roots in (1.1).

The analogous annulus crossing probabilities for the one-arm exponent and the alternating $2k$-arm exponents were derived by Cardy [Car02, Car06] and can be expressed in a similar form as (1.2), except that all the exponents and coefficients in the expansion are rational. Our derivation of (1.2) is based on a method developed in [ARS22]. The same method also recovers Cardy's result for the one-arm case and the alternating two-arm case. In this note, we present the key ideas in deriving $x_B$ and $p_B(r, R)$.

## 2 Conformal radii of SLE and percolation exponents

The first step in our derivation of (1.1) is to encode the backbone exponent using the conformal radius of a domain bounded by an $\mathrm{SLE}_6$ curve. This type of encoding can be done for various percolation exponents. We first demonstrate this using the example of the one-arm exponent $x_1$. On the triangular lattice with small mesh size, consider Bernoulli site percolation in an approximation of the unit disk, as shown in Figure 1, where each site is represented by a hexagon. We set the boundary to be all black, namely, the hexagons not drawn in Figure 1 are black. Then as the mesh size tends to zero, the interfaces separating the black and white clusters converge to a collection of loops, which together form the so-called conformal loop ensemble with parameter $\kappa = 6$ ($\mathrm{CLE}_6$) on the disk [CN06, She09]. Each loop in $\mathrm{CLE}_6$ locally looks like the $\mathrm{SLE}_6$ curve introduced by Schramm [Sch00], which is not self-crossing but has self-intersections.

Let $\eta_0$ be the outermost loop in $\mathrm{CLE}_6$ that surrounds the origin. Let $D$ be the complementary connected component of $\eta_0$ containing the origin, as shown in Figure 1 (right). Let $f$ be a conformal map from the unit disk to $D$ such that $f(0) = 0$. The conformal radius $\mathrm{CR}(0, D)$ of $D$ viewed from the origin $0 \in \mathbb{C}$ is defined as $|f'(0)|$. Let $d(0, \eta_0)$ denote the Euclidean distance between 0 and $\eta_0$, and let $p(r, 1)$ denote the probability that there exists a black crossing of $A(r, 1)$, which is equivalent to the event $d(0, \eta_0) \le r$. Hence, we have

$$\mathbb{P}[d(0, \eta_0) \le r] = p(r, 1) \quad \text{for } 0 < r < 1. \tag{2.1}$$

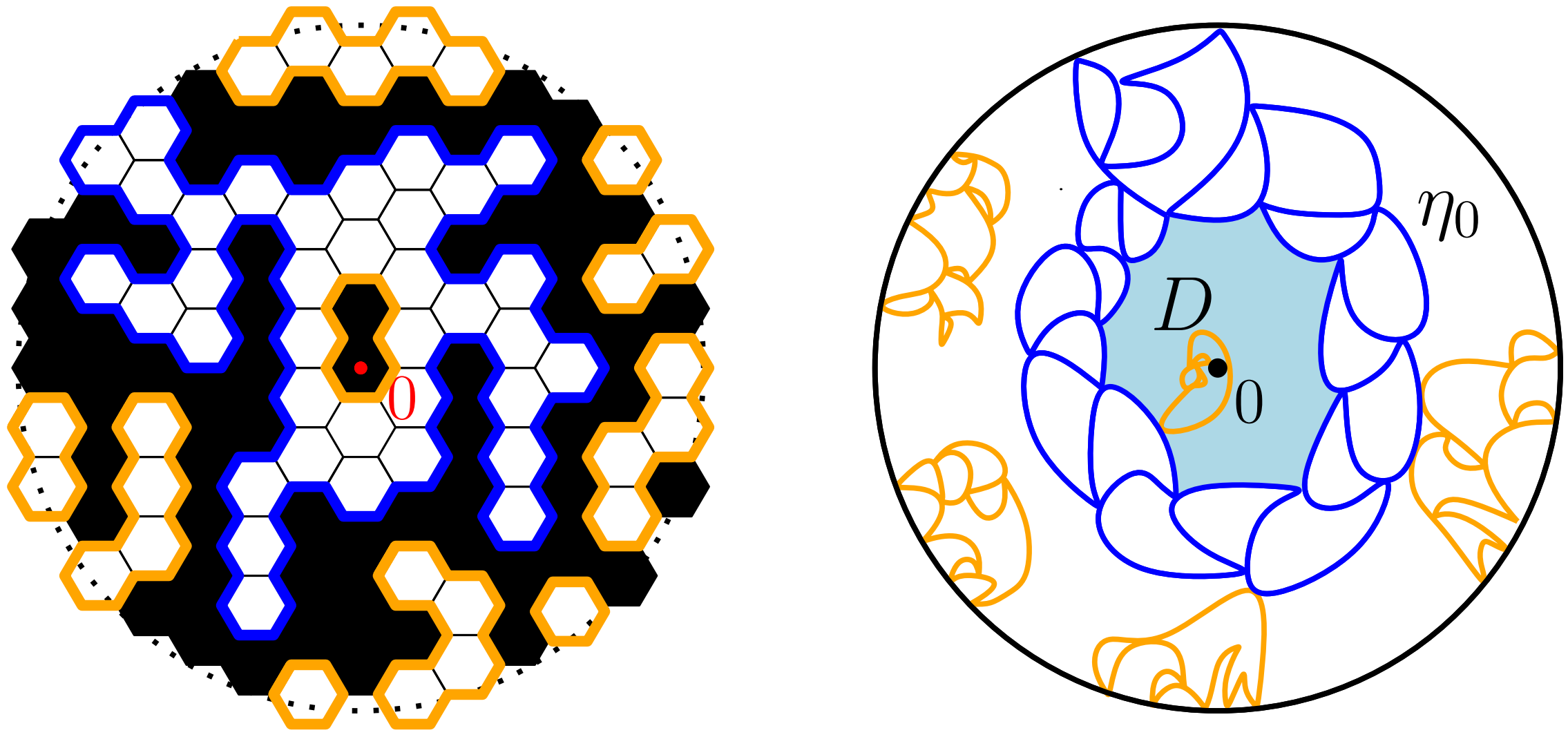

Figure 1: **Left:** Bernoulli percolation on a triangular lattice on the unit disk. The blue loop is the outermost percolation interface surrounding the origin, whose scaling limit is $\eta_0$ on the right. **Right:** $\mathrm{CLE}_6$ on the unit disk. The loops are nested, nonsimple, and may touch each other and the boundary.

The Koebe 1/4 theorem (see, e.g., [Con95]) states

$$d(0, \eta_0) \le \mathrm{CR}(0, D) \le 4d(0, \eta_0). \tag{2.2}$$

Since the one-arm exponent $x_1$ for percolation satisfies $p(r, 1) = r^{x_1 + o(1)}$, it follows from (2.1) that

$$\mathbb{P}[\mathrm{CR}(0, D) \le r] = r^{x_1 + o(1)}. \tag{2.3}$$

Therefore, the one-arm exponent $x_1$ is given by

$$x_1 = \inf\{x \in \mathbb{R} : \langle \mathrm{CR}(0, D)^{-x} \rangle = \infty\}. \tag{2.4}$$

The value $x_1 = 5/48$ was originally established in [LSW02b] by estimating $\mathbb{P}[\mathrm{CR}(0, D) \le r]$. The exact formula $\langle \mathrm{CR}(0, D)^{-x} \rangle = \frac{1}{2\cos(\frac{\pi}{3}\sqrt{12x+1})}$ was further obtained in [SSW09], from which one can also extract $x_1$.

The backbone exponent $x_B$ can be encoded using a similar idea. For each loop in a $\mathrm{CLE}_6$ on the unit disk, its outer boundary is defined to be the boundary of the unbounded component after removing the loop from the plane. Then the outer boundaries of all the $\mathrm{CLE}_6$ loops form a random collection of simple loops, each of which looks like an $\mathrm{SLE}_\kappa$ curve with $\kappa = \frac{8}{3}$ [LSW03]. Let $D_b$ be the domain bounded by the outermost loop surrounding the origin in this loop ensemble. Then

$$x_B = \inf\{x \in \mathbb{R} : \langle \mathrm{CR}(0, D_b)^{-x} \rangle = \infty\}. \tag{2.5}$$

To see (2.5), consider the *filled percolation interfaces*, which are obtained by filling the fjords (i.e., passages with width 1); see the blue loop in Figure 2 (left). If the distance from 0 to the outermost filled percolation interface is less than $r$, then we need to flip at least two black points to disconnect all black crossings in $A(r, 1)$. By Menger's theorem, there exist two disjoint black crossings in $A(r,1)$. In fact, the probabilities of these two events share the same exponent as $r \to 0$; see Appendix A.1. Moreover, these filled percolation interfaces converge to outer boundaries of the $\mathrm{CLE}_6$ loops. This gives (2.5) similar to (2.4).

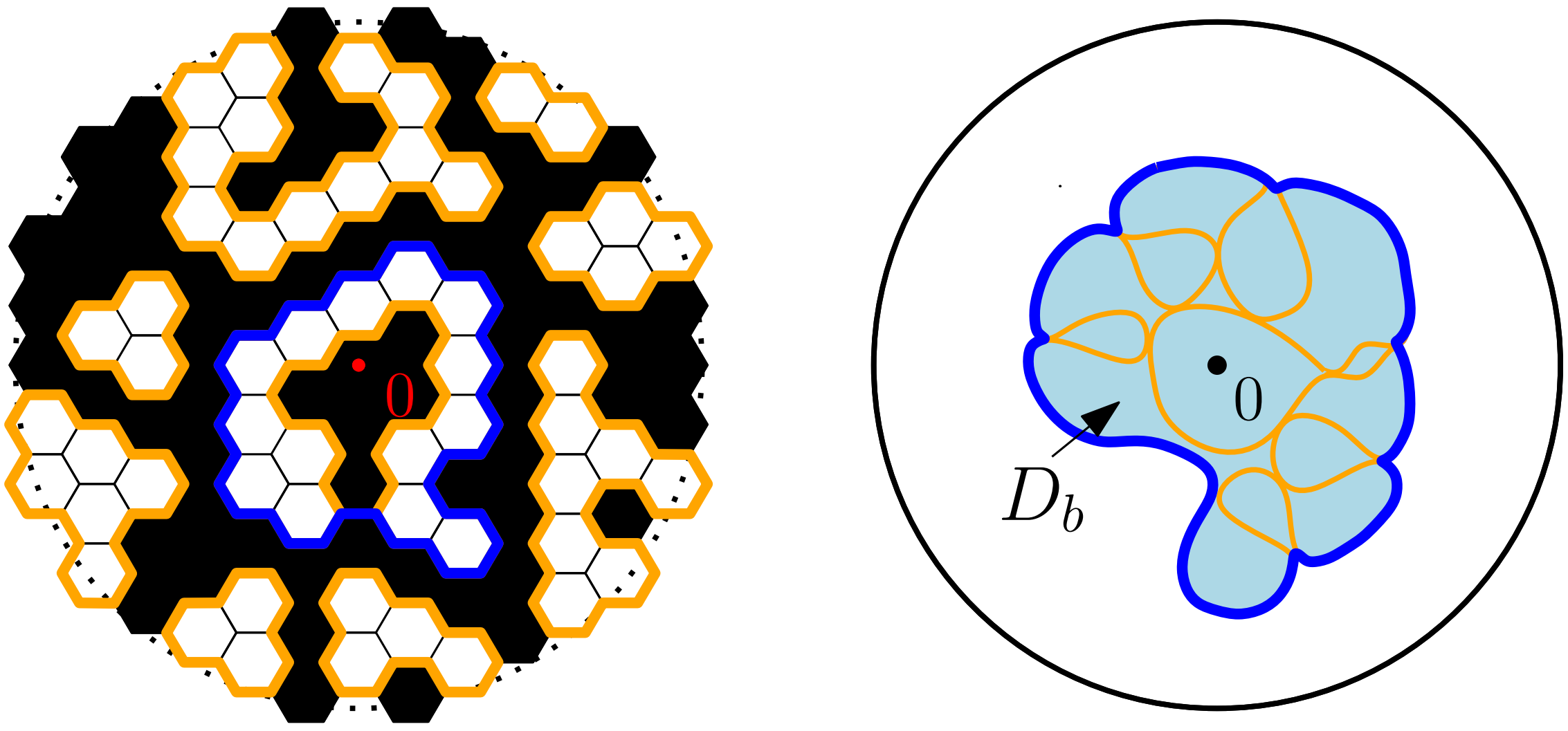

Figure 2: **Left:** percolation interfaces are colored orange. The blue loop is an example of a filled interface, which encloses all black points that can be surrounded by a white cluster together with an additional black point. It is the outermost one surrounding the origin. **Right:** filled percolation interfaces converge to the outer boundaries of $\mathrm{CLE}_6$ loops. The blue loop is the outermost one surrounding 0, and $D_b$ is the domain enclosed by this blue loop.

Now the value of $x_B$ specified by (2.5) follows from the exact formula of $\langle \mathrm{CR}(0, D_b)^{-x} \rangle$,

$$\langle \mathrm{CR}(0, D_b)^{-x} \rangle = \frac{3\sqrt{3}}{4} \times \frac{\sin\left(\pi\sqrt{3s}\right)}{\frac{3}{2}\sqrt{s} + \sin\left(4\pi\sqrt{s/3}\right)} \quad \text{with} \quad s = x + \frac{1}{12}. \tag{2.6}$$

Below we explain how $\langle \mathrm{CR}(0, D_b)^{-x} \rangle$ and $p_B(r, R)$ arise in the 2D quantum gravity framework.

# 3 The quantum gravity approach

We start with an overview of Liouville field theory. The correlation function of Liouville field theory is

$$\int_{\phi:\Sigma\to\mathbb{R}} \prod_{i=1}^{n} e^{\alpha_i \phi(x_i)} e^{-S_L[\phi]} D\phi, \tag{3.1}$$

where $\Sigma$ is a 2D Riemannian manifold, $\alpha_i \in \mathbb{R}$, and $x_i$ are marked points on $\Sigma$. Here, $S_L[\phi]$ is the Liouville action on $\Sigma$. We need the cases where $\Sigma$ is the flat unit disk $\mathbb{U}$ or the cylinder $\mathcal{C}_\tau$ obtained by identifying $[0,\tau] \times \{0\}$ with $[0,\tau] \times \{1\}$ on $[0,\tau] \times [0,1]$. The Liouville action on $\mathbb{U}$ is given by

$$S_L[\phi] = \int_{\mathbb{U}} \left(\frac{1}{4\pi}|\nabla\phi|^2 + \mu e^{\gamma\phi}\right) d^2x + \int_{\partial\mathbb{U}} \left(\frac{Q\phi}{2\pi} + \nu e^{\frac{\gamma}{2}\phi}\right) dl. \tag{3.2}$$

The bulk cosmological constant $\mu$ and the boundary cosmological constant $\nu$ are allowed to vary. For $\mathcal{C}_\tau$, the action $S_L[\phi]$ is given by

$$S_L[\phi] = \int_{\mathcal{C}_\tau} \left(\frac{1}{4\pi}|\nabla\phi|^2 + \mu e^{\gamma\phi}\right) d^2x + \int_{\partial_1\mathcal{C}_\tau} \nu_1 e^{\frac{\gamma}{2}\phi} dl + \int_{\partial_2\mathcal{C}_\tau} \nu_2 e^{\frac{\gamma}{2}\phi} dl \tag{3.3}$$

where $\partial_1 \mathcal{C}_\tau$ and $\partial_2 \mathcal{C}_\tau$ are the two boundaries of $\mathcal{C}_\tau$, and $\mu, \nu_1, \nu_2$ are cosmological constants. The coupling constant $\gamma \in (0,2)$ determines the background charge $Q = \frac{2}{\gamma} + \frac{\gamma}{2}$ and the central charge $c_L = 1 + 6Q^2$. Liouville correlation functions can be explicitly solved under the conformal field theory framework [BPZ84], which is done in physics by [DO94, ZZ96, PT02] and rigorously in math in [KRV20, GKRV24]; see [GKR24] for a review. In our derivation of (2.6) and (1.2), we need the boundary structure constants on the disk solved in [RZ22] and the annulus partition function solved in [Wu22], in the case where the bulk cosmological constant is zero.

Consider Bernoulli site percolation on a random triangulation. We can view the random triangulation as a discrete model for 2D quantum gravity, see e.g. [AW95, BK87, AS03], and percolation as a conformal matter with central charge $c_M = 0$. Assume the random triangulation has the disk topology. Then in the continuum limit, the 2D quantum gravity can be described by the Liouville field theory on the disk with central charge $c_L = 26 - c_M = 26$, and hence $Q = \sqrt{25/6}$ and $\gamma = \sqrt{8/3}$. Consider the field $\phi$ on $\mathbb{U}$ whose distribution is given by $e^{\frac{\gamma}{2}\phi(1)} \times e^{\gamma\phi(0)} \times e^{-S_L[\phi]} D\phi$. The continuum limit of percolation on a random triangulation of the disk can be described by $\mathrm{CLE}_6$ on $\mathbb{U}$ with a random geometric background: the area measure is $e^{\gamma\phi} d^2x$ on $\mathbb{U}$, and the boundary length measure is $e^{\frac{\gamma}{2}\phi} dl$ on $\partial\mathbb{U}$. Here $0 \in \mathbb{U}$ and $1 \in \partial\mathbb{U}$ correspond to one bulk point and one boundary point on the triangulation, which are marked to specify how the random surface is conformally parametrized by $\mathbb{U}$. A version of this is rigorously proved in [HS23].

From now on we set $\mu = 0$ in $S_L[\phi]$ and write it as $S_L^\nu[\phi]$, only allowing $\nu$ to vary.[5] Then,

$$\int_{\phi:\mathbb{U}\to\mathbb{R}} F\Big(\int_{\partial\mathbb{U}} e^{\frac{\gamma}{2}\phi} d\ell\Big) e^{\gamma\phi(0)} e^{-S_L^\nu[\phi]} D\phi \propto \int_0^\infty F(L) e^{-\nu L} L^{-\frac{3}{2}} dL \tag{3.4}$$

for any test function $F$ [HRV18]. (We use $h(F) \propto g(F)$ to indicate that $h(F)/g(F)$ is a universal constant independent of $F$.) On the discrete side, consider a polygon of length $p$. For each triangulation of it with an interior marked point and $n$ faces, we assign weight $a^n b^p$. Then for critical $a$ and $b$, as $p \to \infty$ the total weight grows as $p^{-\frac{3}{2}}$. From both perspectives, we can say that $L^{-\frac{3}{2}}$ is the partition function for the random disk in the pure 2D quantum gravity that has boundary length $L$, one interior marked point, and no area constraint [BM17]; see Appendix A.2 for further background. We denote such a random disk by $\mathrm{QD}_1(L)$. We do not mark any boundary point on $\mathrm{QD}_1(L)$, hence the boundary insertion $e^{\frac{\gamma}{2}\phi(1)}$ does not appear in (3.4).

Suppose $\mathrm{QD}_1(L)$ is conformally realized on $\mathbb{U}$ with the marked point at 0. Consider $\mathrm{CLE}_6$ on $\mathbb{U}$ as in Figure 2 with $D_b$ defined above (2.5), and let $\eta$ be the boundary of $D_b$. The curve $\eta$ cuts $\mathrm{QD}_1(L)$ into two pieces of random surfaces, which are independent once conditioned on the length of $\eta$. This independence in the triangulation setting is clear; see Figure 3. In the continuum, results of this type were pioneered by [She16, DMS21]. A highly nontrivial conclusion from this approach is that the surface inside $\eta$ is another copy of $\mathrm{QD}_1$. In fact, we can write the following equality of partition functions:

$$\int_0^\infty Z(L,\ell) \times \ell \times \ell^{-\frac{3}{2}} d\ell = L^{-\frac{3}{2}}, \tag{3.5}$$

where $Z(L,\ell)$ is the partition function of the random surface bounded by $\partial\mathbb{U}$ and $\eta$ with given boundary lengths, and $\ell^{-\frac{3}{2}}$ is the partition function of $\mathrm{QD}_1(\ell)$ that describes the random surface inside $\eta$. The additional factor $\ell$ is the length of $\eta$ which counts the number of ways in which the two surfaces can be glued together, since we do not mark a point on $\eta$. See Appendix A.2 for more details.

[5] When we set $\mu = 0$, it may be more accurate to view the boundary Liouville theory as a free field theory with conformal boundary conditions. See [?] for further discussion.

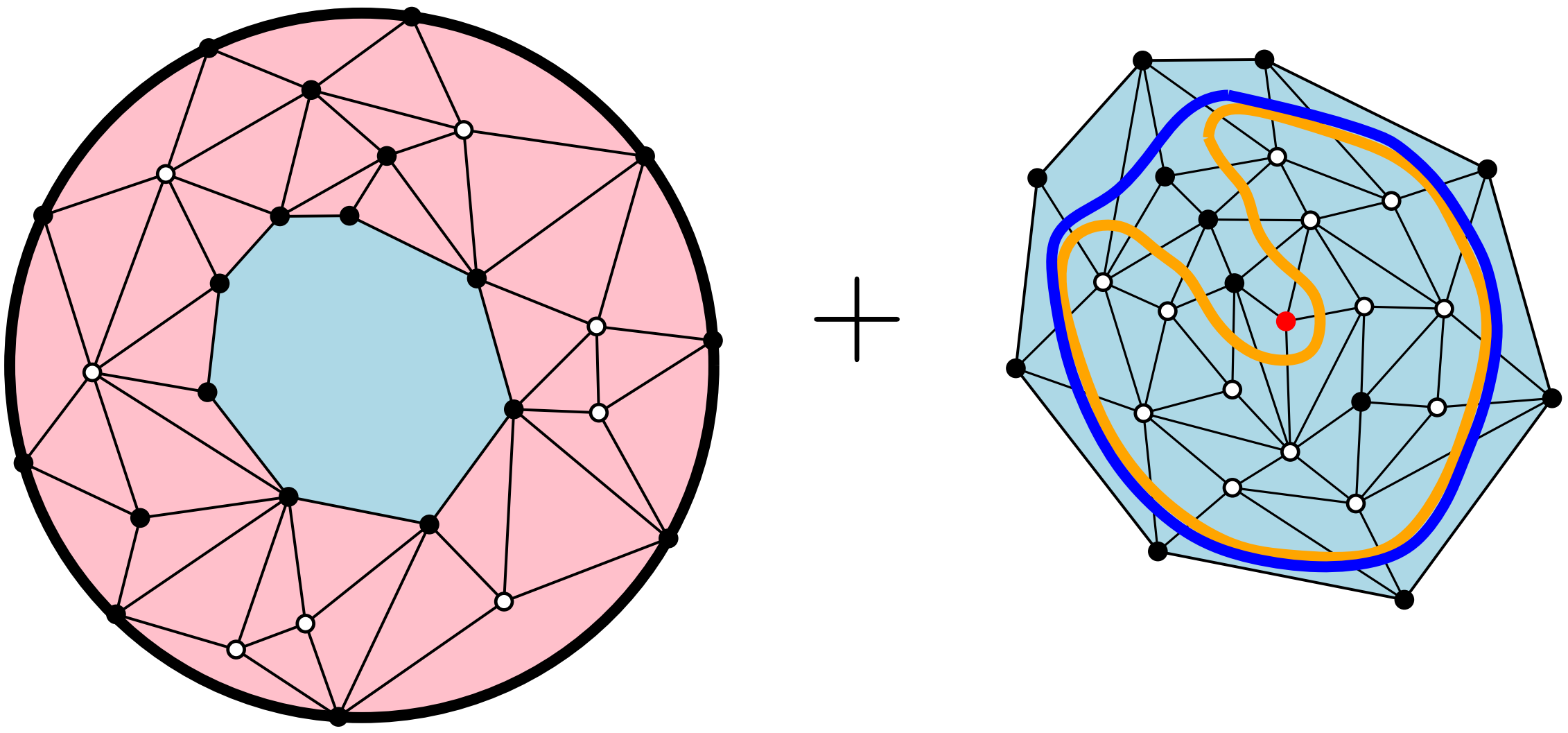

Figure 3: Bernoulli site percolation on a random triangulation of the disk. The outermost filled percolation interface colored in blue divides the random triangulation into two parts. These two parts are independent conditioned on the number of vertices along the filled interface. In the continuum limit, this corresponds to (3.5).

We are now ready to explain the key equation

$$\int_0^\infty Z(L,\ell)\times\ell\times\ell^{-\frac{3}{2}+a}d\ell = L^{-\frac{3}{2}+a}\langle \mathrm{CR}(0,D_b)^{x(a)}\rangle, \tag{3.6}$$

where $x(a) = -\frac{1}{3}a(a-1)$. This is a modification of the surface gluing equation (3.5). The surface bounded by $\partial\mathbb{U}$ and $\eta$ stays the same, while the surface corresponding to $\mathrm{QD}_1(\ell)$ is modified to a new random surface with partition function $\ell^{-\frac{3}{2}+a}$. To describe this new surface, we note that if $e^{\gamma\phi(0)}$ in (3.4) is replaced by $e^{\alpha\phi(0)}$, then (3.4) becomes proportional to $\int_0^\infty F(L)e^{-\nu L}L^{\frac{2}{\gamma}(\alpha-\gamma)-\frac{3}{2}}dL$. This defines a random surface with partition function $L^{\frac{2}{\gamma}(\alpha-\gamma)-\frac{3}{2}}$ just as how $\mathrm{QD}_1(L)$ is defined from (3.4). Taking $\alpha=\gamma+\frac{\gamma}{2}a$ and $\gamma=\sqrt{8/3}$, this defines the random surface in (3.6) with partition function $\ell^{-\frac{3}{2}+a}$.

Once the surface corresponding to $\mathrm{QD}_1(\ell)$ in (3.5) is modified, the surface $\mathrm{QD}_1(L)$ on the right side of (3.5) changes to a surface with partition function $L^{-\frac{3}{2}+a}$ via the same mechanism. The additional factor $\langle \mathrm{CR}(0,D_b)^{x(a)}\rangle$ in (3.6) arises because when modifying the bulk insertion $e^{\gamma\phi(0)}$ to $e^{\alpha\phi(0)}$, we assumed that both the smaller surface $\mathrm{QD}_1(\ell)$ and the larger surface $\mathrm{QD}_1(L)$ are conformally realized on $\mathbb{U}$. Therefore, we first need to apply a conformal map $f$ from $D_b$ to $\mathbb{U}$ while fixing 0. This results in a factor of $|f'(0)|^{x(a)}$ with $x(a)=2\Delta_\alpha-2\Delta_\gamma$, where $\Delta_\alpha=\frac{\alpha}{2}(Q-\frac{\alpha}{2})$ is the scaling dimension of the bulk insertion $e^{\alpha\phi(0)}$ in Liouville theory. By definition, $|f'(0)|^{-1}=\mathrm{CR}(0,D_b)$, which gives the right side of (3.6); see Appendix A.2.

Equations like (3.6) relate the conformal radii of domains bounded by SLE curves with the partition function of random surfaces in 2D quantum gravity with given lengths. As demonstrated in [AHS24], in certain cases, the partition function of the random surfaces can be exactly computed using the structure constants of boundary Liouville theory [RZ22], hence such relations provide the exact formula for the conformal radii. However, since the surface corresponding to $Z(L,\ell)$ has a random conformal modulus, the method in [AHS24] does not readily give a formula for $Z(L,\ell)$. A key step in [NQSZ23] to derive (1.1) is to find an effective variant of $\mathrm{CR}(0,D_b)$ such that it still encodes the backbone exponent; moreover, the analog of $Z(L,\ell)$ is solvable using the method in [AHS24]. The strategy is to consider a chain of adjacent loops connecting the disk boundary to the boundary of $D_b$, which forms a sequence of shrinking domains whose conformal radii can be computed step by step. The precise implementation of this strategy involves a Poisson point process of a special variant of $\mathrm{SLE}_6$ called the $\mathrm{SLE}_6$ bubble [MSW17]. The detail can be found in [NQSZ23, Section 2].

The ratio between $\mathrm{CR}(0, D_b)$ and its effective variant in [NQSZ23] is another conformal radius that can be computed using the method from [AHS24]. This gives (2.6), as done in [SXZ24]. Now $Z(L, \ell)$ can be computed from (2.6) and (3.6). This is the starting point of the derivation of (1.2) for $p_B(r, R)$.

*Quantum gravity on the annulus.*— Depending on whether $\eta$ touches $\partial\mathbb{U}$, the partition function $Z(L, \ell)$ of the random surface bounded by $\eta$ and $\partial\mathbb{U}$ can be decomposed into two terms: $Z(L, \ell) = Z^{\mathrm{nt}}(L, \ell) + Z^{\mathrm{t}}(L, \ell)$, where $Z^{\mathrm{nt}}(L, \ell)$ corresponds to when $\eta$ does not touch $\partial\mathbb{U}$. From the triangulation setting in Figure 3, we see that $Z^{\mathrm{nt}}(L, \ell)$ describes a random annulus in the pure quantum gravity with boundary lengths $L$ and $\ell$, coupled with a percolation configuration for which the monochromatic two-arm crossing occurs. In the Liouville framework, this random annulus can be expressed in terms of three components: the Liouville field, the conformal matter, and the bosonic ghost field [Dav88, DK89]. We conformally parametrize the annulus as the finite cylinder $\mathcal{C}_\tau$ obtained by identifying $[0, \tau] \times \{0\}$ with $[0, \tau] \times \{1\}$ on $[0, \tau] \times [0, 1]$. Then the modulus $\tau$ of the annulus is random and the partition function of a given $\tau$ is the product of the partition function of three components. In our case, the partition function for the conformal matter is simply $p_B(e^{-2\pi\tau}, 1)$. The ghost partition function on $\mathcal{C}_\tau$ is $Z_{\mathrm{ghost}}(\tau) = \eta(2i\tau)^2$ [DP86, Mar03], where $\eta(z) = e^{\frac{i\pi z}{12}} \prod_{n=1}^{\infty}(1 - e^{2ni\pi z})$ is the Dedekind eta function. Therefore, for any test functions $f$ and $g$, we have

$$\iint_0^\infty e^{-\nu_1 L} e^{-\nu_2 \ell} f(L) g(\ell) Z^{\mathrm{nt}}(L, \ell) d\ell dL \propto \int_0^\infty p_B(e^{-2\pi\tau}, 1) \langle f(L_0) g(L_1) \rangle_{\tau, \nu_1, \nu_2} Z_{\mathrm{ghost}}(\tau)\, d\tau, \tag{3.7}$$

where $\langle f(L_0) g(L_1) \rangle_{\tau, \nu_1, \nu_2}$ is averaging over the Liouville theory on $\mathcal{C}_\tau$ with boundary cosmological constants $\nu_1$ and $\nu_2$, and $L_0$ and $L_1$ are the two boundary lengths. Similar to (3.4), we set the bulk cosmological constant to be 0 since there is no area constraint on the surface. See Appendix A.3 for more details on (3.7).

As done in [ARS22], for $\nu_1 = \nu_2 = 0$, solving the Liouville theory on the annulus gives

$$\langle L_0 e^{-L_0} L_1^{ix} \rangle_{\tau, \nu_1 = \nu_2 = 0} = \frac{1}{\sqrt{2}\eta(2i\tau)} \times e^{-\frac{\pi\gamma^2 x^2 \tau}{4}} \times \frac{\pi\gamma x \Gamma(1 + ix)}{2\sinh(\frac{\gamma^2}{4}\pi x)}. \tag{3.8}$$

(The factor $\frac{1}{\sqrt{2}\eta(2i\tau)}$ is not present in [ARS22] due to a different normalization of the Liouville theory on $\mathcal{C}_\tau$.) On the other hand, by (2.6) and (3.6),

$$\int_0^\infty Z(L, \ell) \ell^{ix} d\ell = \frac{3\sqrt{3}}{4} \frac{\sinh(\pi x)}{\sinh(\frac{4\pi x}{3}) + \frac{\sqrt{3}}{2} x} L^{ix-1}. \tag{3.9}$$

The surface corresponding to $Z^{\mathrm{t}}(L, \ell)$ can be analyzed in the probabilistic framework [She16, DMS21], as done in [SXZ24]; see Appendix A.3 for further background. The counterpart of (3.9) is

$$\int_0^\infty Z^{\mathrm{t}}(L, \ell) \ell^{ix} d\ell = \frac{3\sqrt{3}}{4} \frac{\sinh(\frac{\pi x}{3})}{\sinh(\frac{2\pi x}{3})} L^{ix-1}. \tag{3.10}$$

Since $Z^{\mathrm{nt}}(L, \ell) = Z(L, \ell) - Z^{\mathrm{t}}(L, \ell)$, combining (3.7)–(3.10), we arrive at

$$\int_0^\infty p_B(e^{-2\pi\tau}, 1) \eta(2i\tau) e^{-\frac{2\pi x^2 \tau}{3}} d\tau = \frac{\sqrt{3}}{x} \left( \frac{\sinh(\frac{2}{3}\pi x) \sinh(\pi x)}{\sinh(\frac{4}{3}\pi x) + \frac{\sqrt{3}}{2} x} - \sinh(\frac{1}{3}\pi x) \right). \tag{3.11}$$

Taking the inverse Laplace transform of (3.11) with respect to $x^2$ gives $p_B(e^{-2\pi\tau}, 1)\eta(2i\tau)$, which yields (1.2). See Appendix A.3 for details.

# 4 Conclusions

To summarize, we use the conformal radius of a random domain bounded by an SLE curve to encode the backbone exponent (2.5). Using 2D quantum gravity, we relate the conformal radius to the boundary length partition function of certain random surfaces (3.6), which can be computed using the integrability of Liouville conformal field theory. This method is broadly applicable to percolation models whose scaling limits are described by the conformal loop ensemble (CLE). This includes the critical $Q$-Potts random cluster model [FK72] with $Q \in (0, 4]$, where $Q = 1$ corresponds to Bernoulli bond percolation. For general $Q$, the scaling limit is described by CLE with $\kappa = \frac{4\pi}{\pi - \arccos(\sqrt{Q}/2)} \in [4, 8)$. In our derivation of (1.1) in [NQSZ23], we, in fact, treat $\mathrm{CLE}_\kappa$ with $\kappa \in (4, 8)$ uniformly. Under the scaling limit assumption, we obtained the backbone exponent for general $Q$,

$$\sin(\frac{8\pi}{\kappa})\sqrt{\frac{\kappa x}{2} + (1 - \frac{\kappa}{4})^2} - \sin\left(\frac{8\pi}{\kappa}\sqrt{\frac{\kappa x}{2} + (1 - \frac{\kappa}{4})^2}\right) = 0. \tag{4.1}$$

Specializing to $\kappa = 6$, we get (1.1). In [ASYZ24], we derive the nested-path exponent defined in [STZ+22] for the random cluster model. Our method for deriving (1.2) is also broadly applicable. We plan to derive the analog of (1.2) for the random cluster model in a future work.

The conformal field theory (CFT) aspect for percolation is a classical yet active topic [Nie84, DFSZ87, Car92, NRJ24]. The one-arm exponent and the polychromatic arm exponents have a CFT interpretation [HJS20]. It would be highly desirable to find a CFT interpretation for the backbone exponent $x_B$. The expansion (1.2) is reminiscent of the closed channel expansion in boundary CFT, where $\mathcal{S}$ can be thought of as the bulk spectrum. We can also expand in the open channel; see [SXZ24, Equation (1.7)]. In this expansion, a logarithmic structure emerges. We thus suspect that there is an interesting logarithmic CFT that captures $x_B$. The monochromatic $k$-arm exponent for $k \geq 3$ lies strictly between the polychromatic $k$- and $(k+1)$-arm exponents [BN11, JZJ02]. Its precise value remains unknown. Unlike the backbone exponent, we have not found an effective conformal radius encoding that can be computed via the method in [AHS24].

# A Supplementary Material

We provide further background and additional details in our derivation of (1.1) and (1.2). First, we elaborate on the encoding of the percolation exponents using CLE, specifically (2.4) and (2.5). Next, we offer a brief overview of Liouville field theory and 2D quantum gravity and provide further details on (3.5) and (3.6), which ultimately lead to (2.6). Finally, we focus on quantum gravity on the annulus and present the derivation details for (1.2).

## A.1 Percolation exponents and CLE

The conformal loop ensemble (CLE) with parameter $\kappa \in (8/3, 8)$, as introduced by Sheffield [She09], is a random collection of loops, each of which is an $\mathrm{SLE}_\kappa$ curve. Given two loops in a CLE, their enclosed regions are either nested or disjoint. For $\kappa \in (8/3, 4]$, the loops are simple and do not touch each other, whereas for $\kappa \in (4, 8)$, the loops are nonsimple and may touch each other. For $Q \in (0, 4]$, CLE with $\kappa = \frac{4\pi}{\pi - \arccos(\sqrt{Q}/2)} \in [4, 8)$ is conjectured to be the scaling limit of the percolation interfaces in the critical $Q$-Potts random cluster model [FK72]. Specifically, for any $\epsilon > 0$, as the mesh size tends to 0, all interfaces with diameter at least $\epsilon$ are conjectured to converge in law to the CLE loops with diameter at least $\epsilon$. In particular, this conjecture was proved for critical Bernoulli percolation on the triangular lattice [Smi01], which corresponds to $Q = 1$ and $\kappa = 6$, and for critical FK-Ising percolation on the square lattice [Smi10, KS19], which corresponds to $Q = 2$ and $\kappa = 16/3$. Various percolation crossing events can be encoded by CLE loops. For example, let $\eta_0$ be the outermost $\mathrm{CLE}_6$ loop on the unit disk that surrounds the origin, and let $d(0, \eta_0)$ denote the Euclidean distance between 0 and $\eta_0$. Recall the annulus one-arm crossing probability $p(r, 1)$. Then, we have $\mathbb{P}[d(0, \eta_0) \le r] = p(r, 1)$ for $0 < r < 1$.

### Detailed derivation of (2.5)

The monochromatic two-arm event can be encoded using the filled percolation interfaces for the following reason. Fix $0 < r < 1$ and consider the subgraph formed by black vertices in the annulus $A(r, 1)$. We view the inner and outer boundaries of $A(r, 1)$ as two vertices, $x$ and $y$. By the vertex version of Menger's theorem (see, e.g., [Die17]), the maximal number of vertex-disjoint paths from $x$ to $y$, i.e., the number of disjoint black crossings, is equal to the size of the minimum vertex cut, i.e., the minimal number of black vertices needed to remove to disconnect the inner and outer boundaries of $A(r, 1)$. If the outermost filled percolation interface has distance less than $r$ to the origin, then we need to remove at least two black vertices to disconnect the inner and outer boundaries of $A(r, 1)$, and thus, there exist two disjoint black crossings of $A(r, 1)$. In the continuum limit, the filled percolation interfaces converge to the outer boundaries of $\mathrm{CLE}_6$ loops. These outer boundaries are simple loops that locally look like $\mathrm{SLE}_{8/3}$ curves [LSW03], while different loops may touch each other. Let $\eta$ be the outermost such simple loop that surrounds 0. Then, Menger's theorem implies that on the event $d(0, \eta) \le r$, there are two disjoint black crossings of $A(r, 1)$. Hence, we have

$$\mathbb{P}[d(0, \eta) \le r] \le p_B(r, 1) \quad \text{for } 0 < r < 1. \tag{A.1}$$

Let $\widetilde{x}_B$ be the exponent such that $\mathbb{P}[d(0, \eta) \le r] = r^{\widetilde{x}_B + o(1)}$. Then, (A.1) implies that $\widetilde{x}_B \ge x_B$. We now elaborate that $\widetilde{x}_B \le x_B$, and thus the two probabilities in (A.1) share the same exponent as $r \to 0$. If $d(0, \eta) > r$, there may still exist two disjoint black crossings of $A(r, 1)$; see Figure 4. However, in this case, there must be two white crossings separating the two black crossings in the annulus $A(r, d(0, \eta))$, and thus, an alternating four-arm event occurs. The probability of this event is known to be $(\frac{r}{d(0,\eta)})^{x_4 + o(1)}$, where $x_4 = \frac{5}{4}$ [ADA99]. By partitioning the values of $d(0, \eta)$ into dyadic intervals $(2^n r, 2^{n+1} r]$ for $0 \le n \le \lfloor \log_2 r^{-1} \rfloor$, we obtain

$$p_B(r, 1) \le \mathbb{P}[d(0, \eta) \le r] + \sum_{n=0}^{\lfloor \log_2 r^{-1} \rfloor} \mathbb{P}[2^n r < d(0, \eta) \le 2^{n+1} r] \times (2^{-n})^{x_4 + o(1)}. \tag{A.2}$$

Using $\mathbb{P}[d(0, \eta) \le r] = r^{\widetilde{x}_B + o(1)}$ and $\widetilde{x}_B < x_4$, we obtain

$$p_B(r, 1) \le r^{\widetilde{x}_B + o(1)}. \tag{A.3}$$

Therefore, the two probabilities in (A.1) share the same exponent as $r \to 0$. In particular,

$$\mathbb{P}[d(0, \eta) \le r] = r^{x_B + o(1)}. \tag{A.4}$$

Combining this with the Koebe 1/4 theorem $d(0, \eta) \le \mathrm{CR}(0, D_b) \le 4d(0, \eta)$ yields (2.5), similar to (2.4).

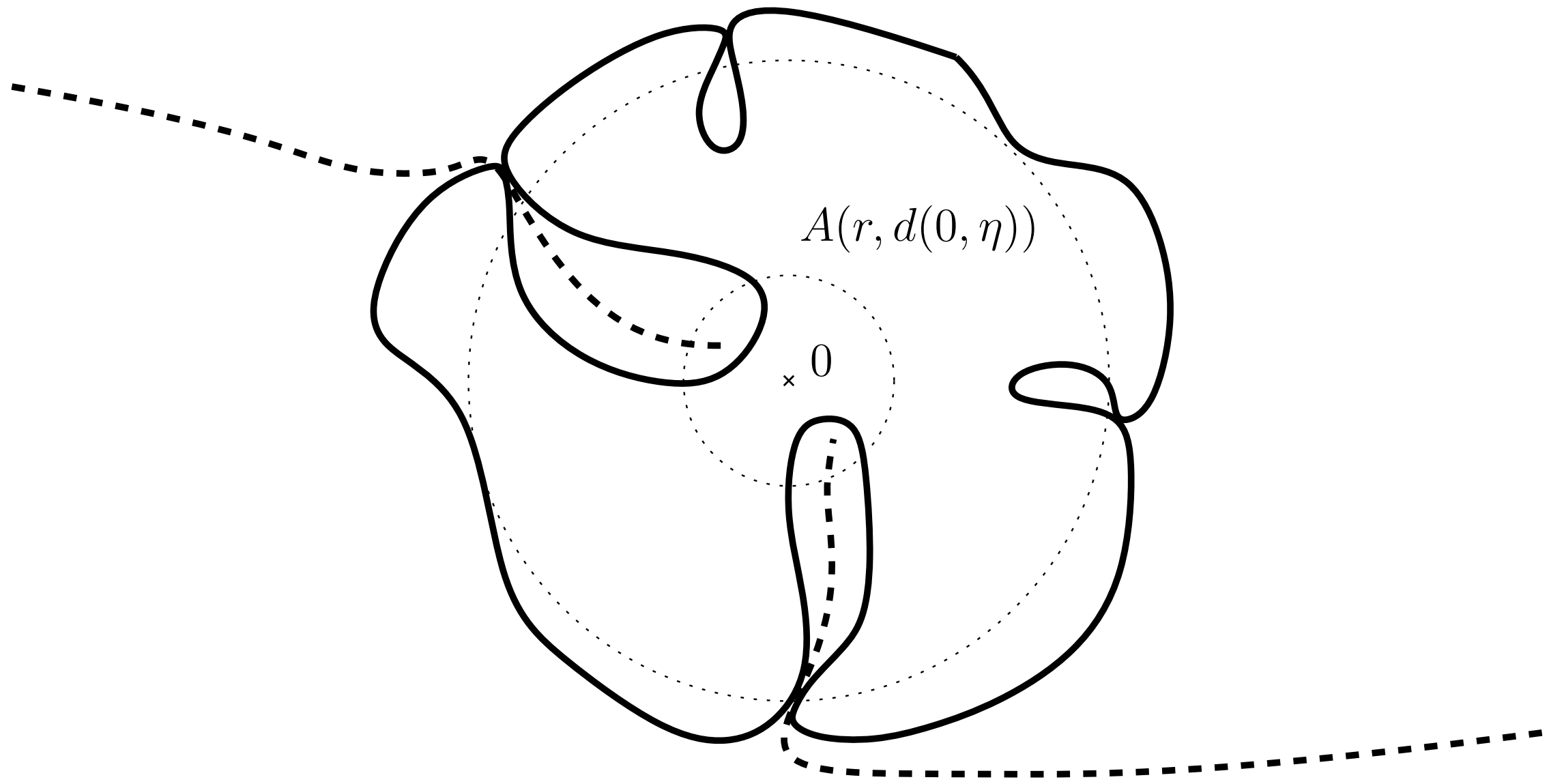

Figure 4: The two dashed lines represent two disjoint black crossings of $A(r,1)$ on the event $d(0,\eta) > r$.

## A.2 Liouville field theory and 2D quantum gravity

We begin with the connection between Liouville field theory and 2D quantum gravity. Next, we review the coupling theory between SLE and Liouville quantum gravity, from which we derive (3.5). We then explain the derivation of the key equation (3.6).

### 2D quantum gravity and random triangulation

Random triangulation is a discrete model for 2D quantum gravity, where the uniform case corresponds to pure gravity; see [LG14, She23] for a mathematical overview. Taking a uniform random triangulation of the sphere with $n$ faces and sending $n$ to infinity, we obtain a random surface with unit area. It is natural to define a surface with unrestricted area by reweighting the area with a random variable $A$ sampled from the infinite measure $A^{-7/2}dA$, since the total number of triangulations with $n$ faces grows like $\alpha^n n^{-7/2}$ for some $\alpha \in (0,\infty)$ [Tut63, Bro64]. Similarly, we can use the random triangulation of the disk to define a random disk with one interior marked point and no area or boundary length constraints. The total number of random triangulations of the disk with one interior marked point, $n$ faces, and perimeter $p$ grows like $\alpha^n \beta^p n^{-3/2} p^{-1/2} e^{-Cp^2/n}$ for $\alpha$ above and some $\beta \in (0,\infty)$ [HRV18]. If we assign the weight $\alpha^{-n}\beta^{-p}$ (known as the critical Boltzmann weight) and sum over $n$, we see that the total weight grows like $p^{-3/2}$ as $p \to \infty$. Therefore, the boundary length of the resulting random disk follows a law proportional to $L^{-3/2}dL$.

2D pure quantum gravity can alternatively be described by Liouville field theory with $c_L = 26$ and $\gamma = \sqrt{8/3}$, where the exponential of field defines a random geometry that can be viewed as the gravitational background. To understand this geometric construction, note that when the bulk and boundary cosmological constants are set to 0, the field $\phi$ sampled from the distribution $e^{-S_L[\phi]}D\phi$ on $\Sigma$ is simply the Gaussian free field (GFF), which is a random generalized function. Using a normalization procedure known as Gaussian multiplicative chaos theory, we can make sense of the area measure $e^{\gamma\phi}d^2x$ on $\Sigma$ and the boundary length measure $e^{\frac{\gamma}{2}\phi}dl$ on $\partial\Sigma$ [DS11]. The potential terms $\int_\Sigma e^{\gamma\phi}d^2x$ and $\int_{\partial\Sigma} e^{\frac{\gamma}{2}\phi}dl$ in the action $S_L[\phi]$ correspond to the total area and boundary length of this random surface. Therefore, Liouville field theory can be understood as reweighting the GFF by the total area and boundary length. Moreover, by Girsanov's theorem, vertex operators $e^{\alpha_i \phi(x_i)}$ correspond to adding the singularity $-\alpha_i \log|\cdot - x_i|$ to the GFF.

For the sphere case, sample $\phi$ from $e^{-S_L[\phi]}D\phi$ with bulk cosmological constant 0. Then, the resulting random sphere describes the scaling limit of the random triangulation of the sphere with no area constraint. In this case, the law of the total area measure of the Liouville field is proportional to $A^{-1-2Q/\gamma}dA$. Now $\gamma = \sqrt{8/3}$ and $Q = \sqrt{25/6}$ give $1 + 2Q/\gamma = 7/2$, which is consistent with the exponent obtained from triangulation enumeration. For the disk case, sample $\phi$ from $e^{\gamma\phi(0)}e^{-S_L[\phi]}D\phi$ on $\mathbb{U}$ with all cosmological constants 0. Then, the resulting random disk describes the scaling limit of the random triangulation of the disk with one interior marked point and no area or boundary length constraints. In particular, the law of the total boundary length of the Liouville field is proportional to $L^{-3/2}dL$, which again matches the triangulation result. From both perspectives, we see that for $\mathrm{QD}_1(L)$ defined below (3.4), its partition function $|\mathrm{QD}_1(L)|$ of is $CL^{-\frac{3}{2}}$ for some constant $C > 0$.

### Derivation of (3.5): cut and glue random surfaces

Sample a critical Bernoulli site percolation on top of a random triangulation of the disk as in Figure 3. A crucial assumption in 2D quantum gravity is that the continuum limit of percolation is still $\mathrm{CLE}_6$, independently of the geometric background. The random geometry affects only the lengths of the loops. A version of this result was rigorously proved in [HS23]. Bernoulli percolation on random triangulations has a greater degree of spatial independence compared to its regular lattice counterpart. A specific example is shown in Figure 3: the outermost filled percolation interface divides the random triangulation into two independent surfaces conditioned on the number of vertices along the filled interface. For $L > 0$, we write $Z_{\mathrm{whole}}(L)$ as the partition function of the original random triangulation with boundary length $L$. Then we have

$$Z_{\mathrm{whole}}(L) = \sum_{\ell} Z_{\mathrm{outside}}(L, \ell) \times \ell \times Z_{\mathrm{inside}}(\ell), \tag{A.5}$$

where $Z_{\mathrm{inside}}, Z_{\mathrm{outside}}$ are Boltzmann-weighted partition functions of the two random triangulations separated by the outermost filled percolation interface, whose length is denoted by $\ell$. The additional factor $\ell$ on the right-hand side counts the number of ways the two surfaces can be glued together to recover the original disk, since there is no marked point on their common boundary.

Equation (3.5) is the exact continuum analog of (A.5). In fact, there is a way to derive such equations directly in the continuum using the quantum zipper method pioneered by Sheffield [She16]. Using this method, we proved in [SXZ24] that similar spatial independence holds when we use the curve $\eta$ to cut the random disk of length $L$. Furthermore, the quantum zipper method shows that the law of the random disk inside $\eta$ is an independent copy of the original random disk given its boundary length. Namely, the continuum analogs of $Z_{\mathrm{whole}}(L)$ and $Z_{\mathrm{inside}}(\ell)$ take the same form, which are $L^{-3/2}$ and $\ell^{-3/2}$ respectively. (Note that the discrete counterpart of this statement is not exactly true, since the surface bounded by the outermost filled percolation interface is a random triangulation of the disk decorated with a percolation configuration subject to complicated constraints.) Writing $Z(L, \ell)$ as the continuum analog of $Z_{\mathrm{outside}}(L, \ell)$, we get (3.5).

### Derivation of (3.6): change of vertex insertion

Recall that the random disk $\mathrm{QD}_1$ can be described by the Liouville field sampled from $e^{\gamma\phi(0)}e^{-S_L[\phi]}D\phi$ on $\mathbb{U}$. For general $\alpha \in \mathbb{R}$ and a field $\phi$ sampled from $e^{\alpha\phi(0)}e^{-S_L[\phi]}D\phi$ on $\mathbb{U}$, the law of the boundary length is proportional to $L^{\frac{2}{\gamma}(\alpha-\gamma)-\frac{3}{2}}$. To obtain (3.6), we change the vertex operator at the origin from $e^{\gamma\phi(0)}$ to $e^{\alpha\phi(0)}$ on both sides of (3.5) while keeping the random surface with partition function $Z(L,\ell)$ unchanged. Take $\alpha = \gamma + \frac{\gamma}{2}a$ such that $\frac{2}{\gamma}(\alpha-\gamma) - \frac{3}{2} = -\frac{3}{2} + a$. This gives the term $L^{-\frac{3}{2}+a}$ on the right-hand side of (3.6).

Since the random disk bounded by $\eta$ is also a sample from $\mathrm{QD}_1$, changing the insertion from $e^{\gamma\phi(0)}$ to $e^{\alpha\phi(0)}$ results in the same shift of the partition function from $\ell^{-\frac{3}{2}}$ to $\ell^{-\frac{3}{2}+a}$, if the random disk were parametrized by $\mathbb{U}$ instead of the domain $D_b$ bounded by $\eta$. This change of domain results in an additional conformal factor, which is exactly $\mathrm{CR}(0,D_b)^{x(a)}$ on the right-hand side of (3.6). To see the effect of the coordinate change, for a simply connected domain $D$ containing 0, let $\langle e^{\alpha\phi(0)}\rangle_D$ be the correlation function $\int_{\phi:D\to\mathbb{R}} e^{\alpha\phi(0)} e^{-S_L[\phi]} D\phi$. For a conformal map $f:(\mathbb{U},0)\to(D_b,0)$, we have

$$\langle e^{\alpha\phi(0)}\rangle_{D_b} = |f'(0)|^{-2\Delta_\alpha}\langle e^{\alpha\phi(0)}\rangle_{\mathbb{U}}, \tag{A.6}$$

where $\Delta_\alpha = \frac{\alpha}{2}(Q-\frac{\alpha}{2})$ is the conformal dimension of $e^{\alpha\phi(0)}$. Since by definition $\mathrm{CR}(0,D_b) = |f'(0)|$, the additional factor is given by

$$\frac{\langle e^{\alpha\phi(0)}\rangle_{\mathbb{U}}}{\langle e^{\alpha\phi(0)}\rangle_{D_b}} \Big/ \frac{\langle e^{\gamma\phi(0)}\rangle_{\mathbb{U}}}{\langle e^{\gamma\phi(0)}\rangle_{D_b}} = \mathrm{CR}(0,D_b)^{2\Delta_\alpha - 2\Delta_\gamma}. \tag{A.7}$$

When $\alpha = \gamma + \frac{\gamma}{2}a$, we have $2\Delta_\alpha - 2\Delta_\gamma = -\frac{1}{3}a(a-1) = x(a)$. This yields (3.6).

## A.3 Quantum gravity on the annulus

Recall that $Z(L,\ell)$ is the continuum analog of $Z_{\mathrm{outside}}(L,\ell)$. In the discrete, $Z_{\mathrm{outside}}(L,\ell)$ can be decomposed into two terms: $Z_{\mathrm{outside}}(L,\ell) = Z^{\mathrm{nt}}_{\mathrm{outside}} + Z^{\mathrm{t}}_{\mathrm{outside}}$, depending on whether the outermost filled interface as in Figure 3 touches the boundary or not (in Figure 3 it does not touch). The continuum analogs of these are respectively $Z^{\mathrm{nt}}$ and $Z^{\mathrm{t}}$. In fact, both of these random surfaces can be described by Liouville field theory, which yields (3.7) and (3.10), respectively.

We first consider $Z^{\mathrm{nt}}$. Recall that $Z^{\mathrm{nt}}_{\mathrm{outside}}$ is the Boltzmann-weighted partition function of a random triangulation of an annulus decorated with a percolation configuration subject to the monochromatic two-arm event. Without the percolation decoration, this random triangulation is the discrete model for pure gravity on the annulus. Its Liouville description is well known in the bosonic string theory [Pol81], which is given by

$$Z(d\tau) = Z_{\mathrm{ghost}}(\tau) Z_{\mathrm{Liouville}}(\tau) d\tau. \tag{A.8}$$

Here, $Z_{\mathrm{ghost}}(\tau)$ is the partition function for the ghost CFT with central charge $-26$, coming from the Faddeev-Popov determinant for the conformal gauge-fixing. For the annulus with modulus $\tau$, it is given in [DP86, Mar03] that $Z_{\mathrm{ghost}}(\tau) = \eta(2i\tau)^2$. The Liouville partition function $Z_{\mathrm{Liouville}}(\tau)$ is given by the Liouville action $S_L[\phi]$ (3.3) on the annulus: $Z_{\mathrm{Liouville}}(\tau) = \int_{\phi:\mathcal{C}_\tau\to\mathbb{R}} e^{-S_L[\phi]} D\phi$.

Now that $Z^{\mathrm{nt}}_{\mathrm{outside}}$ is the random triangulation decorated with a percolation configuration where the monochromatic two-arm event occurs, we can view the decorating percolation as a conformal matter with central charge 0 and partition function $p_B(e^{-2\pi\tau},1)$. For gravity coupled to conformal matter on the annulus, the Liouville description becomes

$$Z^{\mathrm{nt}}(d\tau) = Z_{\mathrm{ghost}}(\tau) Z_{\mathrm{Liouville}}(\tau) Z_{\mathrm{matter}}(\tau) d\tau. \tag{A.9}$$

Applying this to $Z^{\mathrm{nt}}$, we obtain (3.7).

The random surface corresponding to $Z^{\mathrm{t}}$ is quite intriguing. In fact, when the curve $\eta$ touches $\partial\mathbb{U}$, it will touch at infinitely many points. The surface bounded by $\eta$ and $\partial\mathbb{U}$ consists of a countable collection of topological disks. Again based on the quantum zipper method, it was proved in [SXZ24] that the law of this collection of random disks is Poissonian, and conditioning on their boundary lengths, these disks are independent copies of random disks that can be described by Liouville theory on the disk with two boundary insertions. This allows [SXZ24] to derive the exact expression (3.10) for $Z^{\mathrm{t}}$.

**Derivations of (3.11) and (1.2)**

We first provide further background on (3.8). Using the CFT framework [BPZ84], the Liouville annulus partition function was computed in [Wu22]. If we set the bulk cosmological constant to 0 and the two boundary cosmological constants to $\nu_1, \nu_2$, then the partition function becomes $\langle e^{-\nu_1 L_0} e^{-\nu_2 L_1}\rangle_\tau$ where the average is over a Liouville field theory on $\mathcal{C}_\tau$ with all cosmological constants 0, and $L_0$ and $L_1$ are the two boundary lengths. Based on this expression, equation (3.8) was derived in [ARS22].

By (3.6) and (2.6), and taking $b = -\frac{1}{2} + a$, we have

$$\int_0^\infty Z(L,\ell)\ell^b d\ell = \frac{3\sqrt{3}}{4}\frac{\sin(\pi b)}{\sin(\frac{4\pi b}{3}) + \frac{\sqrt{3}}{2}b} L^{b-1}. \tag{A.10}$$

Analytically extending both sides of the equation and setting $b = ix$ yields (3.9). Next, we derive (3.11). On the one hand, by (3.9), (3.10), and $Z^{\mathrm{nt}}(L,\ell) = Z(L,\ell) - Z^{\mathrm{t}}(L,\ell)$, we have

$$\begin{aligned}
&\iint_0^\infty Le^{-L} Z^{\mathrm{nt}}(L,\ell)\ell^{ix} dL d\ell \\
&= \int_0^\infty Le^{-L}\left(\frac{3\sqrt{3}}{4}\frac{\sinh(\pi x)}{\sinh(\frac{4\pi x}{3}) + \frac{\sqrt{3}}{2}x} - \frac{3\sqrt{3}}{4}\frac{\sinh(\frac{\pi x}{3})}{\sinh(\frac{2\pi x}{3})}\right) \times L^{ix-1} dL \\
&= \frac{3\sqrt{3}}{4}\frac{\Gamma(ix+1)}{\sinh(\frac{2\pi x}{3})}\left(\frac{\sinh(\frac{2}{3}\pi x)\sinh(\pi x)}{\sinh(\frac{4}{3}\pi x) + \frac{\sqrt{3}}{2}x} - \sinh(\frac{1}{3}\pi x)\right).
\end{aligned} \tag{A.11}$$

On the other hand, by (3.7) and (3.8),

$$\begin{aligned}
&\iint_0^\infty Le^{-L} Z^{\mathrm{nt}}(L,\ell)\ell^{ix} d\ell \\
&\propto \int_0^\infty p_B(e^{-2\pi\tau}, 1)\langle L_0 e^{-L_0} L_1^{ix}\rangle_{\tau,\nu_1=\nu_2=0} Z_{\mathrm{ghost}}(\tau) d\tau \\
&= \int_0^\infty p_B(e^{-2\pi\tau}, 1)\frac{1}{\sqrt{2}\eta(2i\tau)} \cdot e^{-\frac{\pi\gamma^2 x^2\tau}{4}} \cdot \frac{\pi\gamma x\Gamma(1+ix)}{2\sinh(\frac{\gamma^2}{4}\pi x)} \times \eta(2i\tau)^2 d\tau \\
&= \frac{\pi x\Gamma(ix+1)}{\sqrt{3}\sinh(\frac{2\pi x}{3})}\int_0^\infty p_B(e^{-2\pi\tau}, 1)\eta(2i\tau)e^{-\frac{2\pi x^2\tau}{3}} d\tau.
\end{aligned} \tag{A.12}$$

Combining the above two equations, we obtain (3.11) up to a constant. This constant can be fixed using the condition $\lim_{\tau\to 0} p_B(e^{-2\pi\tau}, 1) = 1$, which gives (3.11).

Finally, we derive (1.2). Replacing $\frac{2\pi x^2}{3}$ in (3.11) with $t$, we obtain $\int_0^\infty e^{-t\tau} p_B(e^{-2\pi\tau}, 1)\eta(2i\tau) d\tau = g(t)$ for any $t \in (0,\infty)$, where $g(t)$ is an explicit function. Applying the inverse Laplace transform, we obtain

$$p_B(e^{-2\pi\tau}, 1)\eta(2i\tau) = \sum_{s\in\mathcal{S}'} \mathrm{Res}(e^{\tau t} g(t), s) \tag{A.13}$$

where $\mathcal{S}'$ is the set of poles of $g(t)$ in $\mathbb{C}$, and $\mathrm{Res}(e^{\tau t}g(t), s)$ is the residue of $e^{\tau t}g(t)$ at $t = s$. Specifically, $\mathcal{S}'$ consists of all the complex solutions to $\sinh(4\pi\sqrt{\frac{t}{6\pi}}) + \frac{3}{2}\sqrt{\frac{t}{2\pi}} = 0$ except 0 and $-\frac{2\pi}{3}$. After simplification, (A.13) reduces to (1.2).

**Acknowledgements.** This article is adapted from the paper with the same title published on *Physics Review Letters* in 2025. We thank Sylvain Rilbault for helpful comments. P.N. is partially supported by a GRF grant from the Research Grants Council of the Hong Kong SAR (project CityU11318422). W.Q. and X.S. are supported by National Key R&D Program of China (No. 2023YFA1010700). Z.Z. is partially supported by NSF Grant No. DMS-1953848.

# Higher-dimensional Conformal Blocks from 2D CFT

**Volker Schomerus**

*Deutsches Elektronen Synchroton DESY, Notkestr. 85, 22607 Hamburg, Germany*
*II. Inst. für Theor. Physik, Universität Hamburg, Luruper Chaussee,*

*E-mail:* `volker.schomerus@desy.de`

Abstract: We investigate conformal blocks in higher-dimensional conformal field theories (CFTs). Our new approach exploits deep connections between d-dimensional conformal blocks and two-dimensional (2D) Wess-Zumino-Novikov-Witten (WZNW) models with affine Kac-Moody symmetry on the conformal group $\mathrm{Spin}(1, d{+}1)$. In this framework, conformal blocks associated with the d-dimensional conformal algebra naturally emerge at the critical level, where Gaudin models and more general Hitchin type integrable systems become relevant. This embedding into 2D WZNW models further uncovers novel links of d-dimensional conformal blocks to semiclassical correlation functions in Liouville field theory and its multifield generalizations.

# Contents

# 1 Introduction

Conformal quantum field theories (CFTs) play an important role for our understanding of phase transitions, quantum field theory and even the quantum physics of gravity, through Maldacena's celebrated holographic duality. Since they are often strongly coupled, however, they are very difficult to access with traditional perturbative methods. Polyakov's famous conformal bootstrap program provides a powerful non-perturbative handle that allows to calculate critical exponents and other dynamical observables using only general features such as (conformal) symmetry, locality and unitarity [1]. The program has had impressive success in $d = 2$ dimensions [2] where it produced numerous exact solutions. During the last two decade, the bootstrap has seen a remarkable revival in higher-dimensional theories with new numerical as well as analytical incarnations. This has produced many stunning new insights, see e.g. [3,4] for reviews and references. Despite these advances, it is evident that significant further developments are needed to extend these techniques in particular to include novel classes of observables, such as boundaries, interfaces and defects, multi-point and thermal correlators etc.

The central tool for CFTs in general and for the conformal bootstrap in particular are conformal partial wave expansions. These were introduced in [5] to separate correlation functions into kinematically determined conformal blocks (partial waves) [1] and expansion coefficients which contain all the dynamical information. For four-point correlators, the relevant blocks are now well understood in any $d$, though only after some significant effort. Here we shall describe a fully coherent approach that allows to extend such results to a wide range of correlation functions with an arbitrary number of insertions of both local and non-local operators. Our approach extends the observation in [6] about a relation between four-point blocks and exactly solvable (integrable) Schroedinger problems.

[1]Let us once stress that conformal blocks and partial waves are actually not the same objects. They both are solutions of the same differential equations, but differ in the choice of boundary conditions. Partial waves form a basis of functions and they can be decomposed into a sum of a block and its shadow. In this paper we shall not distinguish between the two notions and simply use the term *conformal block*.

In order to zoom in on our goal and the methods we shall propose to achieve it, let us begin with the most basic description of conformal blocks, the so-called shadow formalism [7]. The latter provides integral formulas for conformal blocks that are reminiscent of Feynman integrals. Finding analytical expressions in terms of special functions or even just efficient numerical evaluations requires significant technology. One crucial tool in the theory of Feynman integrals is to consider them as solutions of some differential equations. In their important work, Dolan and Osborn followed this same strategy and characterized shadow integrals as eigenfunctions of a set of Casimir differential operators [8]. By studying these differential equations they were able to harvest decisive new results on the conformal blocks [8, 9].

The relation between Casimir equations and integrability was first uncovered in [6] where the Casimir equations for scalar four-point blocks were rewritten as eigenvalue equations for a multi-dimensional exactly solvable Schroedinger problem of Calogero-Sutherland type. The relevance of Calogero-Sutherland models is not restricted to scalar four-point functions but also extends to to setups that involve non-local defect operators, see e.g. [10]. Beyond that, a more general framework is needed. Through the investigation of conformal blocks for correlators with $N > 4$ local field insertions it became clear that Casimir operators may be considered as part of a larger class of commuting Hamiltonians in an integrable Gaudin model [11]. This resulted in a complete set of differential equations for conformal blocks with any number of local insertions.

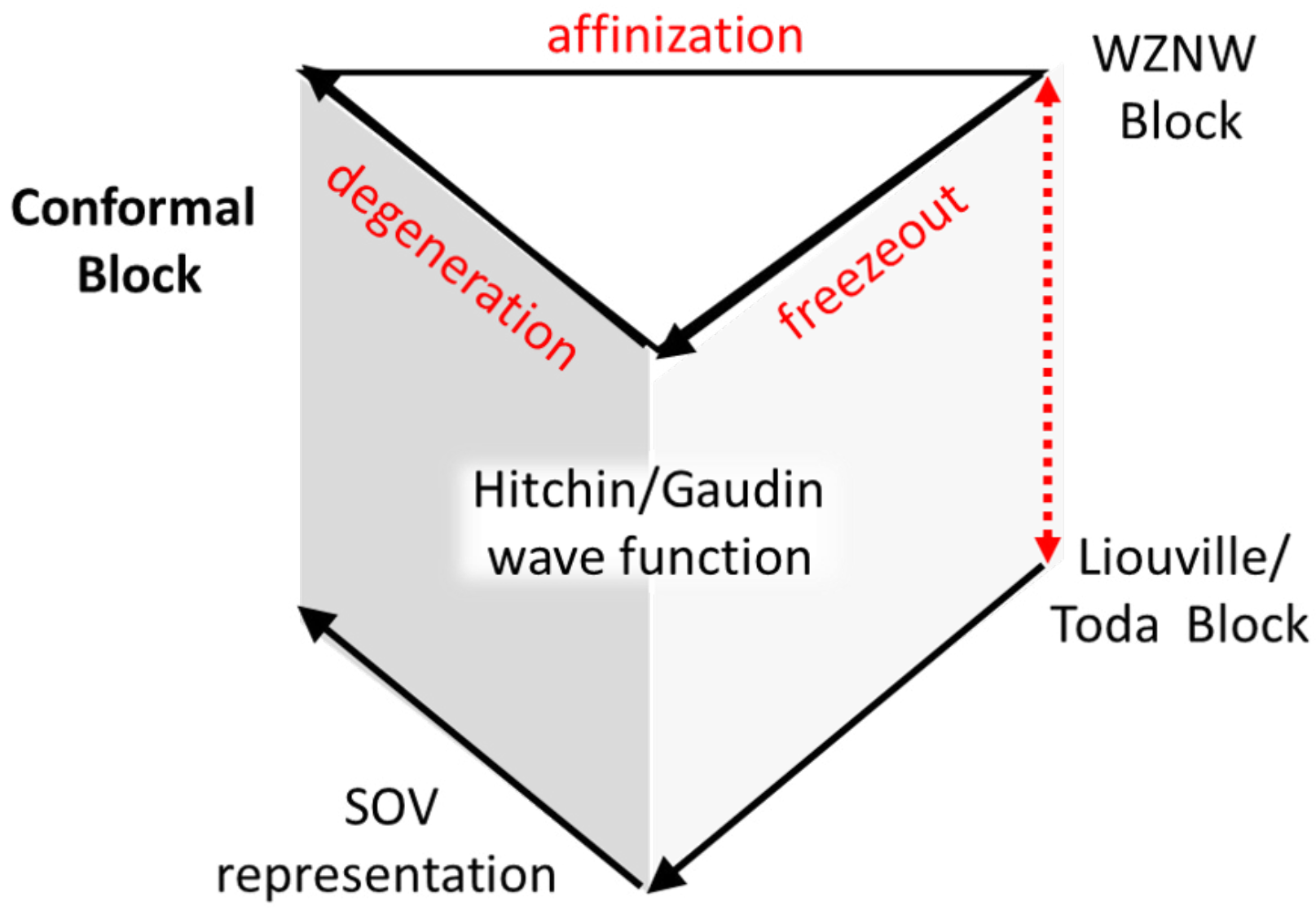

Figure 1: Outline of the content of the paper. The central object are the $d$-dimensional global conformal blocks on the left hand side. The upper triangle illustrates the *affinization* and the way in which conformal blocks are recovered from 2D WZNW models in a two-step process that involves *freezeout* and *degeneration*. The bottom part of the figure shows how a fully factorized SOV representation of (momentum space) blocks can be constructed by using a correspondence between WZNW models in the upper right corner and 2D scalar multi-field models with a Toda-like interaction. For more explanations see the main text.

In this work we will actually carry the relation with integrable models one step further by embedding the theory of global conformal blocks of a $d$-dimensional CFT into the context of affine blocks in a 2D WZNW model on the conformal group $\mathrm{Spin}(1, d + 1)$, see the upper half of Figure 1. The relation is obtained by a process of affinization in which all primary fields of the $d$-dimensional CFT acquire an extra parameter $w_i$ that takes values on some auxiliary surface which we shall often refer to as the "world-sheet". The affinization can be undone in two steps: The first one is the critical level limit of the WZNW model in which the world-sheet parameters $w_i$ of the field insertions freeze out. In the limit, wave functions of a Gaudin model emerge from the affine blocks [12]. The second step involves a degeneration of the underlying surface, i.e. we go to some point on the boundary of its moduli space in which the auxiliary world-sheet degenerates into an arrangement of three-punctured spheres. This can be done in such a way that Gaudin wave functions degenerate to global conformal blocks. In the context of conformal blocks the relevant limit was first described in [11, 13].

But there is yet another layer to our story which is depicted in the bottom half of Figure 1. It can provide explicit factorized expressions for conformal blocks. Quite generally, WZNW models with non-compact target spaces are expected to be dual to multi-field scalar field theories with some exponential interactions. The simplest instance of this correspondence is the relation between the $H_3^+$ WZNW model and Liouville field theory that was uncovered by Ribault and Teschner [14], see also [15] for some earlier work. The original setup was restricted to the sphere, but the extension to surfaces of higher genus, and in particular to the torus, has been obtained with the help of path integral techniques, see [16]. Ribault and Teschner pointed out already that the critical level limit in the WZNW model corresponds to a semiclassical limit of the dual Liouville field theory. Hence, wave functions of Gaudin integrable systems, and their degeneration to conformal blocks, can be calculated by solving the semiclassical equations of motion of Liouville field theory and some multi-field, Toda-like extensions thereof. The resulting semiclassical representation of the wave functions turns out to be rather useful since it leads to manifestly factorized expressions. In addition, upon degeneration of the underlying world-sheet, i.e. in passing from general Gaudin wave functions to global conformal blocks, the individual factors simplify drastically.

Let us briefly outline the content of each of the following sections. In Section 2 we shall begin with a short introduction to conformal symmetry in $d$-dimensions and its representations. This provides most of the relevant background we need from $d$-dimensional conformal field theory (CFT) that is placed on the left upper corner of Figure 1. Section 3 is then devoted to the input from 2D CFT, i.e. to the affinization and freezeout. More precisely, we will give a lightening review to WZNW models and affine Kac-Moody algebras, and we will recall how Gaudin integrable systems emerge at the critical level. In Section 4 we turn to the discussion of Casimir differential equations, starting with the most familiar case of scalar 4-point functions in Section 4.1. The Casimir differential equations provide the diagnostic tools that allows us to identity the conformal blocks after the appropriate degeneration of the underlying 2D surface. The relation of multi-point conformal blocks with Gaudin integrable systems is reviewed in Section 4.2. Sections 4.3 is situated in the lower half of Figure 1. There we explain how solutions of classical equations of motion in certain 2D scalar field theories, such as Liouville field theory and multi-field extensions thereof, can be exploited to construct factorized expressions for d-dimensional conformal blocks in momentum space, in which the variables are separated. We refer to these as SOV representations in Figure 1. While subsection 4.3 are a bit more sketchy than the rest of the paper, it does lay out the path to some intriguing applications that we plan to report on in future work, see also the concluding section.

While the relation between integrable systems and conformal blocks for correlation functions in $\mathrm{CFT}_d$ was well understood in previous work, the embedding of higher dimensional conformal blocks into 2D WZNW models and its application to the construction of conformal blocks is original to this work.

# 2 Conformal Symmetry and its Representations

The purpose of this short section is to review some notations and key notions from higher dimensional conformal symmetry and field theory. After a brief reminder on conformal symmetry and the notion of primary fields we provide a group theoretic construction of the relevant representations of the conformal algebra in the second subsection, see e.g. [17–20] for more detail and further references.

## 2.1 Some notions from conformal field theory

The non-perturbative dynamics of conformal field theory can be probed through a large variety of correlation functions. For most of our discussion we shall focus on (vacuum) correlation functions of the most standard observables, namely of local (primary) fields. The space of fields of a conformal field theory carries an action of the conformal Lie algebra $\mathfrak{so}(1, d+1)$. Each conformal multiplet contains a distinguished set of primary fields $\Phi(x) = \Phi_{(\Delta,\lambda)}(x)$ that possess lowest scaling weight $\Delta$ and transform in some finite dimensional representation $\lambda$ of the rotation algebra $\mathfrak{so}(d)$. We refer to the label $\lambda$ as the spin of the primary field. From the primary fields $\Phi$ all other fields in the conformal multiplet can be obtained by acting with the generators $P_\mu$ of translations, i.e. they are simply derivatives of the primaries. By definition, the transformation behavior of such a conformal primary field with respect to infinitesimal dilations $D$, rotations $L_{\mu\nu}$ and special conformal transformations $K_\mu$ is given by

$$\begin{aligned}
[D, \Phi_\pi(x)] &= (x^\nu\partial_\nu + \Delta)\, \Phi_\pi(x) =: \mathfrak{y}_D \Phi_\pi(x) \\
[L_{\mu\nu}, \Phi_\pi(x)] &= \left(x_\nu\partial_\mu - x_\mu\partial_\nu + \Sigma^\lambda_{\mu\nu}\right) \Phi_\pi(x) =: \mathfrak{y}_{L_{\mu\nu}} \Phi_\pi(x) \\
[K_\mu, \Phi_\pi(x)] &= \left(x^2\partial_\mu - 2x_\mu x^\nu\partial_\nu - 2x_\mu\Delta + 2x^\nu\Sigma^\lambda_{\mu\nu}\right) \Phi(x) =: \mathfrak{y}_{K_\mu} \Phi_\pi(x)
\end{aligned} \tag{2.1}$$

where $\Sigma^\lambda_{\mu\nu} = \pi(L_{\mu\nu})$ denote the representation matrices of the generators $L_{\mu\nu}$ of rotations in the representation $\lambda$ of $\mathrm{Spin}(d)$.

Let us recall that the Euclidean conformal group $G = \mathrm{Spin}(1, d+1)$ acts on $\mathbb{R}^d$ and under this action a point $x_\mu \in \mathbb{R}^d$ is stabilized by a certain subgroup $P_x$. For the origin $x = 0$, this subgroup is simply $P_0 = P = (\mathrm{SO}(1,1)\times\mathrm{Spin}(d)) \ltimes \mathrm{N}_d$, where $\mathrm{N}_d$ denotes the $d$-dimensional abelian subgroup that is generated by special conformal transformations $K_\mu$. The quotient $G/P \cong \mathbb{R}^d$ is the configuration space of a local field, i.e. it is parametrized by the position $x$ at which the field $\Phi$ is inserted. The choice of weight $\Delta$ and spin $\lambda$ may be regarded as the choice of a finite dimensional unitary representation $\pi = (\Delta, \lambda)$ of the stabilizer subgroup $P \subset G$. Indeed, in any such representation the generators of special conformal transformations act trivially and hence finite dimensional unitary representations of $P$ are obtained from finite dimensional unitary representations of the product group $K = \mathrm{SO}(1,1) \times \mathrm{Spin}(d) \subset P \subset G$.

Let us briefly remark that our discussion of local fields can easily be extended to include conformal defect operators $\mathcal{D}_p(X)$ that create some defects of dimension $p < d$. By definition, any such defect is stabilized by the subgroup $K_p = \mathrm{Spin}(1, p+1)\times\mathrm{Spin}(d-p) \subset G$ of the conformal group. The first factor describes conformal transformations along the defect while the second consists of rotations of the directions that are transverse to the defect. We conclude that the configuration space of the defect operator $\mathcal{D}_p$ is given by the quotient $G/K_p$, i.e. the argument $X$ of the operators $\mathcal{D}_p(X)$ denotes a point $X \in G/K_p$.

Defects can also carry a spin quantum number that is associated with the choice of a finite dimensional unitary representation of $K_p$. Note that for $p \neq 0$ such representations must be trivial on elements in the group $\mathrm{Spin}(1, p+1)$ of conformal transformations along the defect and hence the spin of a defect is a transverse spin that describes a representation of $\mathrm{Spin}(d-p)$. The defect operators $\mathcal{D}_p(X)$ transform under infinitesimal conformal transformations is a way akin to equations (2.1) but with a different set of differential operators $\mathfrak{y}$ that depend on the dimension $p$ of the defect and act on the configuration space $G/K_p$. We will not give formulas for these operators here but the discussion in the next subsection shows how such expressions can be obtained.

## 2.2 Elements of conformal group theory

For our discussion below is will be useful to have a clear representation theoretic view on the equations (2.1) that characterize the primary fields in a conformal field theory. It turns out that we can realize the subspace that is spanned by primaries and their descendants in terms of vector-valued functions (or rather sections of some vector bundle) on the quotient $G/P$. In order to do so, let us recall that the index $\pi = (\Delta, \lambda)$ determines a representation of the subgroup $K$ that is generated by dilations and rotations. We denote the corresponding finite dimensional carrier space by $V_\pi$. Then we can introduce the following linear space $\Gamma_\pi$ of sections in a vector bundle over the quotient space $G/P$ as

$$\Gamma_\pi = \Gamma_\pi(G/P) = \{f : G \to V_\pi \,|\, f(gkn) = \pi(k^{-1})f(g)\} \ . \tag{2.2}$$

Here $n$ denotes an arbitrary element from the subgroup $\mathrm{N}_d$ that is generated by special conformal transformations and $\pi(k^{-1})$ is the finite dimensional 'representation matrix' of the element $k^{-1} \in K$. The linear spaces $\Gamma_\pi$ of sections come equipped with the left regular action of the conformal group. More precisely, given any generator $Y$ of the conformal Lie algebra, one can define a first order differential operator $\mathfrak{y}_Y$ as

$$\mathfrak{y}^\pi_Y f(e^{ix^\mu P_\mu}) := f(Y e^{ix^\mu P_\mu}) \quad \text{for all} \quad f \in \Gamma_\pi \ . \tag{2.3}$$

Here we have represented points $x$ in the quotient space $G/P$ through group elements $\exp(ix^\mu P_\mu)$. The argument of $f$ on the right and side may be considered as an element (in an appropriate completion) of the universal enveloping algebra of the conformal Lie algebra. Smooth functions $f$ can be evaluated on such arguments. For generic values of $\Delta$ the resulting representation of the conformal Lie algebra $\mathfrak{g}$ on $\Gamma_\pi$ is irreducible and infinite dimensional.

Let us use the example of the dilation generator to briefly illustrate the relation between our definition (2.2) and the transformation law (2.1) of primary fields. The defining relations of the conformal Lie algebra $\mathfrak{so}(1, d+1)$ imply that

$$D e^{ix^\mu P_\mu} = e^{ix^\mu P_\mu}(D + ix^\mu P_\mu) \ . \tag{2.4}$$

This simple formula can be used to compute the left regular action $\mathcal{L}$ of the dilation operator $D$ on functions $f \in \Gamma_\pi$,

$$\mathfrak{y}_D f(e^{ix^\mu P_\mu}) = f(D e^{ix^\mu P_\mu}) = (x^\mu \partial_\mu + \Delta) f(e^{ix^\mu P_\mu}) \ . \tag{2.5}$$

In the derivation we used the relation (2.4) to commute $D$ through the exponential. Once $D$ is moved to the right, we can employ the covariance law with respect to right action that is defined in eq. (2.2), along with $\pi(D) = -\Delta$ to arrive at the right hand side. The latter agrees with the right hand side of the first commutation relation in eq. (2.1) for primary fields. The remaining relations that describe the transformation law with respect to rotations and special conformal transformations can be treated similarly.

Having constructed a family of irreducible representations of the conformal algebra, we can evaluate arbitrary elements and in particular the conformal Casimir operators, such as e.g. the quadratic Casimir element,

$$Cas_{(2)} = \kappa_{(2)}^{AB} Y_A Y_B \tag{2.6}$$

Here we have introduced a basis $Y_A$ of elements in the conformal algebra with Killing form $\kappa^{AB}$. If we evaluate this element on $\Gamma_\pi$, where $\pi = (\Delta, \lambda = l)$ involves a symmetric traceless tensor representation of rank $l$, we find

$$C_{\Delta,l} = \Delta(\Delta - d) + l(l + d - 2) \ . \tag{2.7}$$

Note that $\Gamma_\pi$ carries an irreducible representation of the conformal algebra so that the Casimir elements are proportional to the identity by Schur's lemma.

Before closing this section let us briefly comment that the construction of representations we outlined here easily extends to the representations of the conformal group that characterize conformal defects of dimension $p$. All one needs to do is to pick a representation $\pi$ of the transverse rotation group $\mathrm{Spin}(d-p)$ with carrier space $V_\pi$. Then one can form the space $\Gamma_\pi(G/K_p)$ of sections in a vector bundle over the configuration space $G/K_p$ in the same way as we did in eq. (2.2) above. By construction, this space carries an (left) action of the conformal algebra. The form of the generators $\mathfrak{y}$ can be worked out explicitly, just as we did for the action of the dilation operator $D$ on the configuration space $G/P = \mathbb{R}^d$ of a local field above. Let us stress that the resulting representations are highly reducible in general, in contrast to the case of local fields.

# 3 WZNW Models, Blocks and the Critical Level

Before we start discussing global conformal blocks we want to provide some relevant background material from 2D CFT. As we shall discuss below, blocks of the $d$-dimensional conformal group may be regarded as limits of conformal blocks of a 2D WZNW model with the affine Kac-Moody algebra $\widehat{\mathfrak{so}}(1, d+1)_k$. The blocks of this affine algebra, which depend on the level $k$, may be characterized through a set of Knizhnik-Zamolodchikov (KZ) equations. When the level $k$ is sent to a certain critical value $k = -h^\vee$, where $h^\vee$ denotes the dual Coxeter number of the conformal Lie algebra, the KZ equations degenerate to a family of Gaudin integrable models, as was first observed by Feigin, Frenkel and Reshetikhin in [12]. In a way we shall see below, the insertions points of the vertex operators on the 2D worldsheet freeze out in the critical limit and rather than dealing with some infinite dimensional integrable system, we are left with a continuous family of finite dimensional quantum mechanical integrable systems that was first constructed by Gaudin [21]. Below we will see that these integrable systems indeed provide a fruitful view on global blocks in higher dimensional conformal field theory.

## 3.1 A brief introduction to WZNW models

WZNW models represent one of the most important classes of 2-dimensional conformal field theories. Their action involves a group valued field $g(w, \bar{w})$ that is defined on some 2-dimensional Riemann surface $\Sigma$. In this section we assume $\Sigma$ to be a 2-sphere $\Sigma = \mathbb{CP}^1$ and we shall let $g(w, \bar{w})$ take values in the $d$-dimensional conformal group $\mathrm{Spin}(1, d+1)$. In order for the theory to be conformal, one needs to add a topological Wess-Zumino term with some coefficient $k$ that is referred to as the level, see e.g. [22] for a detailed discussion of this theory.

The model is well known to possess an affine current algebra symmetry which is generated by the components $J^A$ of an $\mathfrak{so}(1,d+1)$ valued current $J(w) = \sum_A y^A J_A(w)$, where $y^A$ denote the generators of the conformal algebra in the fundamental matrix representation. These components satisfy the following OPE

$$J_A(z)J_B(w) \sim \frac{k\,\kappa_{AB}}{(z-w)^2} + \frac{if_{AB}^{\ \ C} J_C(w)}{z-w} + \text{regular terms} \tag{3.1}$$

where $k$ is the level of the affine Kac–Moody algebra, $\kappa_{AB}$ is the inverse Killing form, and $f_{AB}^{\ \ C}$ are the structure constants of the conformal algebra in the basis we chose. Vertex operators[2] of the 2D WZNW model satisfy a set of Ward identities that follow from the operator product of expansion (OPE) with the currents

$$J_A(w)V(w') \sim \frac{\mathfrak{y}_A^{(V)}}{w-w'} V(w') + \dots \ . \tag{3.2}$$

Here, $\mathfrak{y}_A^{(V)}$ denote some representation matrices of the conformal algebra that depend on the choice of the vertex operator $V$. We have also suppressed any dependence of the vertex operators on the anti-holomorphic variables $\bar{w}$ - this will be irrelevant for what we are about to discuss. The OPE (3.2) implies that

$$\Big\langle J(w) \prod_{i=1}^{N} V_i(w_i) \Big\rangle = \mathcal{L}(w, w_i) \Big\langle \prod_{i=1}^{N} V_i(w_i) \Big\rangle \tag{3.3}$$

where

$$\mathcal{L}(w, w_i) = \sum_{i=1}^{N} \frac{y^A \mathfrak{y}_A^{(i)}}{w - w_i} = \sum_A y^A \mathcal{L}_A(w, w_i) \tag{3.4}$$

is a family of operators that will play an important role later on. But before we go there, let us also recall that the stress tensor $T$ of the WZNW model can be built through the famous Sugawara construction as

$$T(w) = \frac{1}{k + h^\vee} \sum_{A,B} \kappa^{AB} : J_A(w) J_B(w) : \ . \tag{3.5}$$

The modes of $T$ are known to obey relations of the Virasoro algebra. Correlation functions of conformal primary fields satisfy a set of conformal Ward identities

$$\Big\langle T(w) \prod_{i=1}^{N} V_i(w_i, \bar{w}_i) \Big\rangle = \sum_{i=1}^{N} \left( \frac{h_i}{(w-w_i)^2} + \frac{\partial_i}{w - w_i} \right) \Big\langle \prod_{i=1}^{N} V_i(w_i, \bar{w}_i) \Big\rangle \tag{3.6}$$

where $h_i$ are the conformal weights of the 2D vertex operators $V_i$. If we insert the Sugawara construction on the left hand side of the conformal Ward identity and apply the Ward identity (3.3) we obtain a second expression for the left hand side of eq. (3.6). Comparing the first order poles on both sides of the resulting equation gives

$$\left( \partial_i - \frac{1}{k + h^\vee} \sum_{i \neq j = 1}^{N} \frac{\kappa^{AB} \mathfrak{y}_A^{(i)} \mathfrak{y}_B^{(j)}}{w_i - w_j} \right) \Big\langle \prod_{i=1}^{N} V_i(w_i) \Big\rangle = 0 \ . \tag{3.7}$$

[2]Throughout this work we shall refer to the fields of the 2D CFT as vertex operators to distinguish them from the primary fields $\Phi$ of the higher dimensional CFT.

These are the famous Knizhnik-Zamolodchikov (KZ) equations for the blocks of the WZNW model. There exists extensive solution theory for these equations, see [22] for references.

Before we conclude this subsection let us add one comment concerning the comparison between the characterization (2.1) of primary fields $\Phi$ with respect to global conformal transformations and the OPE identities (3.3) in the WZNW model on the conformal group. Let us suppose that we choose the vertex operator $V$ to transform in a representation $\pi = (\Delta, \lambda)$ of the zero mode algebra so that the operators $\mathfrak{y}_A^{(V)}$ agree with the operators $\mathfrak{y}_{Y_A}$ we defined on eq. (2.1). Then the main difference between the characterization of $\Phi$ and $V$ is that the relations for vertex operators contain the additional dependence on the insertion point $w$ on the sphere $\mathbb{CP}^1$. In order to turn this vague comment into a concrete tool for the study of $d$-dimensional conformal symmetry, we will now explain how one can 'freeze out' the auxilary parameter $w$ and in this sense go from vertex operators $V$ in the WZNW model to primary fields $\Phi$ of the global conformal symmetry.

## 3.2 Critical level limit and Gaudin integrable systems

The main idea of how to 'freeze out' the dependence on the worldsheet insertion points $w_i$ in the WZNW model may be inferred from the KZ equation (3.7). There appear two different types of terms in this equation. The first one is the derivative $\partial_i$ with respect to the insertion point $w_i$. The remainder on the left hand side of eq. (3.7) is local on the worldsheet, i.e. it does not contain any derivatives with respect to $w_i$, and it describes how the vertex operators adjust upon changes of their insertion point $w_i$. To freeze out the $w_i$ we should therefore make the second term dominate over the first. This can be achieved by sending the level $k$ to its critical value $k = -h^\vee$.

Obviously, this limiting process is a bit singular. In order make it precise we introduce a small parameter $\epsilon = k + h^\vee$ and multiply the KZ equation by this parameter to obtain

$$\left(\epsilon\partial_i - \sum_{i\neq j=1}^{N} \frac{\kappa^{AB}\mathfrak{y}_A^{(i)}\mathfrak{y}_B^{(j)}}{w_i - w_j}\right)\Psi(w_i) = 0 \; . \tag{3.8}$$

Here we have written $\Psi(w_i)$ to denote the 2D conformal blocks, i.e. the solutions to the KZ differential equations. In order to study the limiting behavior of these equations as we send $\epsilon$ to zero we look for solutions of the form

$$\Psi(w_i) = e^{\frac{1}{\epsilon}S}\psi(w_i) \tag{3.9}$$

with some scalar function $S$ that depends on the parameters $w_i$. Insertion into the KZ equation then gives

$$H_i\psi(w_i) := \sum_{j\neq i=1}^{N} \frac{\kappa^{AB}\mathfrak{y}_A^{(i)}\mathfrak{y}_B^{(j)}}{w_i - w_j}\psi(w_i) = E_i\psi(w_i) \quad \text{where} \quad E_i = \partial_i S \; . \tag{3.10}$$

Hence we conclude that the factors $\psi$ that multiply the semi-classical factor $\exp(S/\epsilon)$ are joint eigenfunctions of the 'Hamiltonians' $H_i$ that are defined through the expression in the middle of the previous equation. The corresponding eigenvalues $E_i$ are derivatives of the function $S$ that captures the 'semiclassical' dependence of the WZNW block. These eigenvalues depend on the parameters of the setup, e.g. on the choice of vertex operators $V_i$ and their insertion points $w_i$.

The answer we obtained actually has a very beautiful interpretation in terms of integrable models. In order to fully appreciate the relation, let us give a bit of background.

The integrable system in question was first written down out by Gaudin [21]. Roughly speaking, it describes a system of spins $\mathfrak{y}^{(i)}$ that are located at $w_i$ and whose spin-spin interaction decreases with the distance $w_i - w_j$, see the definition of $H_i$ in eq. (3.10). Wave functions $\psi(w_i)$ of the Gaudin model take values in the space of spin configurations of the $N$ spins, i.e. they are functions that map $w_i$ to vectors in the tensor product $\bigotimes_i \Gamma_i$ of representations of the underlying symmetry group. In the most standard setup, the spins would take values in finite dimensional representations of the group SU(2). But nothing really prevents from replacing SU(2)= Spin(3) by the conformal group Spin$(1, d+1)$ in $d = 1$ or higher dimensions. Similarly, spins can also take vales in infinite dimensional representations of the symmetry group and these need not be highest weight representations either. Even principal series representations are permitted.

Gaudin noticed that this setup gives rise to an integrable system that possesses as many commuting Hamiltonians as it has degrees of freedom. If the symmetry group is SU(2), a single spin contributes a single degree of freedom, corresponding to the single spin raising operator in the algebra $\mathfrak{su}(2)$. For the higher dimensional conformal group, this number grows with $d$, e.g. in $d = 3$ a single spin in a generic representations contributes four degrees of freedom. The number reduces to three degrees of freedom for the representations that are associated to scalar primary fields since these admit only three non-trivial raising operators, namely the three translation generators $P_\nu$ of the conformal algebra.

Key to the integrable structure is to realize that the family $\mathcal{L}(w, w_i)$ of matrix values first order differential operators that we read off in eq. (3.4) from insertions of currents in the WZNW model serves as a Lax matrix for the Gaudin model model. Using standard methods from the theory of integrable systems, see e.g. [23] for a introduction to integrable models and references to the original literature, one easily obtains a complete set of commuting operators, including the $N$ commuting Hamiltonians $H_i$ of the Gaudin model we have introduced in eq. (3.10). While the $H_i$ alone would not suffice to make the Gaudin model integrable, we can use the components $\mathcal{L}_A$ to build a collection of Hamiltonians, see also [24],

$$\mathcal{H}_{(p)}\left(\omega, \omega_i\right) = \kappa_{(p)}^{A_1 \cdots A_p} \mathcal{L}_{A_1} \cdots \mathcal{L}_{A_p} + \ldots \tag{3.11}$$

where $\kappa_{(p)}$ is a conformally invariant tensor of degree $p$. The $\ldots$ on the right hand side of the expression represent additional quantum corrections that are absent in the case of $p = 2$ which we shall restrict to in our discussion below. It is indeed possible to show that $\mathcal{H}_{(p)}$ commute, i.e.

$$[\mathcal{H}_{(p)}(w, w_i), \mathcal{H}_{(q)}(w', w_i)] = 0 \tag{3.12}$$

for all $p, q = 2, \ldots$ and $\omega, \omega' \in \mathbb{C}$. Moreover, the Hamiltonians also commute with the symmetry generators,

$$[\mathcal{H}_{(p)}(w, w_i), \mathfrak{y}_A^{(1\cdots N)}] = 0 \tag{3.13}$$

where we introduced

$$\mathfrak{y}_A^{(1\cdots N)} := \sum_{i=1}^{N} \mathfrak{y}_A^{(i)} \tag{3.14}$$

to denote the action of $\mathfrak{y}_A$ on the tensor product of the representations $\pi_i$ with $i = 1, \ldots, N$. This implies that the Hamiltonians can be restricted to act on invariants. Due to their commutativity, the Hamiltonians $\mathcal{H}_{(p)}$ provide us with several families of commuting operators, of which only a finite number is independent. There are many ways to extract these commuting operators, as, for example, taking residues at the insertion points $w_i$ on the 2D worldsheet.

We have pointed out above that there exists extensive literature on solutions of the KZ equations at least when $k \neq -h^\vee$. Unfortunately, many of these do not survive the critical level limit. But this does not imply that the embedding of the Gaudin model into WZNW 2D conformal field theory is useless. As we shall review below, there does exist a duality relation between the WZNW models and Toda theories that survives the limit and establishes a useful relation between the critical level limit of the WZNW model and semi-classical solutions of Toda theory. At least for $G = SO(1,2)$ and Liouville theory this relation is well understood and fully explicit.

# 4 Gaudin Integrability of Global Conformal Blocks

In the previous two sections we surveyed some key concepts in higher dimensional CFT on the one hand and a few basic constructions from 2D WZNW models on the other. Our goal now is to bring these two strands together. In the first subsection, we review the concept of Casimir equations for conformal blocks in the standard example of scalar four-point functions. Then we reinterpret these equations in the context integrable models, first in terms of Calogero-Sutherland models and then in the context of the Gaudin integrable systems we saw at the end of the previous section. The latter approach is more general and it is flexible enough to cover many extensions including defects and correlation functions with more than four external fields. At the same time it also brings in some new tools from 2D conformal field theory and supersymmetric gauge theory. Some of these topics will be outlined in the third subsection.

## 4.1 Casimir equations for four-point conformal blocks

In this section we will briefly review some well known material about conformal blocks for correlation functions of four scalar fields. To be specific, let us consider the vacuum correlation function of four scalar conformal primary fields with weight $\Delta_i$, $i = 1, \ldots, 4$ in a $d$-dimensional conformal field theory[3]

$$G_4(x) = \langle 0|\phi_1(x_1)\ldots\phi_4(x_4)|0\rangle = \sum_{\mathcal{O}} C_{12\mathcal{O}}C_{34\mathcal{O}}\, g_4^{(\Delta,l)}(x) \ . \tag{4.1}$$

On the right hand side, we perform a sum over all primary fields $\mathcal{O}$ and $C_{ij\mathcal{O}}$ denote the three-point couplings of the theory. The functions $g_4$ are the conformal blocks, i.e.

$$\langle 0|\phi_1(x_1)\phi_2(x_2)P_{\pi_{\mathcal{O}}}\phi_3(x_3)\phi_4(x_4)|0\rangle =: C_{12\mathcal{O}}C_{34\mathcal{O}}\, g_4^{(\Delta,l)}(x_i) \ . \tag{4.2}$$

Here $P_{\pi_{\mathcal{O}}}$ denotes the projector to the conformal family of the primary field $\mathcal{O}$. Of course, conformal blocks $g_4$ depend on the choice of external fields, i.e. on the weights $\Delta_i$ in the case of scalar fields. We have suppressed this dependence in our notation for the blocks.

Before we analyze this expansion further, let us note that the vacuum correlation functions obey a set of Ward identities, one for each generator $Y_A$ of the conformal algebra. These identities take the form

$$\sum_{i=1}^{4} \mathfrak{y}_A^{(i)} G_4(x) = 0 \tag{4.3}$$

where $\mathfrak{y}_A^{(i)}$ are the first order differential operators that implement the action of the conformal generator $Y_A$ on the primary field $\phi_i$. In order to determine the blocks $g_4^{(\Delta,l)}(x_i)$,

[3]Here we wrote $\langle 0|\cdot|0\rangle$ for the expectation value instead of merely writing $\langle\cdot\rangle_{\mathbb{R}^d}$ in order to distinguish the correlations functions of the $d$-dimensional CFT more clearly from those of the 2D auxiliary model.

Dolan and Osborn wrote down a set of Casimir differential equations. For the quadratic Casimir element (2.6) these take the form

$$Cas^{(12)}_{(2)} g_4^{(\Delta,l)}(x_k) = \sum_{i,j=1,2} \kappa^{AB}_{(2)} \mathfrak{y}^{(i)}_A \mathfrak{y}^{(j)}_B g_4^{(\Delta,l)}(x_k) = C_{\Delta,l} g_4^{(\Delta,l)}(x_k) \tag{4.4}$$

where $C_{\Delta,l}$ are the eigenvalues given in eq. (2.7). It is not difficult to give an explicit formula for this differential operator. In order to do so, we introduce the usual cross ratios

$$\frac{x_{12}^2 x_{34}^2}{x_{13}^2 x_{24}^2} = z\bar{z} \quad , \quad \frac{x_{14}^2 x_{23}^2}{x_{13}^2 x_{24}^2} = (1-z)(1-\bar{z}) \tag{4.5}$$

with $x_{ij} = x_i - x_j$ and redefine the blocks $g_4(x_i)$ as

$$g^{(\Delta,l)}(z,\bar{z}) =: x_{12}^{\frac{1}{2}(\Delta_1+\Delta_2)} x_{34}^{\frac{1}{2}(\Delta_3+\Delta_4)} \left(\frac{x_{24}}{x_{14}}\right)^a \left(\frac{x_{13}}{x_{14}}\right)^b g_4^{(\Delta,l)}(x_i) \tag{4.6}$$

where we also introduced $2a = \Delta_2 - \Delta_1, 2b = \Delta_3 - \Delta_4$. The Ward identities (4.3) imply that the resulting function depends on the cross ratios only. The action of the second order Casimir operator on the blocks $g^{(\Delta,l)}(z,\bar{z})$ takes the form [8],

$$Cas^{(12)}_{(2)} := D^2 + \overline{D}^2 + (d-2)\left[\frac{z\bar{z}}{\bar{z}-z}\left(\bar{\partial}-\partial\right) + (z^2\partial - \bar{z}^2\bar{\partial})\right] \tag{4.7}$$

where

$$D^2 = z^2(1-z)\partial^2 - (a+b+1)z^2\partial \; - abz \; . \tag{4.8}$$

The operator $\overline{D}^2$ is defined similarly in terms of $\bar{z}$. In $d=2$ dimensions the Hamiltonian splits into a sum of two independent pieces and the corresponding eigenvalue equations are straightforwardly related to hypergeometric differential equations. By writing down the equations (4.4), Dolan and Osborn turned the problem of constructing conformal blocks into the problem of solving some partial differential equations. They actually were able to solve the latter through a series expansion and derive many useful properties [9]. In three dimensions, a program that computes series expansions of blocks systematically has been developed, [25].

The way we have written the Casimir differential equations may look a bit uninviting. It was observed in [6], however, that they can be brought into the form of an integrable Schroedinger eigenvalue problem. In order to do so, one needs to change coordinates such that

$$z_i = -\sinh^{-2}\frac{\tau_i}{2} \tag{4.9}$$

and the blocks as

$$\psi^{(\Delta,l)}(\tau_1,\tau_2) := \prod_{i=1}^{2} \frac{(z_i-1)^{\frac{a+b}{2}+\frac{1}{4}}}{z_i^{\frac{1}{2}+\frac{\epsilon}{2}}} |z_1 - z_2|^{\frac{\epsilon}{2}} g_4^{(\Delta,l)}(z_1,z_2) \tag{4.10}$$

where $\epsilon = d-2$ and $a,b$ are defined as before. Now it is easy verify that if $g(z,\bar{z})$ satisfies the Casimir equation of Dolan and Osborn then the associated function $\psi(\tau_1,\tau_2)$ is an eigenfunction of a 2-particle Schroedinger problem with the potential

$$V^{(a,b,\epsilon)}_{\text{CS}}(\tau_1,\tau_2) = \sum_{i=1}^{2}\left[\frac{(a+b)^2-\frac{1}{4}}{\sinh^2\tau_i} - \frac{ab}{\sinh^2(\tau_i/2)}\right] + \frac{\epsilon(\epsilon-2)}{8\sinh^2\frac{\tau_1-\tau_2}{2}} + \frac{\epsilon(\epsilon-2)}{8\sinh^2\frac{\tau_1+\tau_2}{2}} \; . \tag{4.11}$$

The first two terms provide a 1-dimensional single particle potential while the other two describe an interaction between the two particles. We see that the dependence of blocks on the weights of the external fields enters through the parameters of the single particle potentials while the interaction depends on the dimension $d$. In $d = 2, 4$ the interaction vanishes and the resulting quantum mechanical problem factorizes into two non-interacting Poeschl-Teller particles whose wave functions are given by Gauss' hypergeometric function. The interacting case was first proposed by Calogero, Moser and Sutherland in the early 1970s, see [26–28]. Just as the 1-dimensional Poeschl-Teller problem is exactly solvable, so are the higher-dimensional Calogero-Sutherland extensions and hence, by the relation (4.10), the eigenvalue problem for the conformal Laplacian. Much of the solution theory for this Schroedinger problem was developed starting with the seminal work of Heckman and Opdam, see [29] for results in the context of conformal blocks and further references, long before Dolan and Osborn studied the Casimir equations for scalar four-point blocks.

## 4.2 Casimir equations from Gaudin integrable models

The Calogero-Sutherland problem we discussed at the end of the previous sections is actually known to be (super-)integrable. It turns out that integrability is a rather generic feature in the theory of conformal blocks, as we shall see below. There are several ways to approach the integrability of Calogero-Sutherland models. The most universal one that applies to many of the extensions passes through the relation with Gaudin integrable models. Before we go there, we want to slightly enrich the discussion of Casimir differential equations by considering a higher number $N > 4$ of field insertions, i.e. we shall consider correlation functions of the form

$$G_N(x_i) = \langle 0|\phi_1(x_1)\dots\phi_r(x_r)\phi_{r+1}(x_{r+1})\dots\phi_N(x_N)|0\rangle \ . \tag{4.12}$$

It is straightforward to extend our discussion in the previous section to this setting by inserting projectors $P_{\pi_r}$ in between the fields $\phi_r$ and $\phi_{r+1}$. Then we obtain Casimir equations that involve the differential operators

$$Cas^{(1\cdots r)}_{(2)} = \kappa^{AB}_{(2)}\mathfrak{y}^{(1\cdots r)}_A\mathfrak{y}^{(1\cdots r)}_B = \sum_{i,j=1}^{r}\kappa^{AB}_{(2)}\mathfrak{y}^{(i)}_A\mathfrak{y}^{(j)}_B \ . \tag{4.13}$$

where we use the notation for the action on tensor products that we introduced in eq. (3.14). Obviously, the same constructions also apply to higher Casimir operators. Let us note that $N$-point functions with $N > 4$ possess more cross ratios than Casimir differential equations, see e.g. [13] for a detailed discussion. Hence Casimir differential equations do not suffice to characterize multi-point blocks with $N > 4$. We will comment on this below. But let us now first discuss the relation with Gaudin models.

Recall that the Gaudin Hamiltonians are constructed from the Lax connection. The latter gives rise to a multi-parameter family of integrable models that are parametrized by the positions $w_i$ of the punctures on the sphere. The claim is that there exists a limiting configuration of these parameters in which the Gaudin integrable model coincides with the Calogero-Sutherland model and its multi-point extensions. In order to see this, we choose $w_i = \varepsilon^{N-i-1}$ and then send $\varepsilon$ to zero. As we do so, $w_N \to \infty, w_{N-1} \to 1$ and all other points tend to zero but they do so in a hierarchical way. Let us investigate how the Lax connection behaves as we take this limit. If we also rescale the spectral parameter $w$ and the Lax connection, we can actually define $N - 2$ different connections of the form

$$\mathcal{L}^{[r]}(w) := \lim_{\varepsilon\to 0}\varepsilon^{N-r-2}\mathcal{L}(\varepsilon^{N-r-2}w, w_i = \varepsilon^{N-i-1}) = \frac{\sum_A y^A\mathfrak{y}^{(1\cdots r)}_A}{w} + \frac{\sum_A y^A\mathfrak{y}^{(r+1)}_A}{w-1} \tag{4.14}$$

where $r = 1, \ldots, N-2$. It now follows from our general discussion of the Gaudin model and its integrability that

$$\mathcal{H}^{[r]}_{(p)}(w) = \kappa^{A_1 \cdots A_p}_{(p)} \mathcal{L}^{[r]}_{A_1} \cdots \mathcal{L}^{[r]}_{A_p} + \ldots \tag{4.15}$$

pairwise commute for all choices of $p, r$ and the spectral parameter $w$. One can also show that in taking the limit, the number of independent commuting functions is not reduced, i.e. the limit of the Gaudin models remains integrable. The key observation that establishes the relation between conformal blocks and our limiting Gaudin models is that we can actually find all the Casimir operators within the family of commuting operators we constructed,

$$Cas^{(1\cdots r)}_{(p)} = \lim_{w \to 0} w^p \mathcal{H}^{[r]}_{(p)}(w) \ , \tag{4.16}$$

as one can easily see from the explicit expression for the limit (4.14). Indeed, we note that the components $\mathcal{L}_A$ of the Lax connections (4.14) we obtained in the limit are meromorphic functions that take values in first order differential operators with first order poles at $w = 0, 1$. Consequently, the Hamiltonians (4.15) are meromorphic functions that take values in the space of differential operators of order $p$ with singularities of order at most $p$ at $w = 0, 1$. The limit performed on the left hand side of eq. (4.16) singles out the contributions from the first term in the connections (4.14) which involve the action $\mathfrak{y}^{(1\cdots r)}_A$ of the conformal generators on the first $r$ field insertions. Eq. (4.16) then follows immediately from the definition (4.13). Thereby we have now demonstrated that conformal blocks are indeed wave functions of Gaudin integrable models in a certain limit in which the punctures of the Gaudin model are sent to $w - 0, 1, \infty$. This limit is also known as the caterpillar limit since it sends the Lax connection on an $N$-punctured sphere to a chain of $N-2$ Lax connections on 3-punctured spheres, see Figure 2.

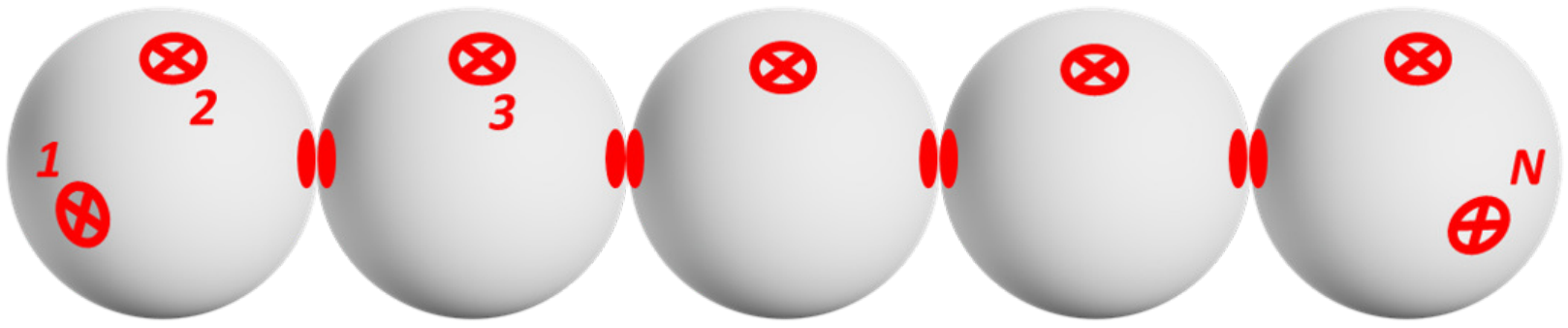

Figure 2: When we perform the OPE limit to pass from Gaudin wave functions to conformal blocks in the so-called comb-channel, the $N$-punctured sphere degenerates to chain of three-punctured spheres. This limit is also known as the caterpillar limit of the Gaudin model.

As simple as these observations may seem, they do have some remarkable consequences [11]. Here we have demonstrated that we can recover the Casimir operators for so-called comb channel blocks from the Gaudin model. The analysis can be extended to include all other channels by considering appropriate limits of the original Gaudin model, see [13]. Moreover, as we commented above, Casimir operators alone do not suffice to characterize conformal blocks for more than $N = 4$ external fields. This may also be seen from the derivation of eq. (4.16) we sketched in the previous paragraph. In fact, the derivation only used the first term of the connection (4.14) which clearly cannot suffice to generate all commuting operators. On the other hand, the Gaudin model does provide us with additional independent commuting operators, as many as there are cross ratios. These are built with contributions from the second term of the term in eq. (4.14). In [30, 31] a particular set of such additional commuting quantities was introduced and studied - these were dubbed *vertex differential operators*.

## 4.3 Conformal blocks from 2D conformal field theory

In the previous subsection we have shown that conformal blocks can be regarded as wave functions of an integrable Gaudin model in a certain limit where the punctures of the Gaudin model collide. In our discussion of the WZNW model on the conformal group we saw that the Gaudin model emerges at the critical level. Putting these two observations together, we conclude that global conformal blocks in a $d$-dimensional CFT may be regarded as limits of affine blocks in a 2D WZNW model. While this seems like an amusing fact, it is not immediately clear that it is any useful, even more so as integral representations for WZNW models are known to be singular when the level is sent to the critical value. The purpose of this short section is to show that some remarkable properties of affine blocks do survive the critical level limit and provide interesting insight into global blocks. We will discuss this here for the special case of $d = 1$, which is best understood. As we shall argue in the concluding section, however, the basic ideas turn out to apply to higher dimensions $d$ and may be turned into concrete construction of blocks, at least for scalar external fields.

Our main goal is to argue that (scalar) conformal blocks may be determined by studying the semiclassical behavior of correlators in certain 2D Toda field theories or, in the case of $d = 1$, in Liouville field theory. The latter case relies on a very remarkable relation between the SL(2) WZNW model and Liouville theory that was uncovered by Ribault and Teschner in [14], through a comparison of singular vector equations. An alternative path integral derivation of the correspondence was given in [16]. The map relates an $N$-point functions in the WZNW model to correlators with $2N - 2$ insertions in Liouville theory, the $N - 2$ additional insertions being Virasoro-degenerate fields. Let us be a little bit more explicit.

The vertex operators in the WZNW model reside in representations that are acted upon by the zero mode algebra $\mathfrak{so}(1, d+1) \cong \mathfrak{so}(1, 2)$ of the currents $J^A$. As we reviewed in section 2, representations of the conformal algebra can be realized in a space (2.2) of functions on $\mathbb{R}^d \cong \mathbb{R}$ that depends on $\pi = (\Delta, \lambda)$. For scalar representations, the only parameter is the conformal weight $\Delta$. Hence, to any function $f \in \Gamma_\Delta$, we can associate a vertex operator $V_\Delta(f|w)$ of the WZNW model. At this point it is advantageous to apply a Fourier transform, i.e. to consider a basis of eigenfunctions $f_p$ of the translations in the space $\Gamma_\Delta$. The corresponding vertex operators of the WZNW model will be denoted by $V_\Delta(p|w)$, where $p \in \mathbb{R}^d = \mathbb{R}$ is a momentum parametrizing states in $\Gamma_\Delta$ and $\Delta \in \mathbb{R}$ is a scaling weight of the representation. The argument $w$ continues to denote the holomorphic coordinate of the insertion point on the 2D worldsheet. Let us note that $\Delta$ is not to be confused with the 2D conformal weight $h$ of the vertex operator, though the two quantities are related as $h = \Delta(\Delta - 1)$.

Now let us pick $N$ such vertex operators that are associated with weights $\Delta_i$ and denote the momenta by $p_i$. Given the set of these momenta, one can compute the $N - 2$ insertion points $y_r, r = 1, \dots, N - 2$ of the degenerate fields through the relation

$$\sum_{i=1}^{N} \frac{p_i}{w - w_i} = u \frac{\prod_{r=1}^{N-2}(w - y_r)}{\prod_{i=1}^{N}(w - w_i)} \ . \tag{4.17}$$

Let us note that the left hand side does not possess a pole at infinity provided the total momentum is conserved. The remaining $N-1$ momentum parameters suffice to determine the $N - 2$ variables $y_r$ and the prefactor $u$ on the right hand side. Up to some relatively simple prefactors, the relation between the correlation functions in the two theories takes the following remarkably simple form

$$\Big\langle \prod_{i=1}^{N} V_{\Delta_i}(p_i|w_i) \Big\rangle^{\text{WZNW}} \sim \delta(\sum_i p_i) \Big\langle \prod_{i=1}^{N} e^{2\alpha_i \varphi(w_i)} \prod_{r=1}^{N-2} e^{-\frac{1}{b}\varphi(y_r)} \Big\rangle^{\text{LFT}} \tag{4.18}$$

where $\varphi$ denotes the Liouville field and the Liouville 'momenta' $\alpha_i$ are related to the weight parameters $\Delta_i$ of the vertex operators in the WZNW model by $\alpha_i = b(1-\Delta_i) + 1/2b$. Finally, the level $k$ of the WZNW model is related to the parameter $b$ through $b^2 = 1/(k-2)$. This parameter $b$ enters the formula $Q = b + 1/b$ for the background charge of the Liouville theory and hence determines the Virasoro central charge $c_{\text{LFT}} = 1 + 6Q^2$.

This correspondence has many remarkable implications. The most relevant feature for us is that it does survive the limit to the critical level of the WZNW model as was first stressed in [14]. In the dual Liouville theory, the associated limit is the semiclassical one in which the parameter $b$ and hence the central charge of Liouville field theory is sent to infinity.[4] In the limit, the $N$ fields that are associated with the original fields of the WZNW model become heavy while the new degenerate field insertions remain light. Hence, the correspondence (4.18) becomes

$$\hat{\psi}(p_i; w_i) \sim \delta(\sum_i p_i)\, e^{-b^2 S_{\text{Liouville}}(\varphi_{\text{cl}})} \prod_{r=1}^{N-2} e^{-\frac{1}{2}\varphi_{\text{cl}}(y_r)} \;. \tag{4.19}$$

Here, on the the left hand side $\hat{\psi}(p_i)$ denotes the Fourier transform of the Gaudin wave function that emerges from the WZNW correlator in the critical level limit. The other side is a rough sketch of the semiclassical correlation in Liouville field theory. In order to compute the Liouville correlator one has to solve the classical Liouville equation of motion in the presence of $N$ heavy field insertions. This problem leads to a Fuchsian differential equation for $\varphi_{\text{cl}}$ with $N$ regular singularities at $w_i$. For $N = 3$ the equation can be solved in terms of hypergeometric functions ${}_2F_1$. The case of $N = 4$ is associated with Heun functions and $N > 5$ some generalizations thereof. In conclusion, we can compute the right hand side of the correspondence (4.19) and hence the wave functions of the Gaudin model by solving some Fuchsian differential equation. On this solution one needs to evaluate the classical Liouville action and multiply with the product of exponentials $\exp(-\varphi_{\text{cl}}/2)$ evaluated at the points $y_r$. Note that on the right hand side, the dependence of the insertions points $y_r = y_r(p)$ is fully factorized, i.e. the correspondence (4.19) realizes the separation of variables for the Gaudin model.

In practice, explicit solutions to the Fuchsian differential equation are not known for $N > 3$ so that the formula (4.19) for the wave functions of the Gaudin integrable model is not that explicit after all. But at this point we recall that we are actually not that interested in general eigenfunctions of the Gaudin Hamiltonians but rather in global conformal blocks that arise in a special limit in which we set $w_i = \varepsilon^{N-i-1}$ and send $\varepsilon$ to zero. If we apply this limit to our formulas (4.17) we find

$$y_r = \varepsilon^{N-r-2} \frac{\sum_{i=1}^{r} p_i}{\sum_{i=1}^{r+1} p_i} + \dots \tag{4.20}$$

This relation shows that the insertion points $y_r$ are also sent to zero, at a rate that is the same as it is for the insertion points $w_{i+1}$. As a consequence, the Fuchsian differential equation we started with splits into $N-2$ hypergeometric differential equations, one for each of the 3-punctured spheres in Figure 2. Hence, after we apply our limiting procedure to pass from wave functions of Gaudin models to conformal blocks, the correspondence (4.19) becomes fully explicit and it represents the Fourier transform of conformal blocks for $d = 1$ as an $(N-2)$-fold product of hypergeometric functions. The existence of such a formula was indeed observed before, see e.g. eq. (2.36) in [32] for the special case of four-point blocks, as well as [33]. It is also consistent with the corresponding position space results [34]. Here we provided a new view on such factorized expressions for momentum space blocks that suggests how to extend them systematically.

[4] Recall that quantum Liouville field theory is self-dual under the replacement $b \to 1/b$ and hence sending $b$ to infinity is equivalent to sending it to zero.

With the understanding of conformal blocks we have reached through the previous sections it is now easy to address various relevant extensions. In particular, with just one simple adjustment in the underlying Lax connection of the Gaudin model it is possible to obtain the Casimir equations for correlation functions involving various non-local insertions such as interfaces or defects of arbitrary dimension, such as line and surface operators.

# 5 Conclusions, Discussion and Outlook

In this paper we have reviewed the integrability approach to conformal blocks in higher dimensional conformal field theories. For $N$-point correlation functions of local fields, conformal blocks are known to be wave functions of a Gaudin integrable model on the $N$-punctured sphere $\mathbb{CP}^1$ in a limit in which the punctures are sent to the three points $w = 0, 1, \infty$ on the sphere. This relation between integrable systems and global conformal blocks is much more general. In particular, it extends to non-local insertions such as defects or interfaces, see e.g. [10] and as well as thermal correlation functions.

The extension of the zero temperature case we discussed throughout this text to thermal correlations was included in my presentation at the Itzykson meeting already. One can indeed show that integrability does not melt when we expose the theory to finite temperature. To be more precise, conformal blocks for correlation functions of local fields in the thermal geometry $S^1 \times S^{d-1}$ turn out to be wave functions of integrable Hitchin systems on a torus. The main difference between the zero temperature and the thermal case is that the sphere $\mathbb{CP}^1$ needs to be replaced by a complex torus $\mathbb{T}$ with modular parameter $\tau$. The associated integrable higher genus extension of the Gaudin model is known as (elliptic) Hitchin integrable model. Once again, to make contact with thermal blocks, one needs to consider a certain limiting configuration of the insertions points on the torus and of the torus modular parameter $\tau$. This will be explained in a forthcoming publication [35].

A new angle that we developed in these notes is the relation with 2D conformal field theory. This relies on the well-known fact that Gaudin and more general Hitchin integrable systems emerge from the Knizhnik-Zamolodchikov(-Bernard) equations for blocks of an affine Kac-Moody algebra at the critical level $k = -h^\vee$. This embedding of higher dimensional conformal blocks into the world of 2D WZNW models can be turned into a powerful new tool if combined with dualities between 2D conformal field theories. We illustrated that here mostly for the simplest case, namely conformal blocks in $d = 1$ dimension and the associated $\mathrm{Spin}(1, 2)$ WZNW model. The latter is known to be dual to Liouville field theory with the critical level corresponding to a semi-classical limit of Liouville field theory. Taken together, these statements imply that $d = 1$ conformal blocks can be constructed by solving the classical Liouville equation of motion in the presence of external sources. This leads to interesting factorized expressions for conformal blocks in momentum space. Let us stress that all this remains true at finite temperature. In particular, KZ-like equations for affine blocks of WZNW models on the torus were proposed by Bernard [36]. Also the duality between WZNW model and Liouville theory can be extended to higher genus, see [16]. The application of these ingredients to the theory of thermal blocks will be worked out in [35].

All of the above is true for conformal blocks in arbitrary dimensions $d$, only that one needs to replace Liouville field theory by some appropriate multi-field extension, i.e. some Toda-like field theory. The precise duality relation needs to be worked out. For generic field insertions in the $\text{Spin}(1, d+1)$ WZNW model this is a challenging task, see [37] however for some promising observations concerning WZNW models with target $\text{SU}(N)$. The main challenge in going from $d = 1$ to $d > 1$ is that the off-diagonal raising operators in the $d$-dimensional conformal group no longer commute with each other and hence they cannot be diagonalized simultaneously. This diagonalization was crucial in our discussion of $d = 1$ conformal blocks above and briefly visible when we passed from coordinate to momentum space. But for special field insertions in the WZNW model, the situation simplifies. In particular, for scalar representations of the conformal group the only relevant raising operators are the $d$ momentum generators which can be diagonalized simultaneously through Fourier transform, as in the case of $d = 1$. Hence, it should not be too difficult to derive a WZNW-Toda duality for such a restricted class of WZNW correlators, using the path integral approach of [16]. Proceeding along these lines one can then obtain product formulas for conformal blocks with $N$ external scalar field insertions in any dimension $d$. The individual factors are determined by some multi-variable version of the ordinary hypergeometric differential equation. This equation can be obtained by studying the classical equations of motion in the Toda theory that emerges from the duality with $\text{Spin}(1, d+1)$ WZNW models. Let us note that the equation is expected to simplify drastically when we pass to the limiting configuration of the 2D auxiliary surface, i.e. the punctured sphere or torus. We saw such simplifications at work in $d = 1$ when we noted the degeneration of Heun and Lamé to a hypergeometric differential equation. We will report on concrete formulas elsewhere.

There are many interesting further extensions of which we only want to briefly mention two. One of them involves going to higher genus. Most of the 2D ingredients of our story extend beyond the sphere and torus to auxiliary surfaces of arbitrary genus, see in particular [16]. It would be interesting to look into applications to higher dimensional conformal field theory. New blocks appear e.g. when we place a CFT on more general surfaces. A concrete example has been discussed in [38] and used to obtain new CFT data. In this context it would be interesting to see whether the relevant blocks are associated with a Hitchin system on an auxiliary surface of genus $g = 2$. A second direction that we have briefly touched upon concerns the inclusion of boundaries, defects and interfaces. In some important instances, the integrability approach to conformal blocks has already been developed for such blocks, see in particular [10]. It is actually quite straightforward to write down Gaudin and Hitchin integrable models for much more general configurations of local and non-local operators. All one needs to do is to replace the first order differential operators $\mathfrak{y}_A$ we used in our Lax connections above by the more general differential operators that describe the transformation of defect operators under conformal transformations, see our comments at the end of section 2. It would be interesting to understand whether such a more general class of Gaudin integrable models can still be embedded into 2D WZNW models, e.g. by adding defects in on the 2D auxiliary worldsheet. If that can be done, it should also be possible to pass the description of $d$-dimensional defect blocks through the 2D duality with Liouville and more generally Toda field theory. We plan to return to these questions in the future.

**Acknowledgements:** I would like to thank Till Bargheer, Ilija Buric, Yasuaki Hikida, Mikhail Isachenkov, Sylvain Lacroix, Francesco Mangialardi, Jeremy Mann, Lorenzo Quintavalle, Francesco Russo, Paul Ryan, Sylvain Ribault, Joerg Teschner and Alessandro Vichi for many discussions and collaboration on various aspects of the program described above. I am also grateful to the organizers and participants of the rencontre Itzykson for questions and feedback. The research that was reviewed here was funded by the German Research Foundation DFG – SFB 1624 – "Higher structures, moduli spaces and integrability" – 506632645 and also under Germany's Excellence Strategy - EXC 2121 Quantum Universe - 390833306.